\documentclass{aa}  
\usepackage{graphicx}
\usepackage{txfonts}
\usepackage{amssymb}
\usepackage{mathtools}
\usepackage{color}
\usepackage{siunitx}
\usepackage{longtable}

\usepackage[dvipsnames]{xcolor}
\definecolor{Rh}{HTML}{228B22}

\usepackage[flushleft]{threeparttable}

\begin{document} 

\title{Ultra diffuse galaxies in the MATLAS low-to-moderate density fields}

\author{ Francine R. Marleau\inst{1}, Rebecca Habas\inst{2}, M\'elina Poulain\inst{1}, Pierre-Alain Duc\inst{2}, Oliver M\"uller\inst{2}, Sungsoon Lim\inst{3}, Patrick R. Durrell\inst{4}, Rubén Sánchez-Janssen\inst{5}, Sanjaya Paudel\inst{3}, Syeda Lammim Ahad\inst{6}, Abhishek Chougule\inst{7,8}, Michal B{\'i}lek\inst{9}, J{\'e}r{\'e}my Fensch\inst{10} }
   
\authorrunning{Marleau et al.}

\institute{
Institut f{\"u}r Astro- und Teilchenphysik, Universit{\"a}t Innsbruck, Technikerstra{\ss}e 25/8, Innsbruck, A-6020, Austria
\and
Observatoire Astronomique de Strasbourg  (ObAS), Universite de Strasbourg - CNRS, UMR 7550 Strasbourg, France
\and
Department of Astronomy, Yonsei University, 50 Yonsei-ro Seodaemun-gu, Seoul, 03722, Republic of Korea
\and
Youngstown State University, One University Plaza, Youngstown, OH 44555, USA
\and
UK Astronomy Technology Centre, Royal Observatory, Blackford Hill, Edinburgh, EH9 3HJ, UK
\and
Leiden Observatory, Leiden University, P.O. Box 9513, 2300 RA Leiden, The Netherlands
\and
Instituto de Astrofísica e Ciências do Espaço, Universidade do Porto, CAUP, Rua das Estrelas, 4150-762 Porto, Portugal
\and
Departamento de Física e Astronomia, Faculdade de Ciências, Universidade do Porto, Rua do Campo Alegre 687, 4169-007 Porto, Portugal
\and
Nicolaus Copernicus Astronomical Center, Polish Academy of Sciences, Bartycka 18, 00-716 Warsaw, Poland
\and
Univ. Lyon, ENS de Lyon, Univ. Lyon 1, CNRS, Centre de Recherche Astrophysique de Lyon, UMR5574, 69007 Lyon, France
}

\date{Received May 31, 2021; accepted September 15, 2021}

 
\abstract
{
Recent advances in deep dedicated imaging surveys over the past decade have uncovered a surprisingly large number of extremely faint low surface brightness galaxies with large physical sizes called ultra diffuse galaxies (UDGs) in clusters and, more recently, in lower density environments. As part of the Mass Assembly of early-Type GaLAxies with their fine Structures (MATLAS) survey, a deep imaging large program at the Canada-France-Hawaii Telescope (CFHT), our team has identified 2210 dwarf galaxies, 59 ($\sim$3\%) of which qualify as UDGs. Averaging over the survey area, we find $\sim$0.4 UDG per square degree. They are found in a range of low to moderate density environments, although 61\% of the sample fall within the virial radii of groups. Based on a detailed analysis of their photometric and structural properties, we find that the MATLAS UDGs do not show significant differences from the traditional dwarfs, except from the predefined size and surface brightness cut. Their median color is as red as the one measured in galaxy clusters, albeit with a narrower color range. The majority of the UDGs are visually classified as dwarf ellipticals with log stellar masses of $\sim 6.5-8.7$. The fraction of nucleated UDGs ($\sim$34\%) is roughly the same as the nucleated fraction of the traditional dwarfs. Only five ($\sim$8\%) UDGs show signs of tidal disruption and only two are tidal dwarf galaxy candidates. A study of globular cluster (GC) candidates selected in the CFHT images finds no evidence of a higher GC specific frequency $S_N$ for UDGs than for classical dwarfs, contrary to what is found in most clusters. The UDG halo-to-stellar mass ratio distribution, as estimated from the GC counts, peaks at roughly the same value as for the traditional dwarfs, but spans the smaller range of $\sim 10-2000$. We interpret these results to mean that the large majority of the field-to-group UDGs do not have a different formation scenario than traditional dwarfs.
}


\maketitle
%

\section{Introduction}

Recent advances in deep dedicated imaging surveys over the past decade have uncovered a surprisingly large number of extremely faint low surface brightness (LSB) galaxies with large physical sizes referred to as ultra diffuse galaxies (UDGs). Although the name was branded recently \citep{vanDokkum2015}, large LSB galaxies have been observed over three decades (e.g., \citealt{Sandage1984,Binggeli1985,Impey1988,Schombert1988,Schwartzenberg1995,Dalcanton1997,Sprayberry1997}). UDGs are defined as having a low central surface brightness, with typical values in the \textit{g}-band on the order of $\mu_{0,g} = 24 - 26$~mag/arcsec$^2$, but unlike the ultra faint dwarf galaxies, they have a large physical size, with effective radii $R_e > $ 1.5~kpc \citep{vanDokkum2015}. Andromeda XIX, a satellite of M31, is an example of a Local Group (LG) UDG with a half-light radius of 1.7~kpc \citep{McConnachie2008}. 

Almost all UDGs discovered until recently resided in dense cluster environments, including the central regions \citep{Koda2015} and outskirts \citep{Kadowaki2017,vanDokkum2015} of the Coma cluster, the Fornax cluster \citep{Munoz2015,Venhola2017}, and other clusters (e.g., \citealt{Mihos2015,vanderBurg2016,Roman2017a}). At the distance of their respective clusters, their estimated physical sizes are $1.5 < R_e < 4.6$~kpc, with the largest ones rivaling the size of the Milky Way ($R_{e,MW} \sim$ 3.6~kpc). For comparison, the typical dwarf galaxies in the same luminosity range have effective radii of a few hundred parsecs \citep{Mueller2019,Chiboucas2013}. Observations indicate that the UDGs found in rich galaxy clusters \citep{vanDokkum2015,Gu2018} are typically old, red ($g-i \sim 0.76$), gas and metal poor, relatively round and featureless, although not all cluster UDGs are devoid of gas and sources of ionizing radiation (e.g., \citealt{Kadowaki2017}). Based on their globular cluster (GC) population, they have a wide range of physical properties with many showing low-mass (dwarf) haloes. However, the most discussed cases have been the ``dark matter deficient'' \citetext{NGC~1052-DF2 and NGC~1052-DF4, \citealp{vanDokkum2018a,vanDokkum2018b,vanDokkum2019,Emsellem2019,Fensch2019a}} and ``massive Milky Way like'' UDGs \citetext{Dragonfly 44, \citealp{vanDokkum2016}; \citealp{Beasley2016b}; \citealp{Toloba2018}} that do not follow the stellar mass-halo mass relationship predicted by the $\Lambda$ cold dark matter ($\Lambda$CDM) cosmological model \citep{Behroozi2013}. A recent report of the lack of X-ray detection suggests that Dragonfly 44 and DF X1 are not failed $\rm L_{\star}$ galaxies \citep{Bogdan2020}. 

The GCs themselves in some UDGs also appear to be too luminous, with them being as bright or brighter than Omega Centauri \citep{vanDokkum2018a,Shen2021}. The recent analysis of the GC population in the UDG MATLAS-2019, in the NGC~5846 group, using deep --- sampling most of the globular cluster luminosity function (GCLF) --- high resolution Hubble Space Telescope (HST) data \citep{Mueller2021} has revealed that the brightest GCs in this UDG are consistent with the normal bright end of the GCLF at the group distance, indicating there are no over-luminous GCs present in this galaxy, contrary to what is found for NGC~1052-DF2 and -DF4 \citep{Danieli2019,Shen2021}. Rather, the GCLF is best explained by a large population of GCs for a galaxy of this luminosity. Recent results for 33 UDGs in the Coma cluster based on HST observations similarly suggest that UDGs can reach a high specific frequency \citetext{$0<S_N<50$, \citealp{Lim2018}}, with a mean $S_N$ higher for UDGs than for classical dwarfs. In this cluster, \citet{Forbes2020} find that the rich GC systems tend to be hosted in UDGs of a lower luminosity, smaller size, and fainter surface brightness, with a similar trend for the normal dwarfs. The GC systems of Virgo UDGs have a wide range in specific frequency, with a higher mean $S_N$ than normal Virgo dwarfs, but a lower mean $S_N$ than Coma UDGs at a fixed luminosity \citep{Lim2020}. In the Fornax cluster, \citet{Prole2019a} find very few UDGs with a high $S_N$. Rather their GC numbers are consistent with those of other dwarf galaxies of a comparable luminosity. The different abundances of GCs, especially the demarcation between UDGs with and without an excess of GCs as compared to the normal dwarfs, may point to different origins.

Recently, a growing number of UDGs have been identified in nearby groups using blind optical surveys (\citealt{Merritt2016,Bennet2017} in the M101 group with the Dragonfly Telephoto Array and as part of the Canada-France-Hawaii Telescope Legacy Survey (CFHTLS), \citet{Roman2017b} in the Sloan Digital Sky Survey (SDSS) Stripe 82, \citet{Greco2018} in the Hyper Suprime-Cam Subaru Strategic Program (HSC-SSP), \citet{Forbes2019,Forbes2020} in the VST Early-type GAlaxy Survey (VEGAS)) and blind HI surveys (\citealt{Du2015} using ALFALFA+SDSS, \citealt{Leisman2017} using ALFALFA+SDSS,WIYN), as well as in low density environments such as the Pisces-Perseus supercluster (DGSAT I, \citealt{Martinez-Delgado2016}), the MATLAS (Mass Assembly of early-Type GaLAxies with their fine Structures) survey \citep{Habas2020}, the SDSS Stripe 82 \citep{Barbosa2020} and other isolated spiral and early-type spiral galaxies \citep{Roman2019,Mueller2018,For2019,Crnojevic2014}. The UDGs found in groups are typically blue ($g-i \sim 0.45$) and irregular \citep{Roman2017b}. The UDGs found in the low to moderate density environments of the MATLAS survey \citep{Habas2020} do not appear to form a distinct group in the scaling relations but are simply an extension of the dwarf population toward larger radii and fainter surface brightnesses. X-Ray observations of multiple UDGs \citep{Kovacs2019} also support the fact that the majority of UDGs are consistent with being normal dwarfs even though this does not exclude the possibility that some UDGs may have be failed Milky Way galaxies. The field UDG candidates of \citet{Barbosa2020} --- as they have no distance estimates so are given a redshift by association to the nearby overdensity of normal galaxies --- appear to have stellar masses and metallicities that are similar to those observed in clusters. Other works on field UDGs have shown that they can contain HI gas \citep{Poulain2021a,Leisman2017} and are predominantly blue and star forming \citep{Prole2019b}.

Various scenarios have been put forward to explain the formation of UDGs, based on the properties of the population of UDGs in galaxy clusters. It may be that they are ``failed'' high halo mass galaxies, prevented from building a normal stellar population due to, for example, extreme feedback from supernovae and young stars, gas stripping, or AGN feedback. On the other hand, UDGs may be ``genuine'' dwarf galaxies with correspondingly low halo masses but anomalously large sizes. They could represent the most rapidly spinning tail of the distribution of dwarf galaxies \citep{Amorisco2016}, or feedback and outflow could create an expansion of both the dark matter and stellar component of dwarf galaxies \citep{DiCintio2017}. Several UDGs show highly elongated shapes, suggestive of tidal disruption \citep{Merritt2016,Mihos2015}. Another scenario is that some UDGs could be tidal dwarf galaxies (TDGs), that is galaxies formed from gas expelled from a massive galaxy after an interaction. TDGs are known to be devoid of dark matter \citep{Lelli2015}, have a higher metallicity and radius than dwarfs of similar mass \citep{Weilbacher2003,Duc2014} and their GCs are less massive \citep{Fensch2019b}. The old ones have the same properties as UDGs \citep{Duc2014}.

A recent survey for dwarf galaxies in the MATLAS low to medium density fields has revealed 2210 low surface brightness galaxies \citep{Habas2020}. In this paper, we identify the UDGs in the MATLAS sample and present their detailed properties. In Section\,\ref{section:dwarfs} we discuss the observations and data analysis of the MATLAS sample as well as the UDG selection procedure. In Section\,\ref{section:environment}, we examine the types of environments in which the MATLAS UDGs are located. The photometric and structural properties of the UDG, as well as their compact central nuclei when present, are presented in Section\,\ref{section:properties}. Their derived stellar masses are also discussed in this section. In Section\,\ref{section:GCs}, we determine the globular cluster populations of the UDGs and compute their estimated halo masses. We search for possible tidal features and present the results in Section\,\ref{section:tidalfeatures}. The UDGs with HI detections are discussed in Section\,\ref{section:HI}. Finally, in Section\,\ref{section:conclusion}, we summarize our results.

\begin{figure*}
\centering
\includegraphics[width=0.95\textwidth]{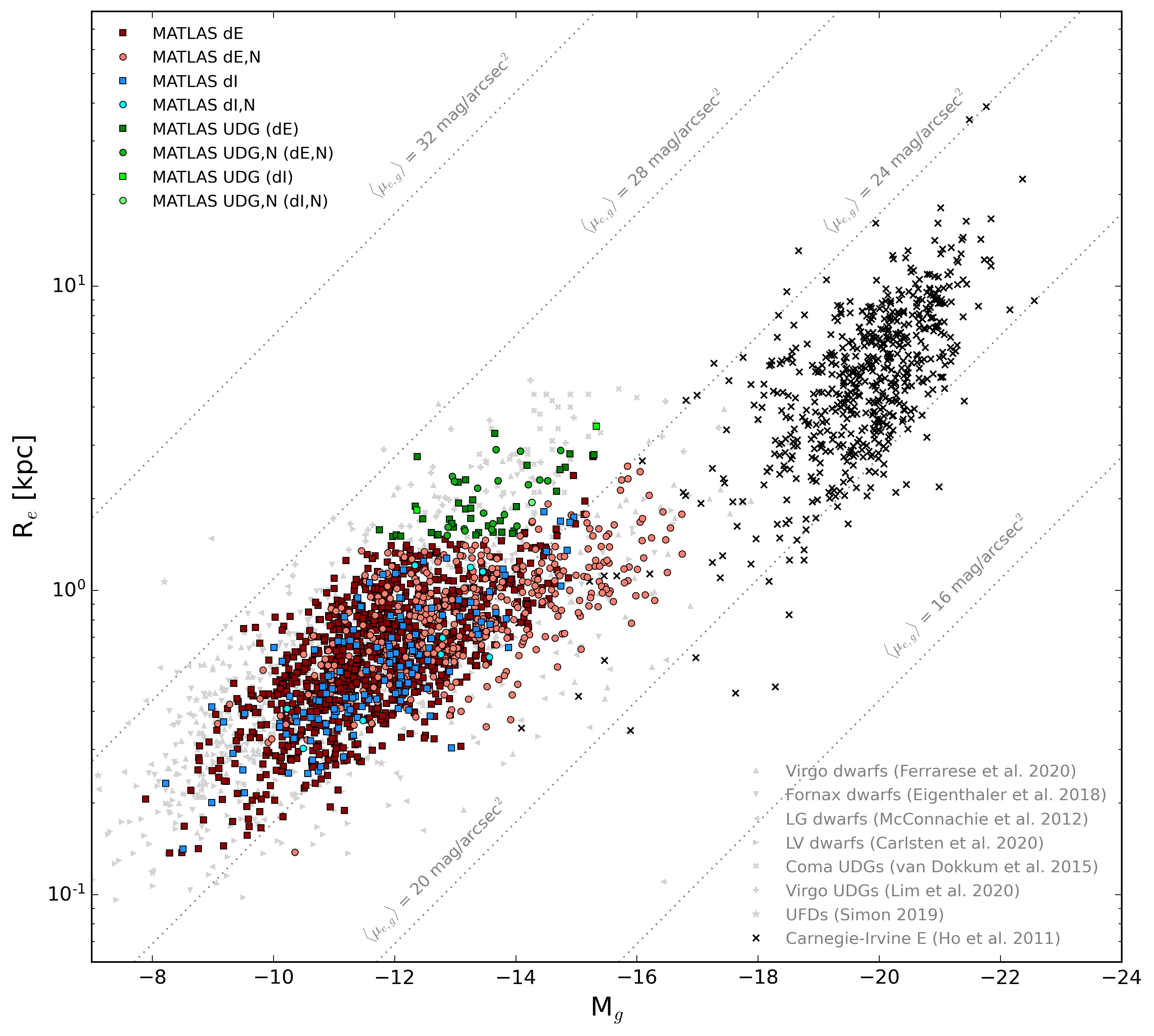}
\caption{Scaling relation M$_g$ versus $R_e$ for the MATLAS dwarfs and UDGs: The plotted quantities are based on parameters returned by the \textsc{Galfit} modeling, and we have used the estimated distance to the dwarf when available and otherwise assumed that the dwarfs are at the same distance as the central ETG in the field. The nuclei of the dE,N have been fitted with a two-component (diffuse component and nucleus) model, and only the properties of the diffuse component of the galaxy are used here. The MATLAS dwarfs occupy the same parameter space as dwarfs identified in the NGVS (gray triangles; \citealt{Ferrarese2020}) and NGFS (inverted gray triangles; \citealt{Eigenthaler2018}) images, as well as Local Group dwarfs (rotated left gray triangles; \citealt{McConnachie2012}) and Local Volume (rotated right gray triangles; \citealt{Carlsten2020}). The MATLAS UDGs occupy the same parameter space as the UDGs found in the Coma (gray cross symbols; \citealt{vanDokkum2015}) and Virgo cluster (gray plus symbols; \citealt{Lim2020}). The population of UDGs appear to simply be an extension in parameter space to the traditional dwarf population.}
\label{fig:scalingrel}
\end{figure*}

\section{MATLAS dwarf galaxy sample}
\label{section:dwarfs}

\subsection{Observations and catalog}
\label{section:cat}

The UDG candidates were identified from the deep ($\mu_g \sim 28.5$~--~$29$~mag/arcsec$^2$) optical imaging of the 
Mass Assembly of early-Type gaLAxies with their fine Structures (MATLAS) large observing program \citep{Duc2014,Duc2015,Duc2020,Bilek2020}. MATLAS was designed to study the low surface brightness outskirts of massive early-type galaxies (ETGs) in low- and moderate- density environments in the nearby ($z<0.01$) Universe. The target galaxies were taken from the ATLAS$^{3D}$ legacy survey \citep{Cappellari2011}, which compiled a complete sample of galaxies (the ``parent'' sample) with distances between $\approx$ 10 -- 45~Mpc, declinations obeying $|\delta - 29^\circ | < 35^\circ $, Galactic latitudes $|b| > 15^\circ$, and \textit{K}-band absolute magnitudes $M_K < -21.5$. The galaxies were morphologically reclassified by the ATLAS$^{3D}$ team for consistency, and only the 260 elliptical and lenticular galaxies were included in the final sample. The Next Generation Virgo Cluster Survey (NGVS; \citealt{Ferrarese2012}) survey obtained deep optical imaging of the 58 ATLAS$^{3D}$ ETGs within the virial radius of the Virgo cluster, while MATLAS planned to target the remaining (non-Virgo) ETGs.

The MATLAS (and NGVS) images were taken with MegaCam on the 3.6~meter Canada-France-Hawaii Telescope (CFHT). MATLAS ultimately imaged 150 $1^\circ \times 1^\circ$ fields, encompassing 180 ETGs and 59 late-type galaxies (LTGs),with an ETG typically at (or very near) the center of the image. This field of view corresponds to physical scales ranging from approximately 175~kpc ($d=$10~Mpc) to 785~kpc ($d=$45~Mpc). Select fields in moderate density environments have some overlap, however, thus the total coverage of the survey is 142 square degrees. The fields were preferentially observed first in the \textit{g}-band (150 fields), followed by the \textit{r}-band (148 fields), the \textit{i}-band (77 fields), and the nearest fields ($d_{ETG} < 20$~Mpc) in the \textit{u}-band (12 fields). All magnitudes presented in this paper were calibrated in the MegaCam AB magnitude system. Details of the observing strategy, data reduction, and image quality can be found in \citet{Duc2015}, while a full list of targeted fields in presented in Table 1 of \citet{Habas2020}, and \citet{Bilek2020} presents a comparison with other recent deep imaging surveys. 

The depth and image quality of the MATLAS survey is ideal for identifying new low surface brightness galaxies. The systematic search for dwarfs in the MATLAS images is described in detail in \citet{Habas2020}, but we briefly outline the selection criteria here for completeness. We first compiled a catalog of dwarf candidates from a visual inspection of every MATLAS image; this was then used to test various {\textsc{SExtractor}} \citep{Bertin1996} detection parameters, which were tweaked until $\sim$90\% of the visually identified dwarfs were successfully extracted. The visual catalog was then supplemented with additional candidates from {\textsc{SExtractor}}, identified by their surface brightness $\mu_g$, \textit{g}-band magnitude, and size. This joint catalog underwent two round of visual classification and cleaning, resulting in the final catalog of 2210 dwarfs based on the majority opinion of the final five classifiers. Further details, comparisons with overlapping catalogs, and discussions of the limitations and biases of the catalog can be found in \citet{Habas2020}.

\subsection{MATLAS dwarf distances}
\label{subsection:dwarf_distances}

Distances are essential to characterise many properties of dwarf galaxies, for example, absolute magnitudes, effective radii in physical units, and determining the local environment of the dwarf. We have distance estimates for a fraction (14.7\%; 325) of the MATLAS dwarfs from various sources: spectroscopic redshifts from SDSS DR13 \citep{Blanton2017,Albareti2017} and the Catalog of Visually Classified Galaxies (CVCG; \citealt{Ann2015}), HI velocities from the ALFALFA catalog \citep{Haynes2018} and the Westerbork Synthesis Radio Telescope (WSRT) imaging data of the ATLAS$^{3D}$ targets \citep{Serra2012,Poulain2021b}, distances from the NEARGALCAT \citep{Karachentsev2013}, as well as stacked MUSE spectra and a distance estimate from the GCLF for MATLAS-2019 \citep{Mueller2020,Mueller2021}.

Previous work has demonstrated that these galaxies are all dwarf-like systems with absolute magnitudes $M_g \geq -18$ and that $>90$\% of them are satellites of the nearest massive galaxy in the ATLAS$^{3D}$ parent (ETG $+$ LTG) sample with relative velocities $|\Delta v | < 500$~km~s$^{-1}$ \citep{Habas2020,Poulain2021a}. It should be noted that 21 (6\%) of these galaxies, however, have distance estimates $>45$~Mpc, so we would not expect to find a good host among the massive ATLAS$^{3D}$ galaxies. This further strengthens the argument that the dwarfs in our sample are overwhelmingly expected to be satellite galaxies. Thus, if we can identify the likely host of the dwarfs without known distances, we are able to use the distance of the host as the distance for these dwarf with some degree of confidence. \citet{Habas2020} used the subsample of dwarfs with known distances to test the accuracy of the host identification using other methods --- by assuming either the MATLAS targeted ETG (often the most massive and most central ETG in the image) is the host, or that the ATLAS$^{3D}$ massive galaxy (ETG or LTG) with the smallest on-sky separation is the host. The accuracy of both tests was $\sim$80\%.

We opted to proceed under the assumption that all of the dwarfs are satellites of a massive galaxy, and adopted the distance of the MATLAS targeted ETG when no other distance estimates are available.



\subsection{MATLAS dwarf structural parameters}

The structural and photometric properties of each MATLAS dwarf galaxy were extracted using \textsc{Galfit} \citep{Peng2002,Peng2010}. The nonnucleated dwarfs were fitted with a single S\'ersic profile in the \textit{g}, \textit{r} and \textit{i} bands, measuring their effective radius ($R_e$), S\'ersic index ($n$), total magnitude, and axis ratio. The \textit{g}-band surface brightnesses $\mu_{0,g}$ (central), $\mu_{e,g}$ (at $R_e$) and $\langle \mu_{e,g} \rangle$ (within $R_e$) were then calculated using the equations in \citet{Graham2005} taking as inputs the total magnitude and effective radius returned by \textsc{Galfit}. The galaxy model $g-i$ and $g-r$ colors were computed when images were available in these bands (see Section~\ref{section:cat}). When hosting one or multiple nuclei, the galaxies were fitted instead with a S\'ersic profile (for the galaxy) and a PSF/King profile (for each nucleus) simultaneously. For nucleated galaxies, the properties reported for the dwarf relate to the galaxy (S\'ersic profile). The detailed properties of the nucleus/nuclei are discussed in Section~\ref{section:properties}. The \textsc{Galfit} output parameters were derived for a total of 1589 MATLAS dwarf galaxies, yielding a morphological type mix of 1022 dE, 415 dE,N, 142 dI, and 10 dI,N \citep{Poulain2021b}. \textsc{Galfit} modeling of 292 nuclei were also obtained.

\begin{figure*}
\centering
\includegraphics[width = 0.88\textwidth]{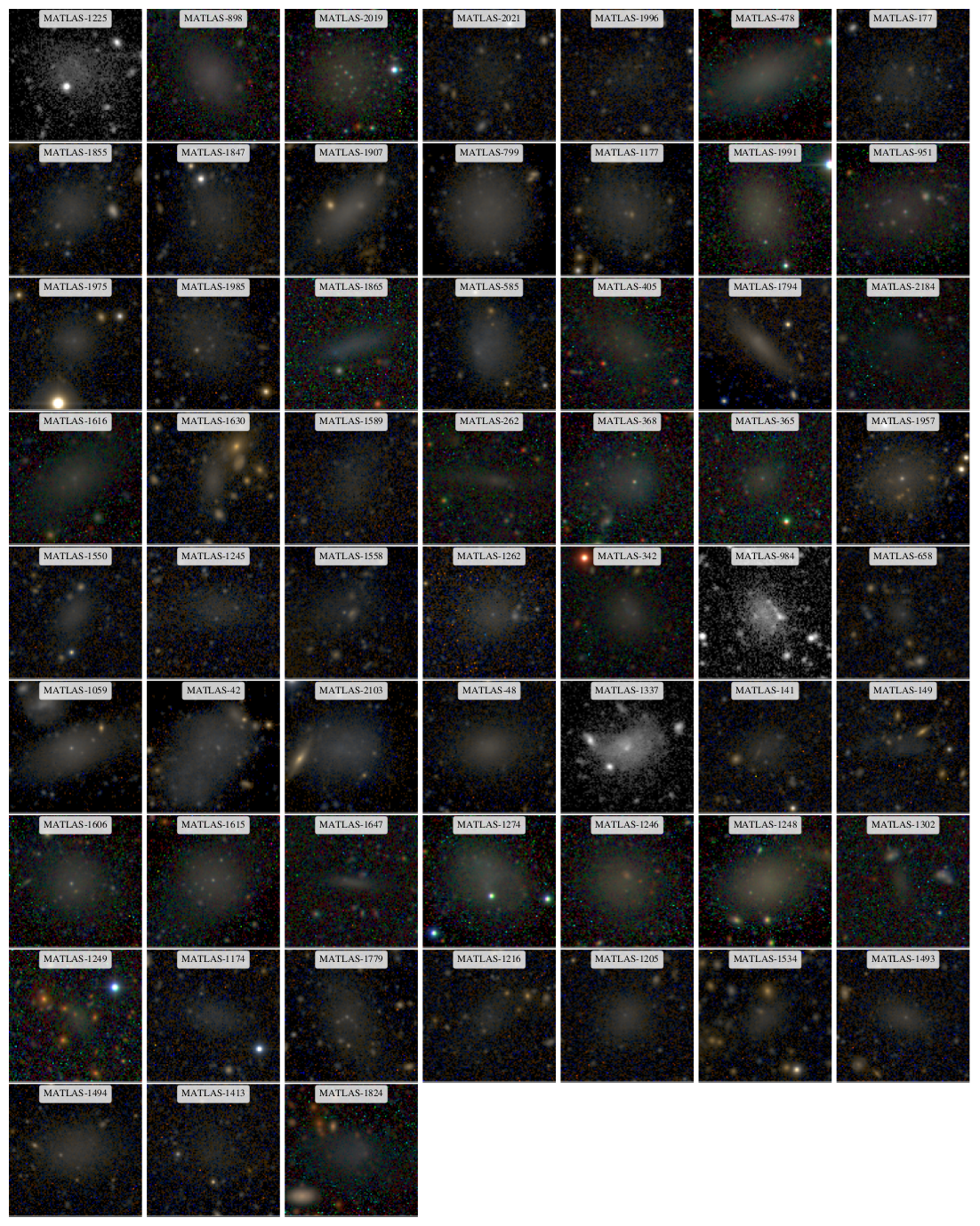}
\caption{Color ($g,r,i$) images of the 59 MATLAS UDGs in our final sample sorted by distance (from {\it top left} to {\it bottom right}, 19.1 to 46.3~Mpc). Each image is $\sim 1$ sq.arcmin.\ in size. North is up and east is left. }
\label{fig:UDGimages1}
\end{figure*}

\section{MATLAS ultra diffuse galaxy sample}
\label{section:UDGs}

\subsection{UDG Selection}
\label{section:UDGselection}

Using the structural properties discussed in the previous section, the UDGs were selected using a cutout in effective radius of $R_e \geq 1.5$~kpc and surface brightness of $\mu_{0,g} \geq 24.0$~mag/arcsec$^2$. This definition is adopted with no prejudice on whether they are a real class of objects or not, in order to allow comparison to the other similarly selected UDG samples. However, as can be seen already in Figure~\ref{fig:scalingrel}, these selection criteria are somewhat arbitrary: the UDGs (green symbols) do not appear to represent a distinct sequence of galaxies but rather an extension of the dwarf population in terms of effective radius and surface brightness. 

Using this definition, we identify a total of 59 UDG candidates: 37 dE, 19 dE,N, 2 dI and 1 dI,N. Averaging over the surveyed area, we estimate that there are $\sim$0.4 UDG per square degree. We note that this selection depends on the dwarf galaxy having a successful and reliable \textsc{Galfit} result. Hence, it is possible that there are some UDGs amongst the MATLAS dwarfs that were not successfully modeled using \textsc{Galfit}, in particular those 1) with an extremely low surface brightness (e.g., MATLAS-1830, \citealt{Duc2014}; see Section~\ref{section:tidalfeatures}) and 2) classified as dIs since only 27\% of the complete sample of MATLAS dIs were successfully modeled with \textsc{Galfit} (compared to 87\% of dEs) due to the presence of irregular features \citep{Poulain2021b}.

\subsection{UDG distances}
\label{section:UDGdistances}

The identification of a UDG sample strongly depends on the accuracy of the distances to the candidates. Four UDGs in the sample defined above have available distance estimates: three have velocity measurements from HI data (one from the ATLAS$^{3D}$ HI data cubes and two from the ALFALFA catalog) and one, MATLAS-2019, has a distance estimate based on the GCLF \citep{Mueller2021}.


It is worth noting, given the general uncertainties in the distances to the remaining UDGs, that three of these four UDGs would still be classified as such if we adopted the distance to the assumed host, as defined in Section~\ref{subsection:dwarf_distances}. For the other UDG, the assumed host had a distance of 19.1~Mpc, but the UDG actually lies at a distance of 35.9~Mpc according the velocity extracted from the ALFALFA HI catalog. Conversely, we also have exactly one dwarf galaxy that would be classified as a UDG given the assumed distance of the host (37.1~Mpc), but had its physical size revised downward due to the distance estimate in the NEARGALCAT (10~Mpc).


Thus, in total, 4/59 of our UDG candidates have a distance estimate confirming their effective radii $R_e \geq 1.5$~kpc. Color ($g,r,i$) images of the 59 MATLAS UDGs in our final sample, ordered by increasing distance ($19.1 - 46.3$~Mpc), are shown in Figure~\ref{fig:UDGimages1}. The effective radius of the MATLAS UDGs ranges from 1.5 (our cut) to 3.5~kpc (see Figure~\ref{fig:scalingrel}). As the distances of the UDGs are critical, we further explore the potential for misidentified host galaxies for the full UDG sample in Appendix~\ref{section:appendix_hosts}.


\section{UDG environments}
\label{section:environment}
\begin{figure*}
\includegraphics[width=0.45\textwidth]{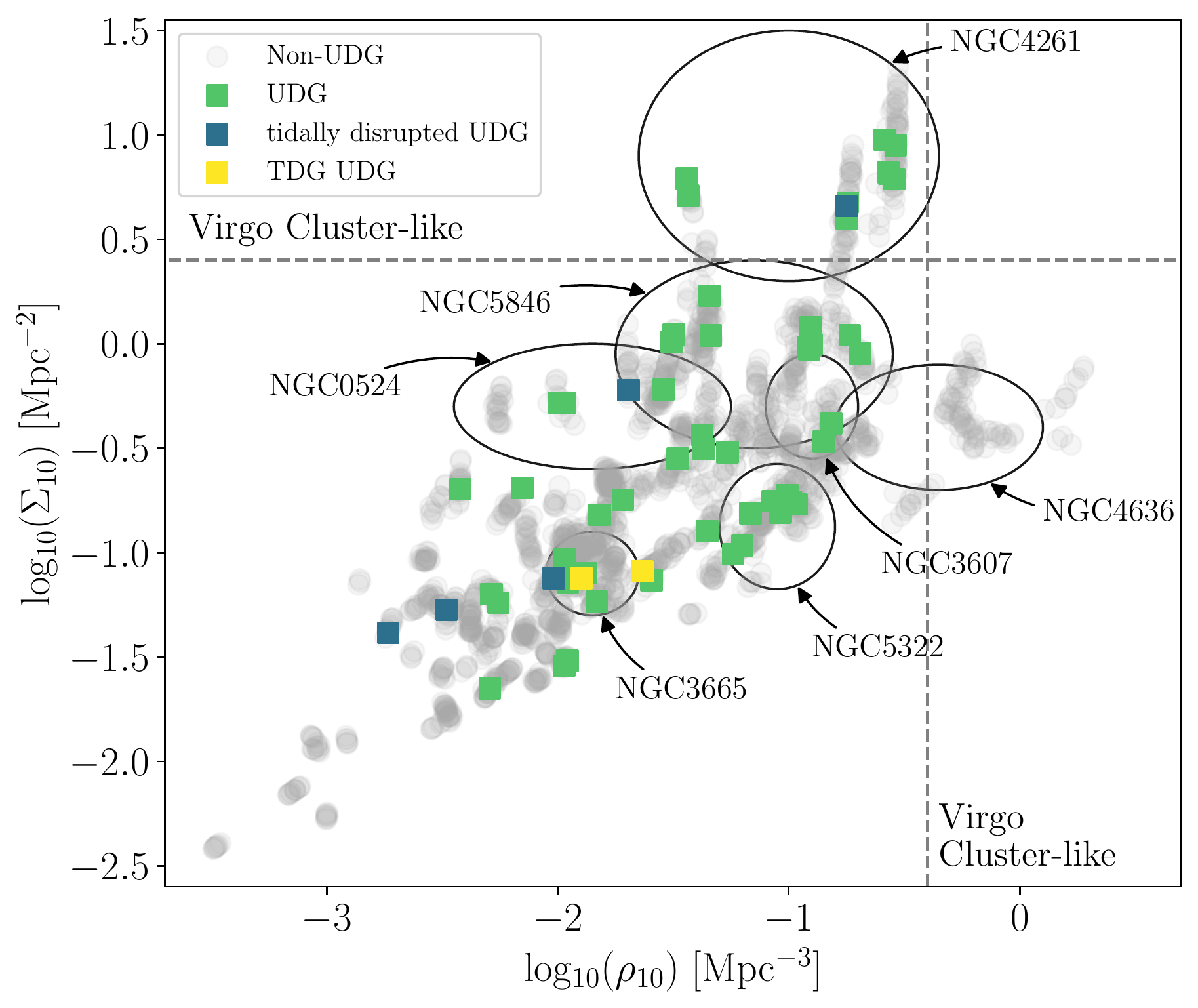}
\includegraphics[width=0.5\textwidth]{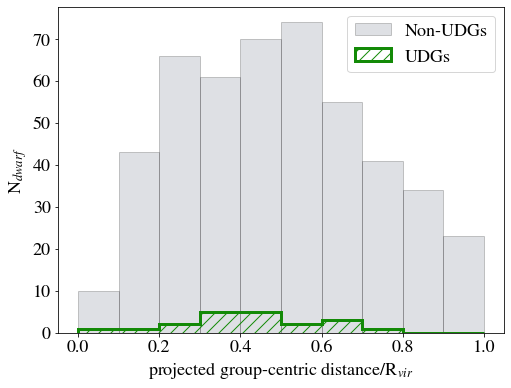}
\caption{{\it Left}: Estimates of the local volume density ($\rho_{10}$) versus the local surface density ($\Sigma_{10}$) for the MATLAS traditional dwarf (non-UDG) galaxies ({\it gray circles}) and UDGs ({\it green squares}). The tidally disrupted UDGs and TDG UDGs discussed in Section~\ref{section:tidalfeatures} are marked in {\it turquoise} and {\it yellow}, respectively. The density estimates were calculated based on the tenth-nearest neighbor, using only the massive galaxies from the ATLAS$^{3D}$ parent sample.  \citet{Cappellari2011b} noted that nearly all galaxies in the Virgo cluster have $\log_{10} \rho_{10}$ > -0.4 and $\log_{10} \Sigma_{10}$ > 0.4; these boundaries are marked with dashed lines. The open ovals indicate where the dwarfs in select groups fall in the plot. {\it Right}: The radial distribution of traditional dwarf (non-UDG) galaxies ({\it filled gray}) and UDGs ({\it hashed green}) in the group environments defined in Table~\ref{tab:group_properties}. The projected group-centric distances have been normalized such that $R_{vir} = 1$. A KS test returns a p-value of 0.35, thus we cannot reject the null-hypothesis that the two samples originate from the same parent sample.}
\label{fig:locdensity}
\end{figure*}

MATLAS specifically targeted the ATLAS$^{3D}$ sample outside the Virgo cluster, which includes ETGs in groups, pairs, and isolated galaxies. A detailed study of the galaxy associations in the MATLAS fields is currently in progress (Smith et al.\ 2021)\nocite{Smith2021}. For now, we calculate the local volume density ($\rho_N = 3N/4 \pi r_{3D}^3$) and the surface density ($\Sigma_N = N / \pi r_{2D}^2$)
using the $N=10^{th}$ nearest neighbor --- in 3-dimensional and projected on-sky space, respectively --- of each dwarf as a proxy for the local environment. We note that we considered only the massive galaxies from the ATLAS$^{3D}$ parent catalog (ETGs and LTGs) when determining $N$; to avoid large uncertainties for the 14 dwarfs known to lie at distances $>50$~Mpc, well beyond the ATLAS$^{3D}$ sample, we have removed these few galaxies from the plot. The distribution is shown in Figure~\ref{fig:locdensity}. 

In this plot, we have assumed that the dwarfs are at the same distance as the host galaxy, when no independent distance measurement is available. To estimate the error this introduces into $\log_{10}\rho_{10}$, we incorporated an offset of $\pm500$~km/s in velocity space, following the cut we used to define likely satellite/non-satellite dwarfs in Section~\ref{subsection:dwarf_distances}, and propagated this through the calculations. We concatenated the $\pm$ error estimates, and binned the data into four bins containing nearly equal number of datapoints in $\log_{10}\rho_{10}$ space: $\log_{10}\rho_{10} \leq -2.0$, $-2.0 < \log_{10}\rho_{10} \leq -1.58$, $-1.58 < \log_{10}\rho_{10} \leq -1.0$, and $\log_{10}\rho_{10} > -1.0$. In these four bins, the median errors are 0.01, 0.18, 0.28, and 0.4, respectively.

The UDGs can be found throughout the densities probed by the MATLAS sample, except for the lowest density regions, and follow a similar distribution as the dwarfs. Applying a 
two-sample Kolmogorov–Smirnov (KS) test, we find no statistically significant deviation in $\rho_{10}$ between the sample of UDGs and the remainder of the MATLAS dwarfs sample, with a returned p-value of $p=0.21$.

A number of groups with previously detected X-ray emission fall within the MATLAS fields and are also marked in the figure, to help gauge the local densities at various points in the plot. We note that while these open ovals enclose the dwarfs and UDGs associated with each group, the ovals themselves are illustrative and may enclose points that are not associated with the labeled group. The groups span a range of values in both $\rho_{10}$ and $\Sigma_{10}$, depending in part on the relative position of the dwarf to the group center and the presence of substructures or nearby systems. For example, contamination from foreground galaxies in the Virgo cluster increases $\Sigma_{10}$ for NGC~4261 ($d\sim31$~Mpc), while the proximity of NGC~4636 ($d\sim16$~Mpc) to the Virgo cluster has increased $\rho_{10}$ to cluster-like values for this group. In total, 19 UDGs (32.2\%) lie within the virial radii of the groups with X-ray emission, while 484 dwarfs (22.5\%; UDGs removed) lie within the same region.   

\subsection{Groups}

Here we comment on the properties of those individual, and presumably massive, groups detected in X-rays. The group properties are summarized in Table~\ref{tab:groups}. Spatial maps of all fields containing at least one MATLAS UDG are presented in Appendix~\ref{section:appendix2}. \\

\begin{table*}
	\centering
	\caption{Summary of known groups with X-ray detections that fall within the MATLAS images. The columns are: (1) the group name; (2) whether the X-ray emission is on group scales ($G$) or galaxy halo scales ($H$), taken from \citet{Osmond2004}; (3) the richness of the sample: L$=$low richness, H$=$high richness, from \citet{OSullivan2017}; (4) the distance to the group in Mpc, (5) projected virial radius in Mpc and (6) dynamical mass from \citet{Kourkchi2017}; (7) the fraction of the area within $R$ covered by the MATLAS footprint; (8) the number of UDGs identified within $R_{vir}$ --- for two groups, we increased the search radius by 0.15~Mpc in order to catch the UDGs just outside the cluster bounds, which are included in brackets; (9) the number of predicted UDGs using the dynamical mass in Column (6) and the relation described in \citet{vanderburg2017} --- we note that this $N_{UDG,pred}$ value is nearly identical to the value returned by the relation in \citet{Janssens2019}; (10) the number of MATLAS dwarfs (including UDGs) within $R_{vir}$ --- the numbers in brackets include the additional dwarfs found in a search radius increased by 0.15~Mpc; (11) the fraction of early-type galaxies in the group. For consistency, this fraction was calculated using the group membership and morphologies from \citet{Kourkchi2017}. The groups are ordered by increasing mass. }
	\label{tab:group_properties}
	\begin{tabular}{lrrrrrrrrrr} 
		\hline
Group & T & R & dist & $R_{vir}$ & $\log(M_{dyn}/M_\odot)$ & $f_{cover}$ & $N_{UDG}$ & $N_{UDG,pred}$ & $N_{dwarfs}$ & $f_{ETG}$ \\
 &  &  & [Mpc] & [Mpc] &  &  &  &  &  & \\
\hline
NGC~3665 &G& L & 23.99 & 0.288& 12.45&1.00 &1 &$0.36^{+0.15}_{-0.29}$ & 44 & 0.28 \\
NGC~4636 &G & -- & 16.20 & 0.472& 12.93&0.41 & 0 &$1.23^{+0.39}_{-0.29}$ &60 & 0.51 \\
NGC~3607 &G& -- & 22.14 & 0.367& 13.31&0.81 & 1 &$3.23^{+0.76}_{-0.29}$ & 37 & 0.41 \\
NGC~0524 & H & L & 31.19 & 0.388& 13.13 &0.98 &0[+2] & $2.06^{+0.56}_{-0.29}$ & 62[+12] &  0.33 \\
NGC~5322 &H& L& 23.28&0.386 &13.34&0.72 &3[+2] &3.5$^{+0.8}_{-0.29}$ & 49[+19] & 0.38 \\
NGC~5846 &G& H &26.74&0.596 &13.75&0.64 & 7  &$9.95^{+1.51}_{-0.29}$ & 143 & 0.77 \\
NGC~4261 &G& H & 31.39 &0.649 &13.94&0.35 & 7 &$16.22^{+1.89}_{-0.29}$ & 89 & 0.54 \\
		\hline
	\end{tabular}
\label{tab:groups}
\end{table*}

\begin{figure}
\centering
\includegraphics[width=\linewidth]{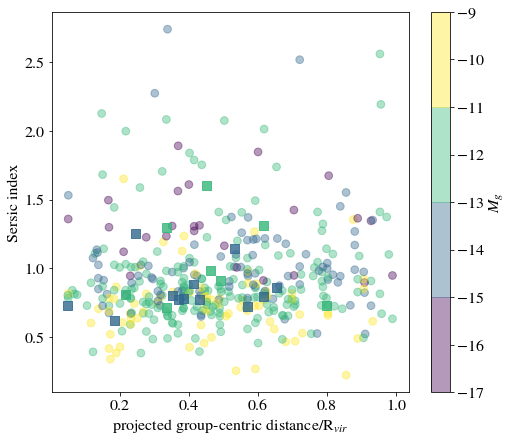}
\includegraphics[width=\linewidth]{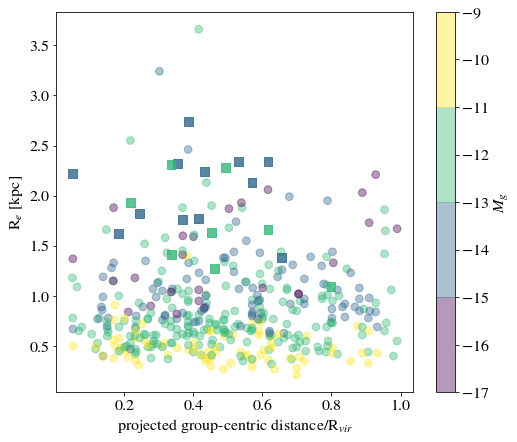}
\caption{The S\'ersic index ({\it top}) and  physical size in kpc ({\it bottom}) as a function of group-centric distance for the traditional dwarf (non-UDG) galaxies ({\it circles}) and UDGs ({\it squares}). The colorbar represents the absolute $g$-band magnitude of the galaxies. }
\label{fig:group_radial_properties_sersic_size}
\end{figure}

\begin{figure}
\centering
\includegraphics[width=\linewidth]{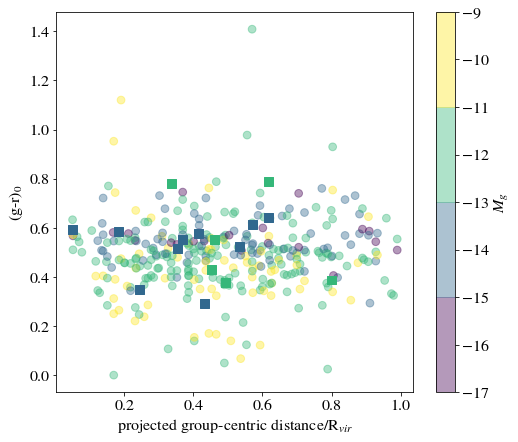}
\includegraphics[width=\linewidth]{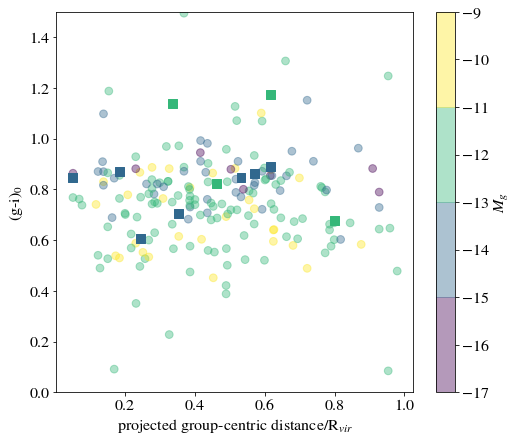}
\caption{The ($g-r$)$_0$ color ({\it top}), and ($g-i$)$_0$ color ({\it bottom}) as a function of group-centric distance for the traditional dwarf (non-UDG) galaxies ({\it circles}) and UDGs ({\it squares}). The colorbar represents the absolute $g$-band magnitude of the galaxies. }
\label{fig:group_radial_properties_colors}
\end{figure}

\noindent{\em The NGC~3665 group}: \hfill \\
The NGC~3665 group is a loose association of approximately 10 galaxies, mainly late-type, with an anomalously low velocity dispersion \citep{Helsdon2005}. Approximately 40 new, potential dwarf group members have been identified in the MATLAS images, and we find one UDG located near the virial radius $R_{vir}$. \\

\noindent{\em The NGC~4636 group}: \hfill \\
Located just south of the Virgo cluster at a distance of $d\sim16$~Mpc \citep{Kourkchi2017}, the NGC~4636 group is difficult to separate from the massive cluster. However, extended X-ray emission indicates that this group is a distinct structure. An analysis of the velocity dispersion of estimated members, and the position of the ETG NGC~4636 near the centroid of the group, suggest that it is a dynamically mature system \citep{Brough2006}. We identified 60 dwarfs within $R_{vir}$ of the group, but none meet the selection criteria to be classified as a UDG. Nor are any UDGs found just outside $R_{vir}$. \\

\noindent{\em The NGC~3607 group}: \\
Also known as the Leo~II group or the Leo Cloud, this group is centered around two ETGs, NGC~3607 and NGC~3608. The group contains $\sim$16 massive group members with total blue magnitudes brighter than $B \approx 16$, most of which are LTGs \citep{Afanasiev2007,Giuricin2000}. The morphologies of the group members, the observation of hot X-ray gas maxima around both NGC~3607 and NGC~3608  within the common gaseous envelope \citep{Mulchaey2003}, and the presence of optical low-surface brightness features indicative of recent mergers within the group \citep{Bilek2020} all suggest that this group is relatively young \citep{Afanasiev2007}. We identify one UDG among the 37 dwarfs within the MATLAS footprint, but it should be noted that we do not have full coverage of the group within $R_{vir}$. The UDG is in close projection to the group center.  \\   

\noindent{\em The NGC~0524 Group}: \\
The NGC~0524 galaxy group is a relatively isolated, low-richness group \citep{OSullivan2017} with a higher-than-average fraction of early-type galaxies. The group appears to be dynamically mature \citep{Brough2006} and a detailed study of the low surface brightness outskirts of the group members indicates that most are relaxed. Only NGC~0502 and NGC~0518 show evidence of past mergers \citep{Bilek2020}. We identified 62 dwarfs within the projected virial radius, but no UDGs. However, two UDGs can be found just outside the $R_{vir}$ boundary and are likely associated with the group. Interestingly, both UDGs lie to the southeast of the group, in the direction of NGC~0502 --- one of the two galaxies to have undergone a recent merger, although this may be a coincidence.  \\ 

\noindent{\em The NGC~5322 group}: \\
Optically, the NGC~5322 group is similar to  NGC~4261, with similar $B$ magnitudes, $R_e$ estimates, colors, color gradients, and the brightest group galaxy in each is classified as a Low-Ionization Nuclear Emission-line Region (LINER) galaxy. However, the NGC~5322 group exhibits a much fainter X-ray component, which has been interpreted as disruption of the X-ray halo of a single galaxy due to a recent galaxy interaction or merger. The ETG NGC~5322 contains a counter-rotating core, further support of a recent merger \citep{Finoguenov2006} and suggesting a younger group. We identified 49 dwarfs and three UDGs in this group. Like NGC~0524, however, there are an additional two UDGs a little beyond the virial radius, which we have noted in Table~\ref{tab:group_properties}.\\

\noindent{\em The NGC~5846 group}: \\
The NGC~5846 group is a massive and relatively isolated group at $d\sim26$~Mpc. \citet{Mahdavi2005} note that the group has the third largest over-density of ETGs in the local Universe, after the Virgo and Fornax clusters. The group appears to have significant substructure, with two distinct subgroups centered on the ETGs NGC~5846 and NGC~5831. Nevertheless, the mass, high fraction of ETGs, and velocity dispersion of the group members suggest that this is a dynamically evolved group. It should be noted that low luminosity members of the group have been studied previously \citep{Mahdavi2005,Eigenthaler2010}, and Mahdavi et al.\ noted that a few of the dwarfs in their sample are very low surface brightness objects, five of which are part of our UDG sample. In total, we have identified seven UDGs from the 143 dwarfs in the MATLAS footprint. One UDG candidate, MATLAS-2019, has been confirmed as a group member, and hence is a robust UDG, in individual case studies \citep{Mueller2020,Mueller2021,Forbes2019,Forbes2021}.\\

\noindent{\em The NGC~4261 Group}: \hfill\\
Also known as the W Cloud, NGC~4261 appears close in projection to the Virgo cluster, although it is more distant at $d\sim31$~Mpc. It is thought that the group may be part of a filament more or less along the line of sight, which is feeding into the cluster \citep{Kourkchi2017}. The light profile of the group is unusual, with approximately half of the light found at radii $> 0.75R_{vir}$; \citet{Helsdon2003} have interpreted this as evidence that galaxies are currently falling into the group. The central ETG, NGC~4261, also has signs of a minor merger in the past $1-2$~Gyr \citep{Bilek2020,OSullivan2011}. Taken together, these suggest that the group in fairly young.

The proximity of the Virgo cluster makes it difficult to determine group membership, and the association of the 82 traditional dwarf (non-UDG) galaxies and 7 UDGs may also be suspect. We can estimate the contamination from the LSB Virgo cluster members in this region of the sky statistically, however. Figure~4 from \citet{Roberts2007} shows the number density of dwarfs as a function of cluster-centric distance, measured in two directions that avoid the various substructures in the cluster. Taking the larger of the two estimates gives a number density of roughly 10 LSB dwarfs per square degree at the distance of NGC~4261. Thus, we would expect approximately 15 of our dwarfs (18\%) to be members of the foreground Virgo cluster rather than the NGC~4261 group. We note that if any of our UDG candidates are at the distance of the Virgo cluster, they would be classified as normal dwarf galaxies. The largest of the UDG candidates would have a recalculated $R_e = 1.16$~kpc. Without further confirmation, however, we retain the current distance estimate for these dwarfs, that is, $d \sim$31~Mpc. \\

The traditional dwarfs and UDGs within the projected virial radii of the known X-ray groups can be stacked to better understand trends of these systems within groups. In Figure~\ref{fig:locdensity}, ({\it right}), we plot the distribution of projected group-centric distance for both the traditional dwarf (non-UDGs) galaxies and the UDGs, where the virial radius of each group has been scaled such that $R_{vir} = 1$. The traditional dwarf (non-UDG) galaxies are spread fairly uniformly throughout the groups. The distribution of UDGs is similar, and we cannot reject the null hypothesis that the two distributions arise from the same parent distribution via a two sample KS test with a p-value $p=0.35$.

The distribution of UDGs in individual groups varies greatly. For example, in NGC~5846, many of the UDGs appear close in projection to the center of the group, while the two UDGs associated with NGC~0524 are found just outside the virial radius. The expected distribution of UDGs is not clear. \citet{vanderBurg2016} found that the projected radial distribution of UDGs in clusters is well fit by an Einasto profile, with a steep profile in the cluster outskirts that flattens toward the cluster center. It should be noted that the regular dwarfs in that study do not follow the same distribution, at least in the central regions. However, \citet{Lim2020} found that UDGs in the Virgo cluster are more centrally concentrated than other galaxies with similar magnitudes. In a study of UDGs in the Frontier Fields \citep{Janssens2019}, the authors note that the UDGs in individual clusters often show lopsided distributions, with the exception of the most relaxed cluster in the sample, where the distribution is roughly uniform. Given the uncertainties in the expected UDG distribution, and whether that distribution would apply equally to groups as clusters, we have made no attempt to correct the number of UDGs in the X-ray groups for incomplete MATLAS coverage. However, this fraction is noted in Table~\ref{tab:group_properties}.

In Figures~\ref{fig:group_radial_properties_sersic_size} and ~\ref{fig:group_radial_properties_colors}, we plot the S\'ersic index, $R_e$, and colors of the dwarfs and UDGs as a function of the group-centric distance and absolute magnitude. The effective radius and absolute magnitudes were recalculated assuming the group distance given in Table~\ref{tab:group_properties}; this was applied to the dwarfs/UDGs with independent distance estimates as well, to avoid uncertainties in the distance due to relative motions of the satellites. It should be noted that this results in minor changes to the sample; the recalculated effective radii of four UDGs (Figure~\ref{fig:group_radial_properties_sersic_size} {\it bottom}, {\it squares}) is lower than $R_e = 1.5$, while 16 traditional dwarf (non-UDGs) galaxies  (Figure~\ref{fig:group_radial_properties_sersic_size} {\it bottom}, {\it circles}) with $\mu_{0,g} \geq 24.0$~mag/arcsec$^2$ now have $R_e \geq 1.5$~kpc; we note, however, that these 16 galaxies are not distinguished from the other traditional dwarfs with brighter $\mu_0$ in the plot. None of the 20 galaxies which would be reclassified as a UDG or non-UDG have robust distance measurements available. As the association of the galaxies with the groups has not been proved, the group distance estimate is no more certain than the central ETG in the field distance estimate previously assumed, and we will not update our sample of 59 UDGs until more accurate distances are obtained.

The distribution of the UDGs and traditional dwarfs is similar in both Figures~\ref{fig:group_radial_properties_sersic_size} and ~\ref{fig:group_radial_properties_colors}. We do not see a systematic offset between the UDGs and the traditional dwarfs, aside from a separation in $R_e$ between the UDGs and the bulk of the group dwarfs (Figure~\ref{fig:group_radial_properties_sersic_size}, {\textit{bottom}}), which is simply by the construction of the UDG definition; even here, however, there are a number of traditional dwarfs with UDG-like sizes, but which have central surface brightness $\mu_{0,g} < 24.0$~mag/arcsec$^2$. We also do not observe any trends with group-centric distance among the four parameters, except for a possible color shift toward the blue with increasing distance from the group center. It should be noted, however, that using the projected distances may erase the signature of any radial trends. A more rigorous analysis must wait until distances estimates for a large number of the traditional dwarfs and UDGs are obtained.

\subsection{Other galaxy associations}
In addition to the groups with observed X-ray emission discussed above, the MATLAS images contain a number of groups and galaxy associations that have been identified in other wavelengths. To identify other associations, we searched for groups in the \citet{Kourkchi2017} groups catalog. This catalog is based on an all-sky sample of $\sim$15,000 galaxies with $V_{LS} < 3500$~km~s$^{-1}$, where $V_{LS}$ is the radial velocity relative to the Local Sheet. The groups were identified through a galaxy linkage program using known scaling relations. These authors placed a particular emphasis on small groups, and the catalog contains a number of groups with fewer than three members. The X-ray groups discussed above are all contained within the catalog, and when we expand the search to include all groups with $R_{vir} < 1.5\deg$, to focus on the local environment, we find that 61\% of the UDGs fall within the virial radii of one of the \citet{Kourkchi2017} groups, and 81\% are enclosed within $1.5R_{vir}$. In comparison, 58\% of the non-UDG dwarfs lie within $1R_{vir}$ and 70\% lie within $1.5R_{vir}$.

These additional optical groups each have between $0-3$ UDGs. As with the X-ray groups, we find that the more massive groups are more likely to host a UDG, in agreement with previous work by \citet{Janssens2019} and \citet{vanderburg2017}. In particular, we note a rough break at $\log(M_{dyn}/M_\odot) \sim 12.5$; below this value, only 4/43 groups contain at least one UDG, while above this cut 16/35 groups have $1+$ UDGs within $R_{vir}$. At the high mass end, however, we are likely underestimating the fraction of groups with associated UDGs; the more massive groups typically also have larger $R_{vir}$, and the MATLAS footprint may not be large enough to observe UDGs on the outskirts or a lopsided distribution.  The least massive system with at least one UDG within the estimated virial radius, the NGC~772/NGC~770 pair, has $\log(M_{dyn}/M_\odot) = 11.88$ and has two UDGs. 

The MATLAS images also include the Hickson~44 group, which is dominated by three LTGs and one ETG, although there is some debate if the ETG is part of the group or an isolated background galaxy. The morphologies of the group members, along with a lack of X-ray emission or intra-galactic light, suggests that this is a dynamically young system \citep{Aguerri2006}. \citet{SmithCastelli2016} identified two UDG candidates in this field, although neither is included in our sample; one is in the halo of the ETG, while the other is near a bright star. Neither UDG could be deblended from the surrounding emission on the MATLAS images, either by eye or with {\textsc{SExtractor}}. This group is mentioned here for completeness, and to highlight another dynamically young group with UDGs.  

\subsection{Isolated hosts}
Defining a galaxy as isolated without a full kinematic analysis of its nearest neighbors is somewhat subjective, and a number of UDGs are located near likely galaxy pairs and triplets that are not part of the Kourkchi \& Tully catalog. Nevertheless, a small number of UDGs appear to be associated with relatively isolated ETGs --- ETGs that are not associated with a Kourkchi \& Tully group and have no massive galaxies (ETGs or LTGs from the ATLAS$^{3D}$ parent sample) in their vicinity. Examples include: IC~0560, NGC~1248, NGC~5493, and UGC~4551. Each of these galaxies has at most one associated UDG in our sample. Two of the four UDGs show evidence for tidal interactions (see Section~\ref{section:tidalfeatures}), while only one of the hosts (NGC~5493) shows clear evidence that it has undergone an old major merger; the other hosts are relaxed, although IC~0560 is surrounded by bright stellar halos that may have masked such faint features \citep{Bilek2020}. These ETGs have stellar masses, estimated from the K-band luminosity, that span the range $10.15 < \log(M_\ast/M_\odot) < 11.25$.

There is one potentially isolated ETG, NGC~4690, which has two UDGs in close projection. However, all three galaxies are near the $R_{vir}$ boundary of two other groups, thus the `isolated' classification of these galaxies is suspect. NGC~4690 shows evidence for a recent minor merger; neither of the UDGs display any tidal features. 

\begin{figure*}
  \centering
    \includegraphics[scale=0.44]{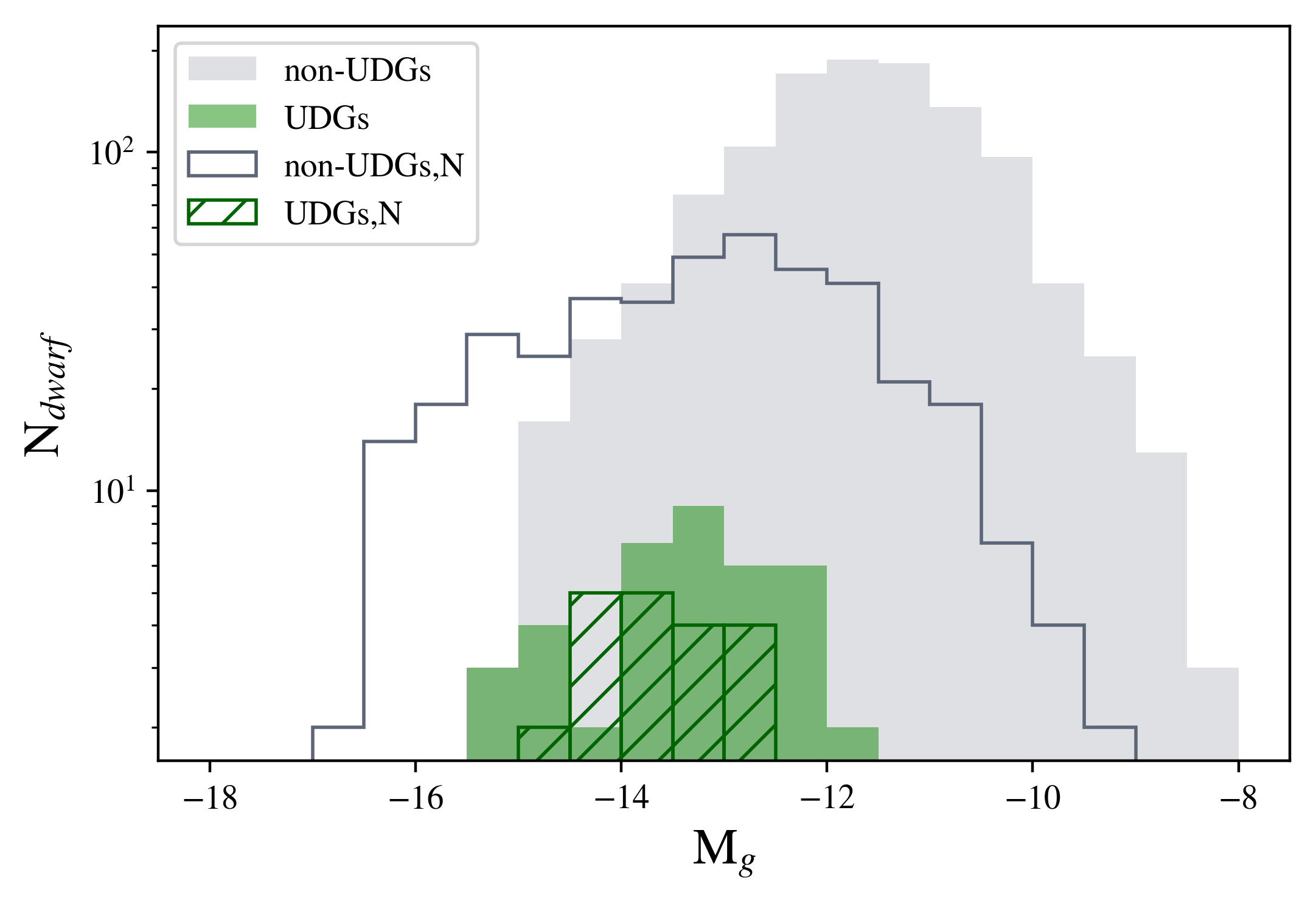}
    \includegraphics[scale=0.44]{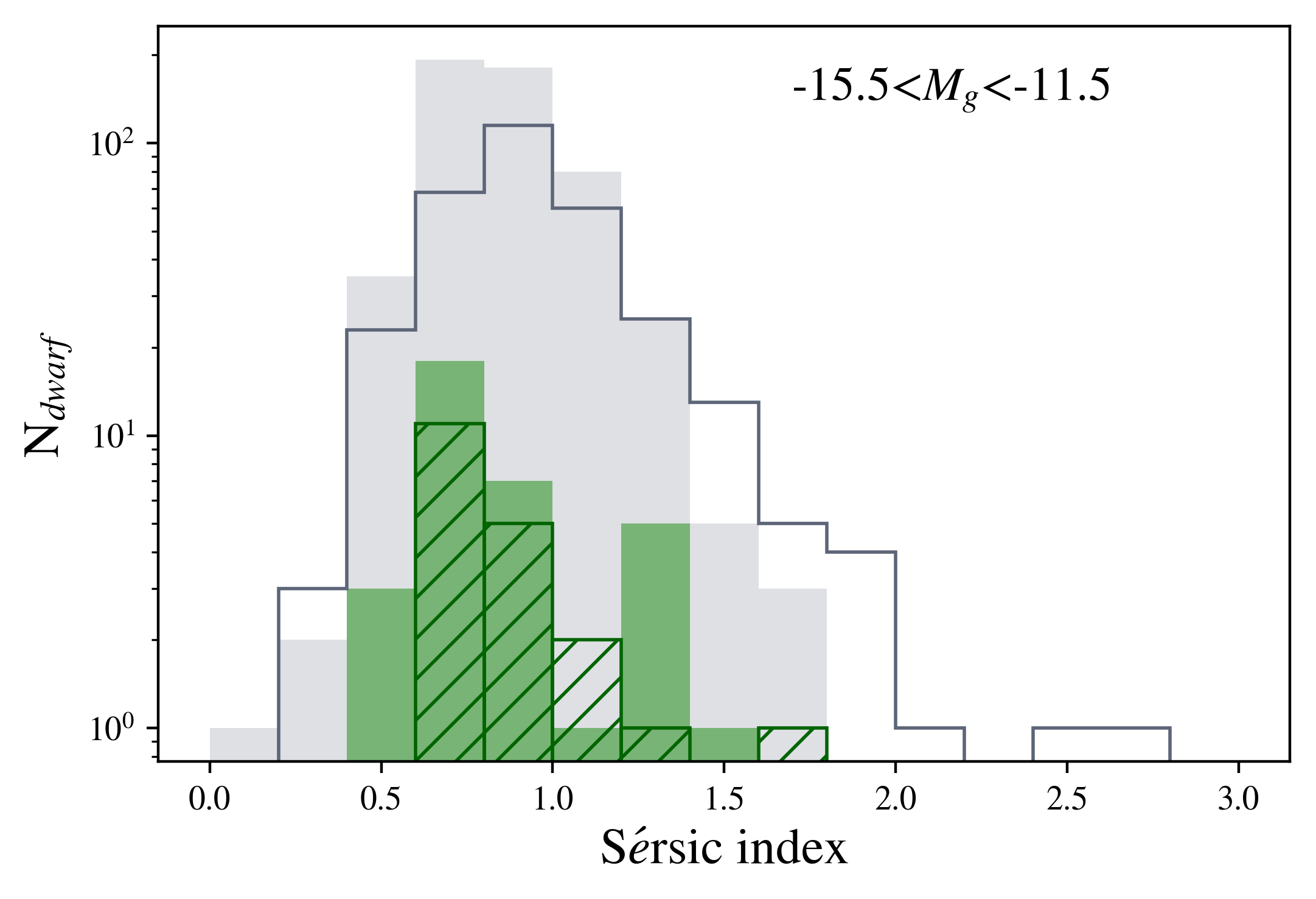}
    \includegraphics[scale=0.44]{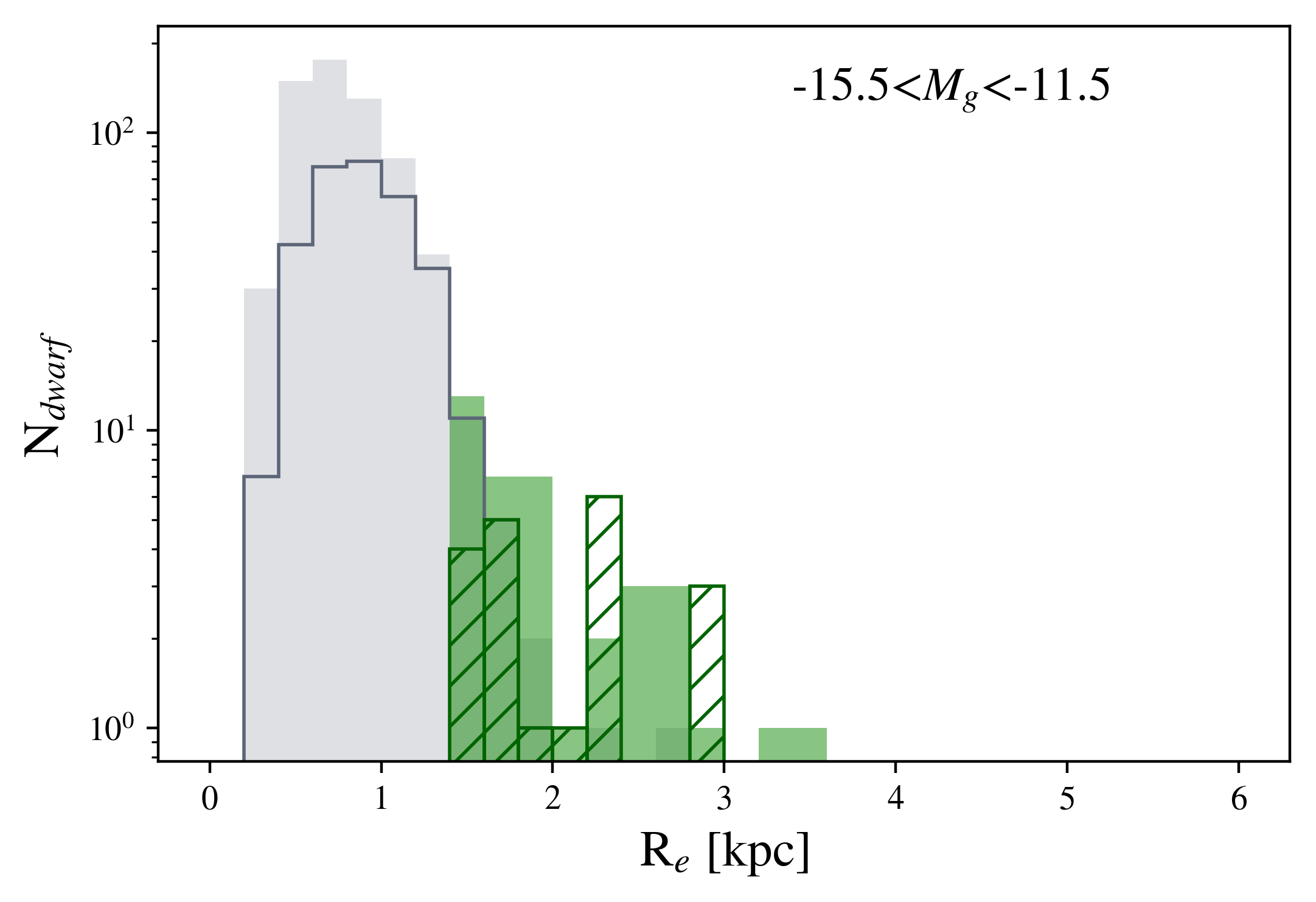}
    \includegraphics[scale=0.44]{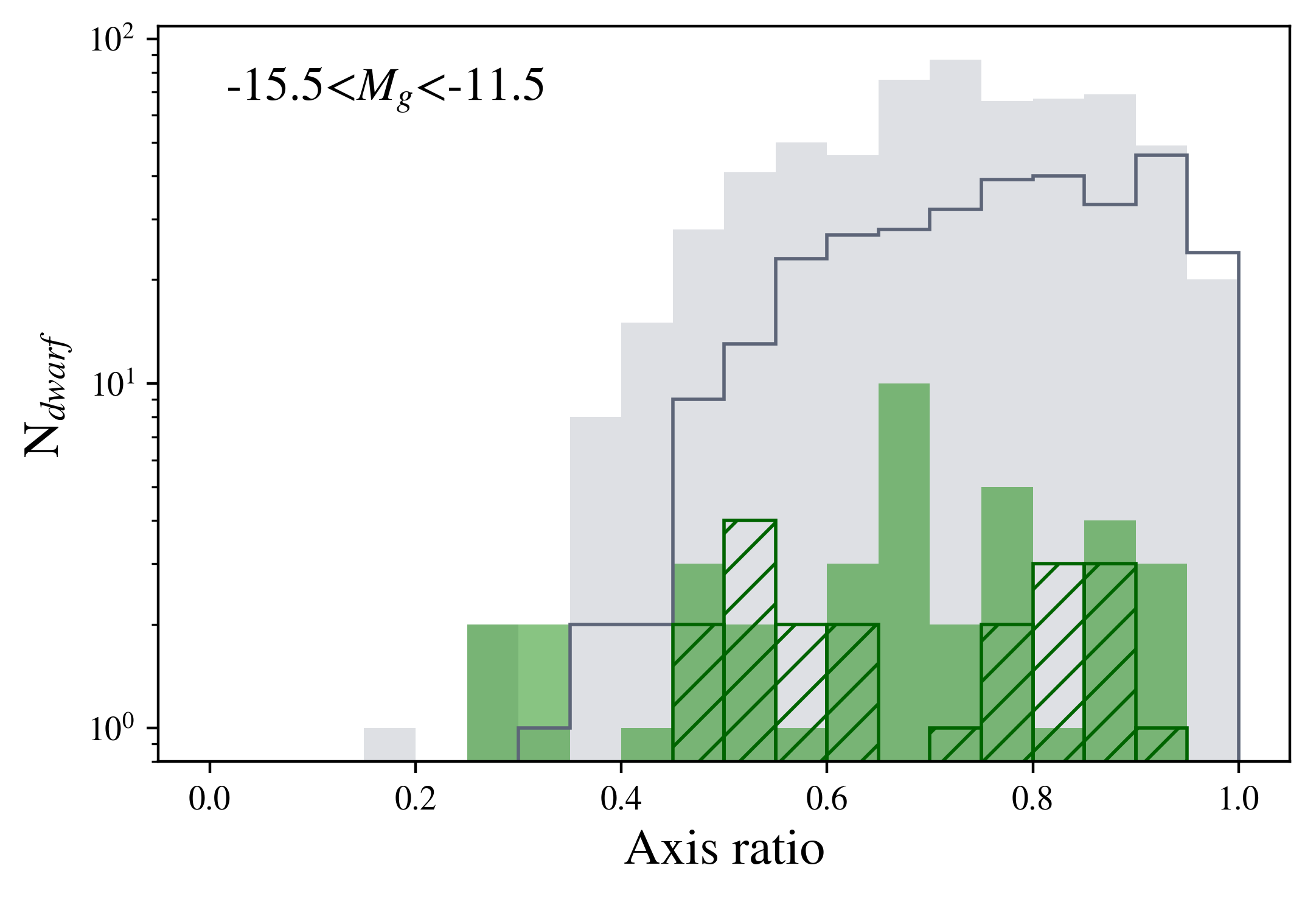}
    \includegraphics[scale=0.44]{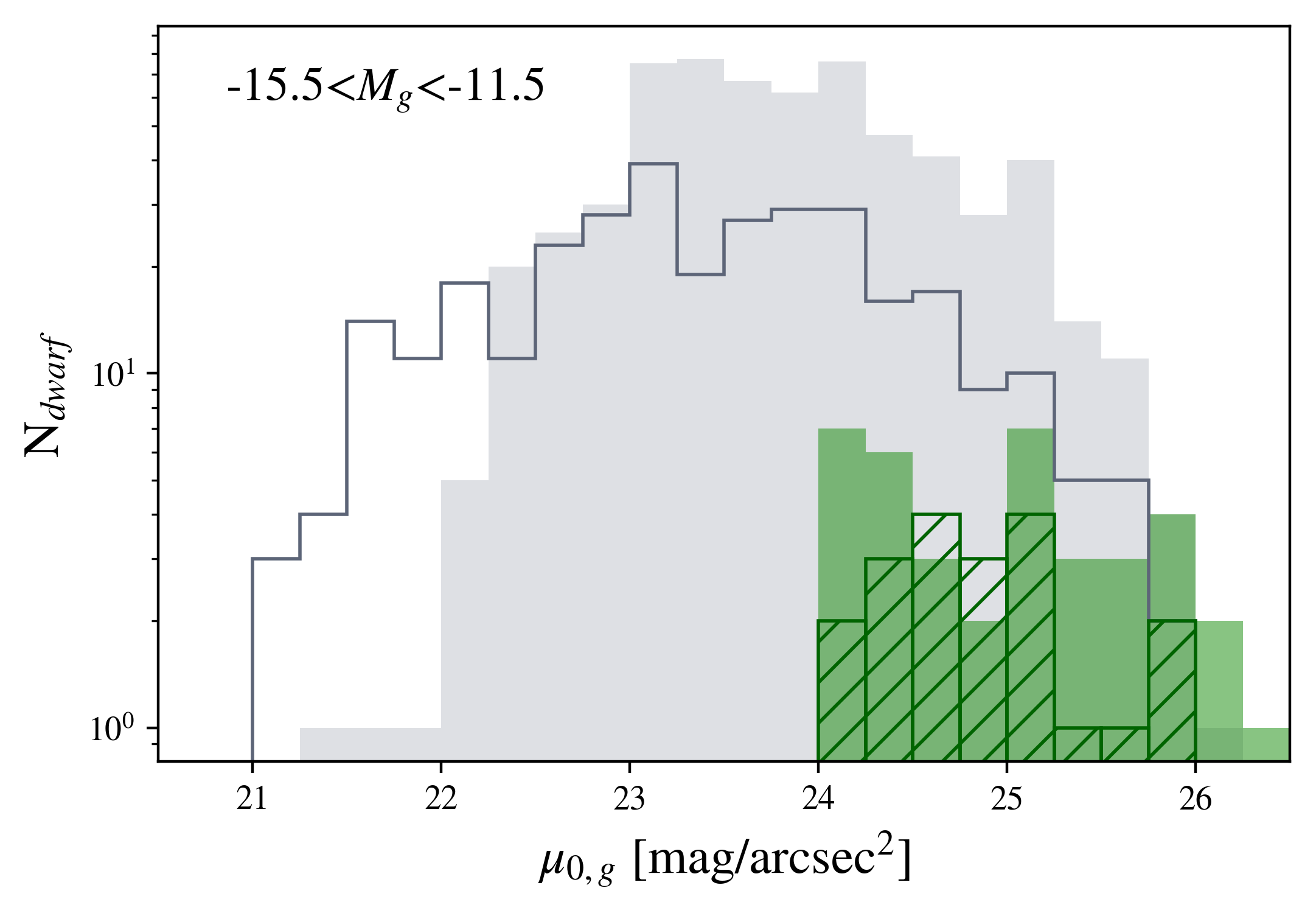}
    \includegraphics[scale=0.44]{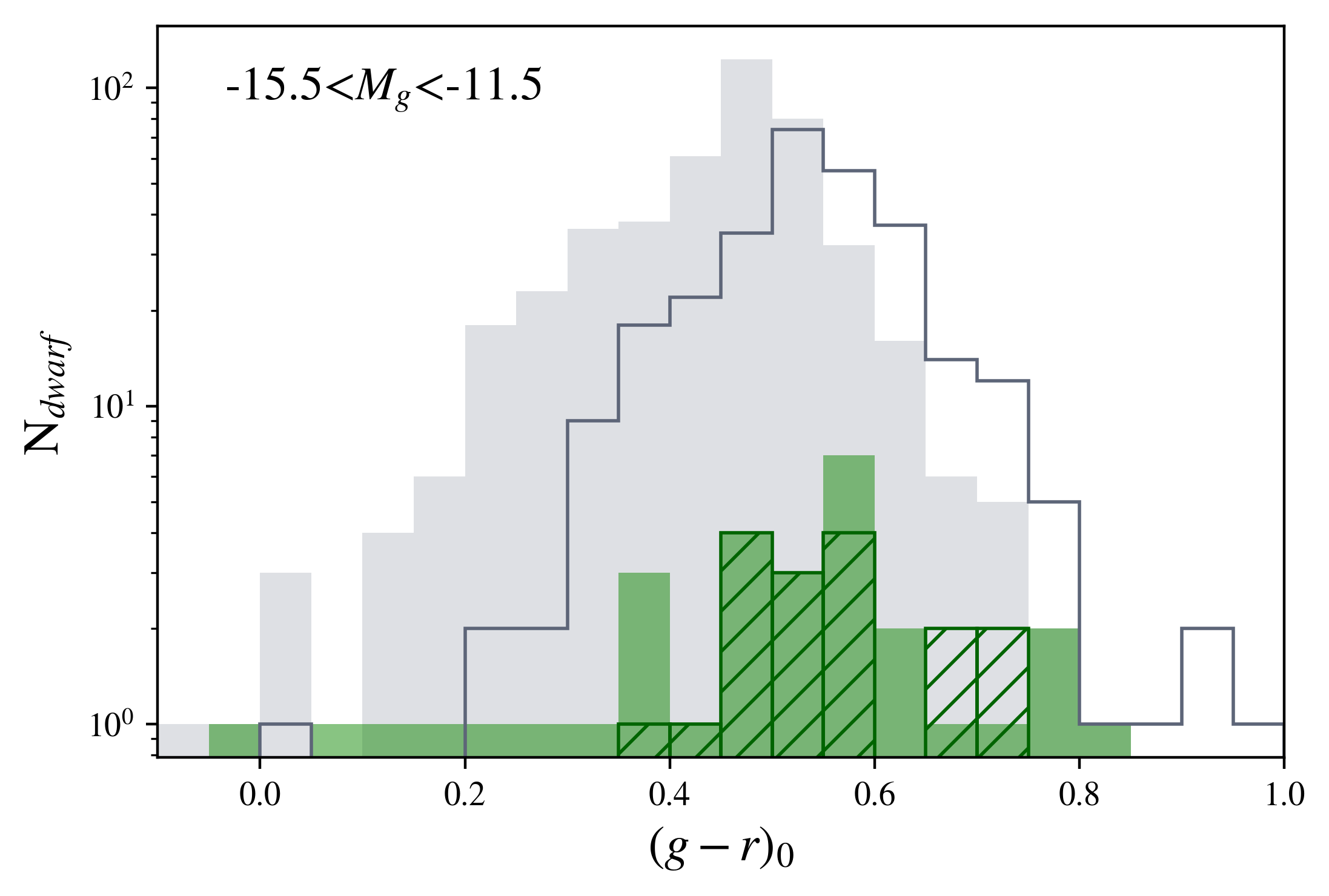}
  \caption{Comparison of distribution of photometric and structural properties --- absolute magnitude $M_g$, S\'ersic index $n$, effective radius $R_e$, axis ratio, central surface brightness $\mu_{0,g}$, and $(g-i)_0$ color --- between the UDGs ({\it green}) and the non-UDGs ({\it gray}) in the MATLAS dwarfs sample. The samples are divided by nucleation status: nucleated ({\it empty gray/hashed green}) and nonnucleated ({\it filled gray/green}). All dwarfs are shown for M$_g$ while only galaxies in the range of the UDGs luminosity, $-15.5<M_g<-11.5$, are considered for the other parameters.}
  \label{fig:propdwarfs}
\end{figure*}

\begin{figure*}
  \centering
    \includegraphics[scale=0.6]{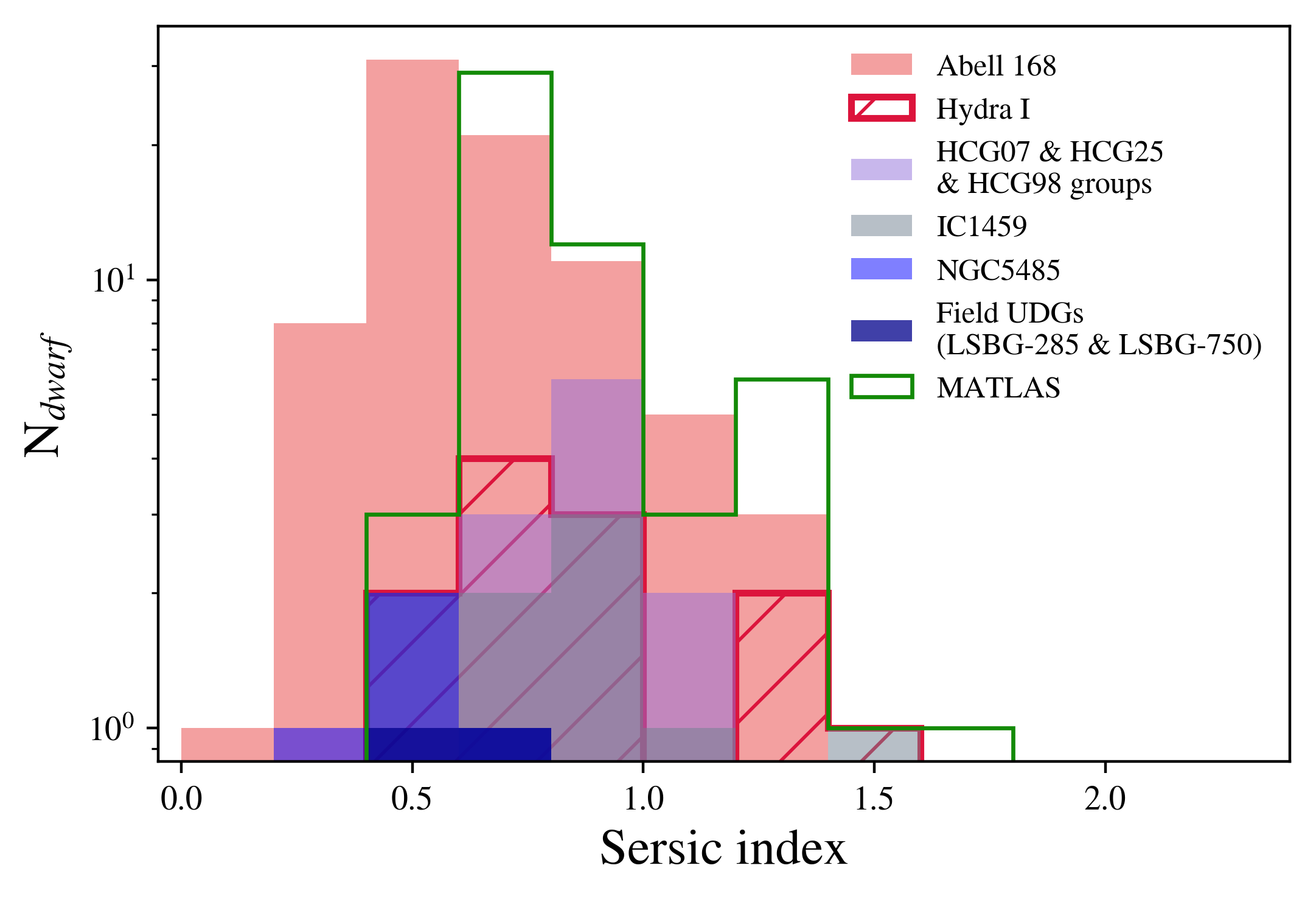}
    \includegraphics[scale=0.6]{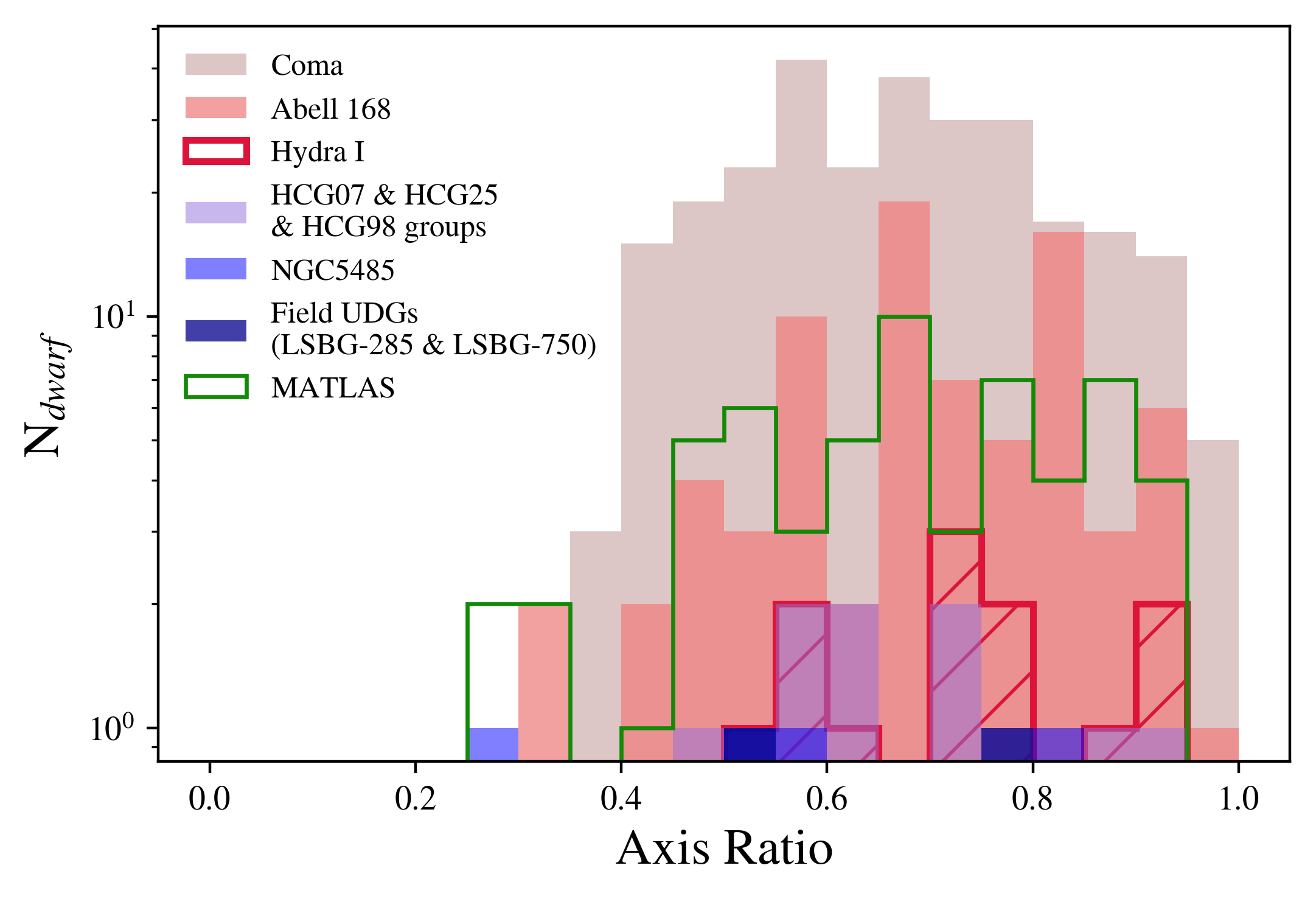}
    \includegraphics[scale=0.75]{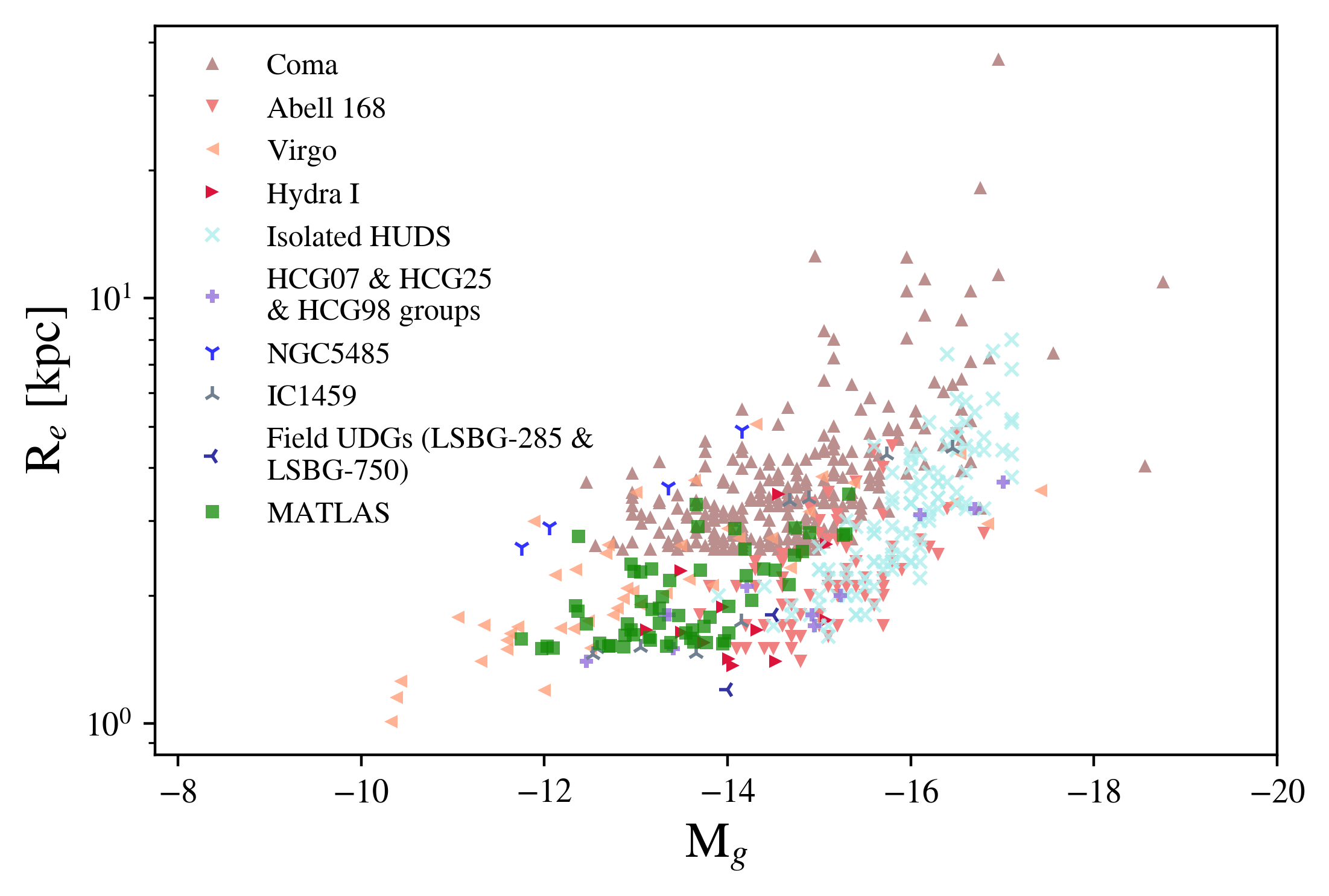}
  \caption{Structural properties of the MATLAS UDGs as compared to the ones of UDGs from the Coma, Abell~168, Virgo and Hydra I clusters \citep{Zaritsky2019,Roman2017a,Lim2020,Iodice2020}, from the HCG~07, HCG~25, HCG~98, IC~1459 and NGC~5485 groups \citep{Roman2017b,Forbes2020,Merritt2016} as well as of isolated UDGs (HI-bearing UDGs sample, LSBG-285 and LSBG-750, \citealt{Leisman2017,Greco2018}). We display the UDGs from high density (cluster) environments with shades of red and triangular markers, the UDGs from moderate (group) to low (field) density environments with shades of blue and cross-like markers and the MATLAS UDGs with a green color and squares. {\it Top}: distributions of S\'ersic index ({\it left}) and axis-ratio ({\it right}). {\it Bottom}: scaling relation $M_g$ vs.\ $R_{e}$. We note that a different cut in size and surface brightness was applied to select the UDG samples in \citet{Greco2018,Lim2020,Zaritsky2019,Roman2017b,Leisman2017}.}
  \label{fig:propUDGs}
\end{figure*}

\section{UDG photometric properties}
\label{section:properties}

\subsection{Photometry and structural parameters}

In Figure~\ref{fig:propdwarfs} we display the S\'ersic index, effective radius, absolute magnitude, axis ratio and  $(g-i)_0$ color distribution of the UDGs, as compared to the traditional dwarf (non-UDGs) galaxy sample. We note that when we compare the photometric and structural properties of these two samples, we compare galaxies of similar luminosity and therefore we only consider the range of luminosities where the two populations overlap, that is, $-15.5 < M_g < -11.5$. Other than the effective radius and the central surface brightness, which, by definition shows that the UDGs are selectively larger in size and fainter, the UDGs reach smaller values of S\'ersic index but have similar range of axis ratios and colors than that of the traditional dwarf (non-UDGs) galaxy sample.

We compare in Figure~\ref{fig:propUDGs} the structural parameters of our UDGs with those located in the high density environments of clusters (Coma; \citealt{Zaritsky2019}, Virgo; \citealt{Lim2020}, Abell~168; \citealt{Roman2017a} and Hydra I; \citealt{Iodice2020}) and in moderate to low density environments (around NGC5485: \citealt{Merritt2016}, around IC~1459: \citealt{Forbes2020}, around HCG~07, HCG~25 and HCG~98: \citealt{Roman2017b}, and in the field: \citealt{Greco2018,Leisman2017}). For the comparison, we use the absolute magnitude range of $-15.1 < M_g < -13.0$ that is common to all clusters. We note that a different cut in size and surface brightness was applied to select the UDG samples in \citet{Greco2018,Lim2020,Zaritsky2019,Roman2017b,Leisman2017}.
Focusing on the Sérsic index, the MATLAS UDGs have a larger minimum value than the Abell~168 cluster UDGs but similar to Hydra I cluster and the low density environment UDGs. This highlight a possible trend for the UDGs in low density environments to have a brighter center than the UDGs in the high density environments. \citet{Pina2019} studied a sample of 442 UDGs located in eight nearby clusters ($z$<0.035). We compare the ranges of Sérsic index found for all eight clusters and find a similar tendency to the Abell 168 cluster, as the MATLAS UDGs minimum value is larger than the minimum indices of five clusters. And we also observe this trend by comparing our sample to the UDGs located in the Abell 370 cluster \citep{Lee2020}. However, we do not observe any relation between the Sérsic index and the local volume density $\rho_{10}$ of the MATLAS UDG. Now, looking at the distributions for the axis ratio, we observe similar ranges for all the samples. Concerning the luminosity and the size of the UDGs, the MATLAS UDGs have $R_e$ below 3.5~kpc. For a similar range of M$_g$ ($-15.5$ to $-11.5$), the Coma cluster has a few larger UDGs, reaching 12.6~kpc, and two UDGs, one in the Virgo cluster and one around NGC~5485, have $R_e \sim$ 5~kpc while the other samples show ranges in $R_e$ similar to MATLAS.

The detection of the dwarfs was based solely on the \textit{g}-band imaging, but magnitudes were obtained using \textsc{Galfit} on the \textit{g}-, \textit{r}- and \textit{i}-band images, allowing us to extract $g-r$ and $g-i$ colors for 1307 and 782 MATLAS dwarfs, including 49 UDGs, respectively. We extracted the magnitudes in the \textit{r}- and \textit{i}-bands of all the dwarfs having an available observation and a good model in the \textit{g}-band with the use of \textsc{Galfit}. We used the same model as in the \textit{g}-band, leaving only the magnitude free to change for the initial input parameters of \textsc{Galfit}. For the nucleated galaxies, the galaxy and nucleus was modeled and photometry was extracted separately \citep{Poulain2021b}.

The magnitudes were corrected for Galactic extinction that were obtained from the IRSA database at the coordinates of each dwarf, using the reddening values from \citet{Schlafly2011} and assuming $R_V = 3.1$. The median \textit{g}, \textit{r}, and \textit{i}-bands extinction corrections are 0.09, 0.06, and 0.04, respectively. Applying these corrections, we measure median colors ($g-r$)$_ 0$ = 0.48 and ($g-i$)$_0$ = 0.73 for the full dwarf sample and ($g-r$)$_ 0$ = 0.49 and ($g-i$)$_0$ = 0.74 for the traditional dwarf (non-UDG) galaxies.

\begin{figure}
\centering
\includegraphics[width=\linewidth]{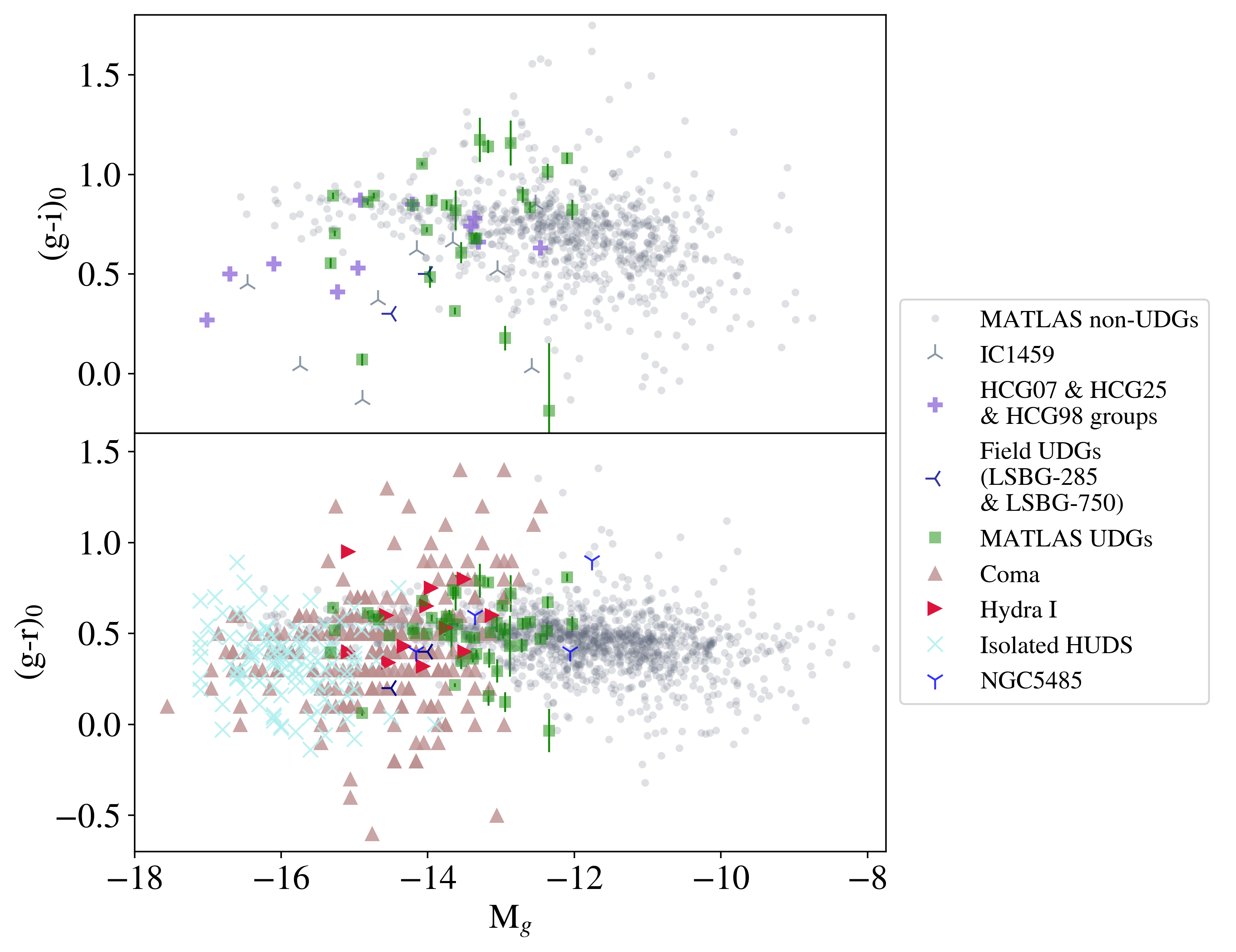}
\caption{{\it Top}: ($g-i$)$_0$ color as a function of the $M_g$ for the MATLAS non-UDGs ({\it gray dots}) and UDGs ({\it green squares}) compared to UDGs located in groups and in the field ({\it cross-like blue markers}). {\it Bottom}: ($g-r$)$_0$ color as a function of $M_g$ for the MATLAS non-UDGs ({\it gray dots}) and UDGs ({\it green squares}) as compared to UDGs located in the Coma and Hydra I clusters ({\it red triangles}), in groups and in the field ({\it cross-like blue markers}). The lines of markers visible for Coma's $g-r$ color are due to the rounded absolute magnitudes used for the calculation.}
\label{fig:color}
\end{figure}

The color of UDGs have been previously reported to depend on their environment. In the Coma cluster, they appear to follow the red sequence \citep{Koda2015}, while they show bluer colors in the outskirts of groups \citep{Roman2017b} as well as blue colors and star forming activity in the field \citep{Prole2019b}. In Figure~\ref{fig:color} we compare the colors of our UDGs to the MATLAS traditional dwarf (non-UDGs) sample and to the samples of UDGs from Figure \ref{fig:propUDGs} with available $g-r$ or $g-i$ color. Our UDGs have colors in the same range as the MATLAS traditional dwarf (non-UDGs) galaxies, which show red to blue colors, with ($g-r$)$_0=-0.1$ to $0.8$ and ($g-i$)$_0=-0.2$ to $1.2$. No UDG catalogs in the cluster environment had available $g-i$ color, allowing us to compare our UDG sample (Figure~\ref{fig:color}, green squares) with low density environment UDGs only (Figure~\ref{fig:color}, various blue markers). We find that some MATLAS UDGs are as blue as the bluest UDGs (in IC1459) while other MATLAS UDGs are redder than the reddest UDG of the low density environment samples. 

We can compare the MATLAS UDGs (Figure~\ref{fig:color}, green squares) to the one in the Coma and Hydra I clusters (Figure~\ref{fig:color}, various red markers) using $g-r$ color, in addition to samples of UDGs from low density environments (Figure~\ref{fig:color}, various blue markers). In this color-magnitude diagram, the MATLAS UDGs show a similar range of color to the ones from other low density environments while we find some bluer and redder UDGs in the Coma cluster. The median color of UDGs in the MATLAS sample, ($g-r$)$_0 = 0.36$, is as red as the one measured in Coma, ($g-r$)$_0=0.40$, but the range of colors is narrower. The Hydra I cluster median color is redder, with ($g-r$)$_0=0.57$, and the UDGs are as red as the red UDGs from low density environments while they do not show as blue UDGs. 

In summary, other than the predefined size and surface brightness cut, the MATLAS UDGs do not appear to have significantly different photometric and structural properties than the traditional dwarf (non-UDGs) galaxies. They also show similar colors than the MATLAS traditional dwarf (non-UDGs) galaxies in groups and in the field. Their median color is as red as the one measured in galaxy clusters such as Coma, but the range of colors is narrower. A table of the photometric properties of the MATLAS UDGs can be found in Appendix~\ref{section:appendix1}.

\subsection{Nucleated}

\begin{figure}
\centering
\includegraphics[width=\linewidth]{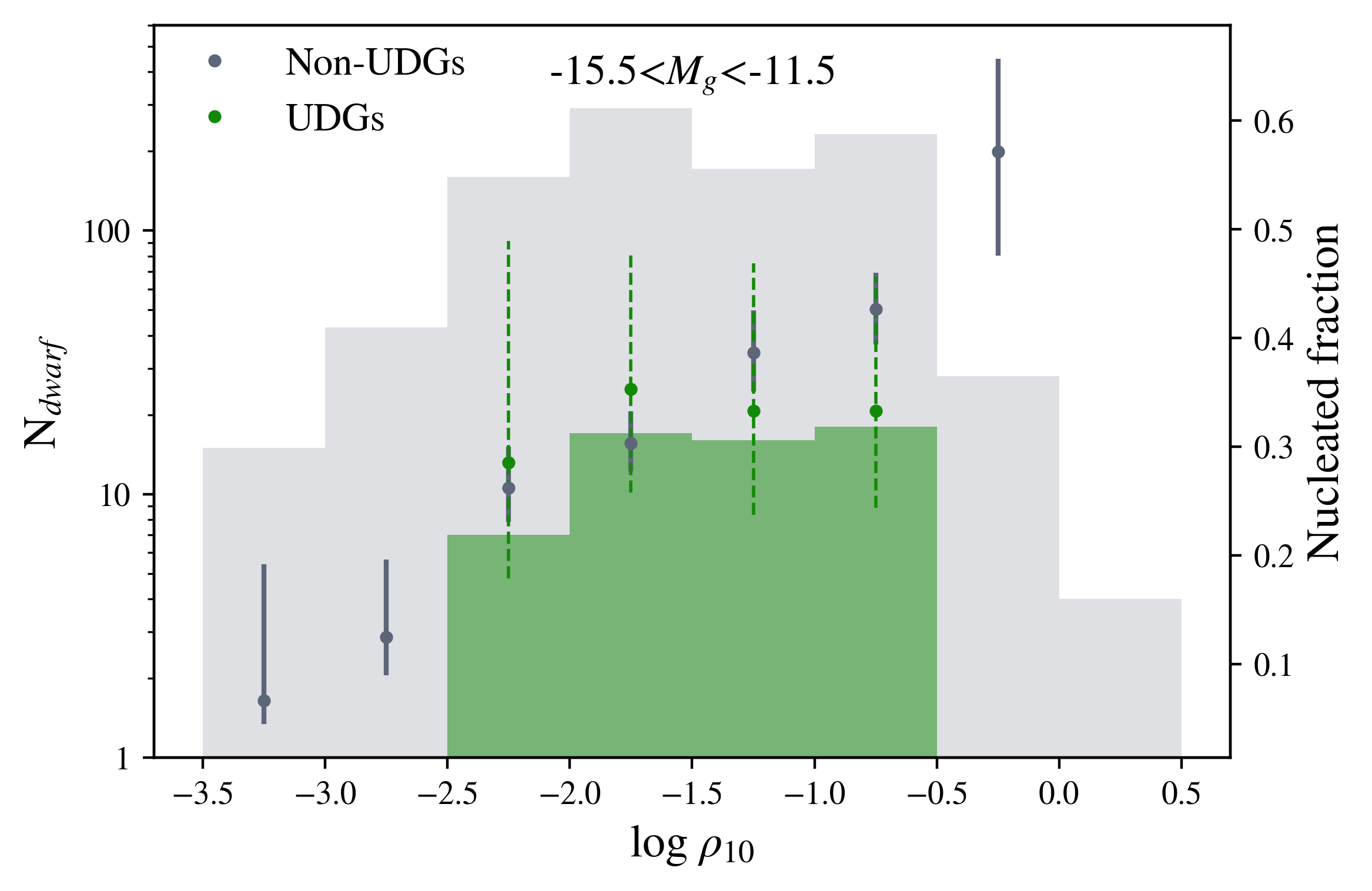}
\caption{Histograms: UDG ({\it green}) and non-UDGs ({\it gray}) counts in each $\rho_{10}$ bin. We computed the fraction for bins of at least five galaxies. Points: Nucleated fraction as a function of the local volume density $\rho_{10}$ of the UDGs ({\it green squares}) and non-UDGs ({\it gray dots}). Error bars: 1$\sigma$ binomial confidence intervals. Only galaxies in the range of the UDGs luminosity, $-15.5<M_g<-11.5$, are considered in this analysis.}
\label{fig:nucfraction}
\end{figure}

During the visual classification of the 2210 MATLAS dwarf galaxies, 507 nucleated dwarfs (22.9\%) were identified. Considering only those with \textsc{Galfit} parameters, 425 (26.7\%) are nucleated. We define a nucleus as a compact source within $\sim$ 0.5 $R_e$ of the dwarf photocenter, that appears to be the brightest compact source within the dwarf’s effective radius. In the UDG sample, we find 20 nucleated, leading to a nucleated fraction of 33.9\%. One of these is irregular in morphology while the others are ellipticals. The nucleated fraction of the UDGs is similar to the one found in the Coma and Hydra I clusters \citep{Lim2018,Iodice2020}. Looking at the dwarfs with $M_g$ in the range of the UDGs luminosity ($-15.5$ to $-11.5$), the nucleated fraction of non-UDGs is 33.8\%, similar to the nucleated fraction of the UDGs. The nucleated fraction of dwarf galaxies depends on the galaxy mass and environment \citep{Lim2018,Janssens2019b}. We know that the nucleated fraction of the MATLAS dwarfs is higher in more massive galaxies and that, for a similar stellar mass, the fraction is systematically lower than the nucleated fraction found in the Virgo cluster \citep{Poulain2021b}. We investigate the effect of the environment on the nucleated fraction of the MATLAS dwarfs non-UDGs and UDGs by using the local volume density $\rho_{10}$. We show the relation between the nucleated fraction and $\rho_{10}$ in Figure \ref{fig:nucfraction}. We can see, as expected, an increase of the fraction toward higher local density environment for the non-UDGs. The same increase is seen only in the lowest two bins for the UDGs. \citet{Lim2018} found that the fraction of nucleated UDGs is lower than the one of the dwarfs at the center of the Coma cluster. We observe a tendency for the nucleated fraction of the UDGs to be larger than the one for the non-UDGs in the lowest density environments while it appears to be lower in the highest density environments. We note however that if we consider the error bars, the fractions of both samples are in agreement in most of the density bins.

\begin{figure}
\centering
\includegraphics[width=\linewidth]{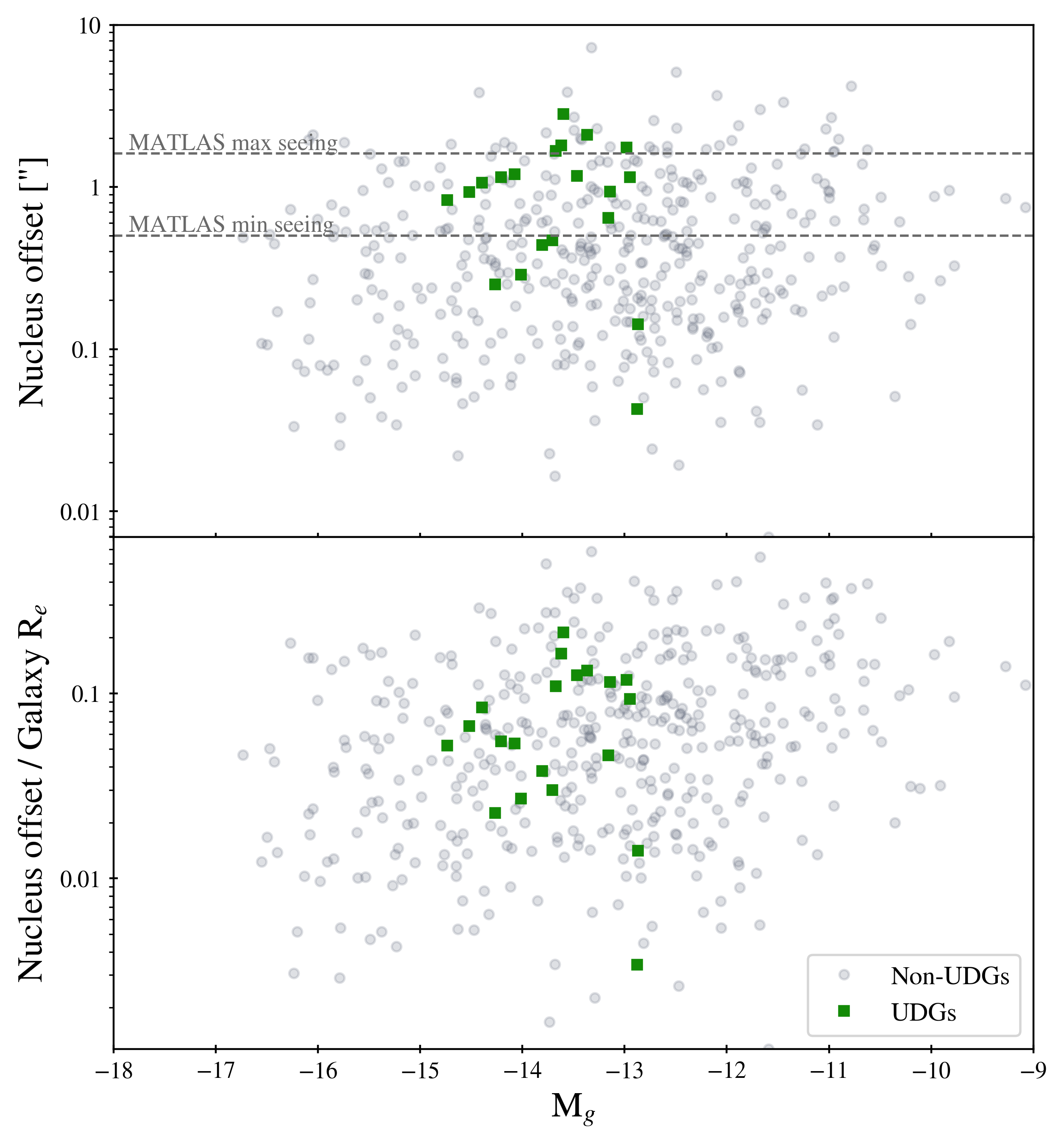}
\caption{Distribution of the estimate offset distance between the photometric center of the host galaxy and the nucleus as a function of the absolute magnitude $M_g$ of the host. {\it Top}: offset in arcseconds. {\it Bottom}: offset in fraction of $R_e$ of the host.}
\label{fig:nuc}
\end{figure}

\begin{figure}
\centering
\includegraphics[width=\linewidth]{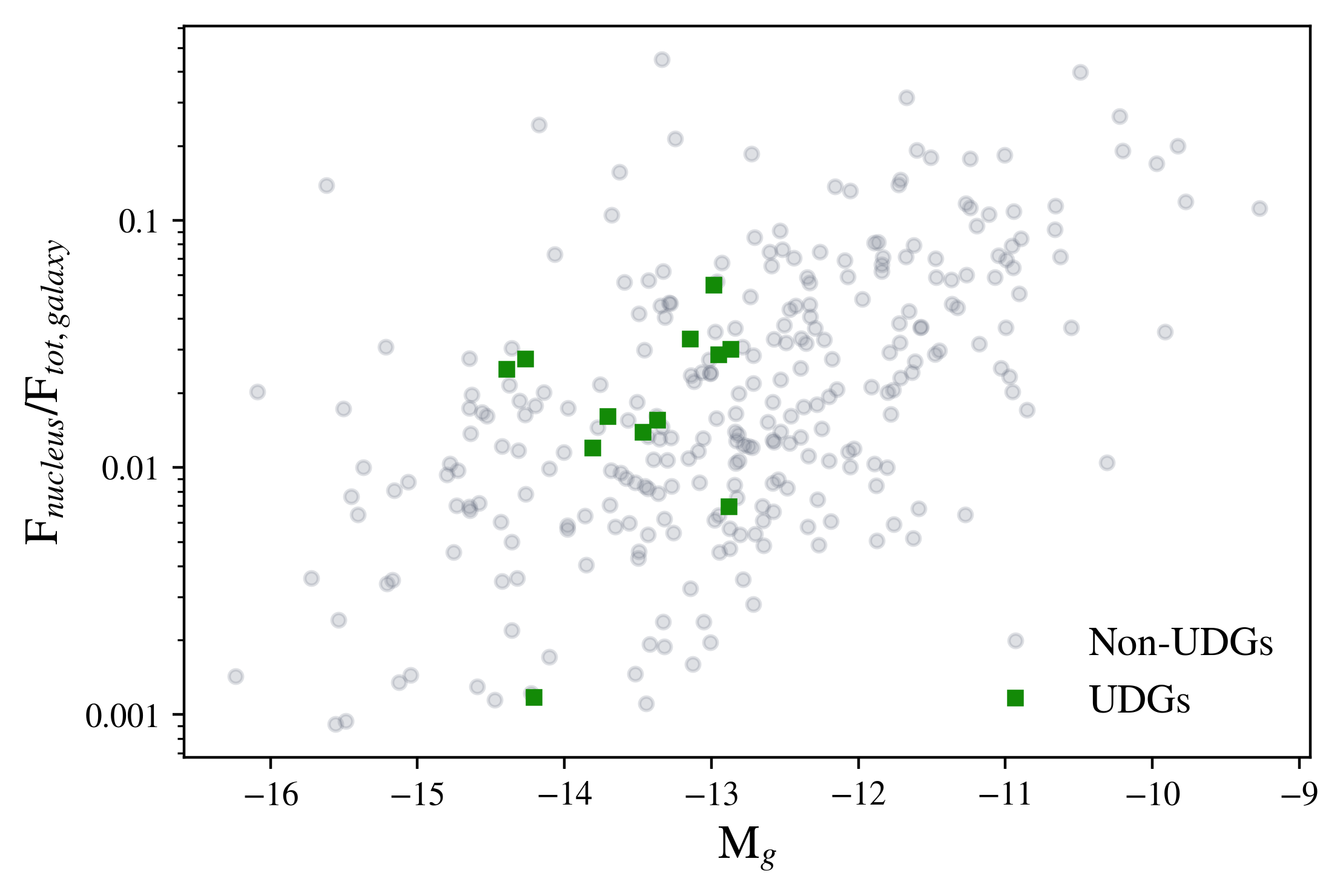}
\includegraphics[width=\linewidth]{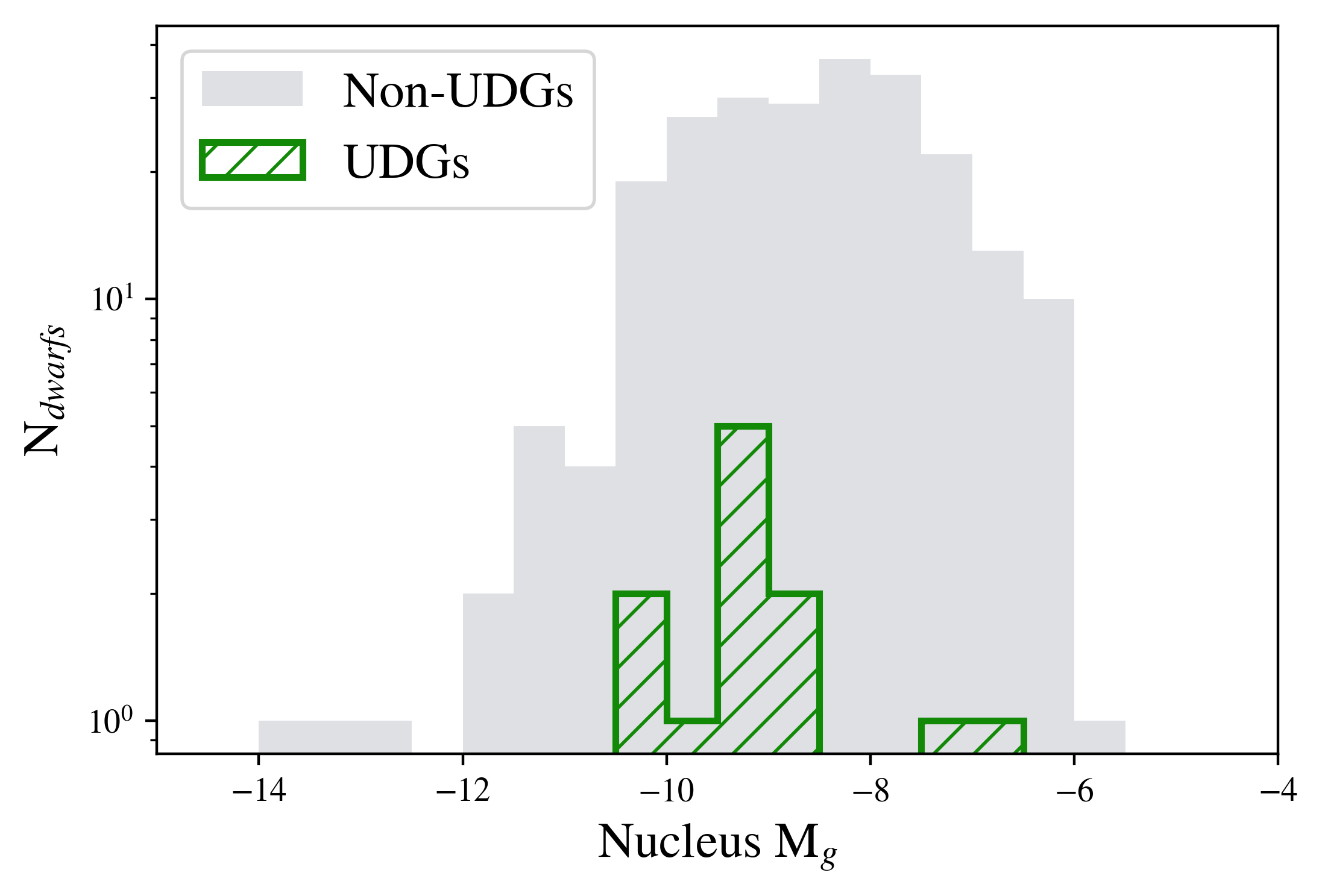}
\caption{Luminosity of MATLAS UDGs nuclei as compared with the ones of the MATLAS dwarfs. Top: contribution of the nucleus to the total luminosity of the galaxy as a function of the galaxy absolute magnitude. Bottom: distribution of absolute magnitude of the nuclei of the MATLAS dwarfs and UDGs. No difference in luminosity is visible between the UDGs and dwarfs nuclei.}
\label{fig:fractionfluxnuc}
\end{figure}

\begin{figure}
\centering
\includegraphics[width=\linewidth]{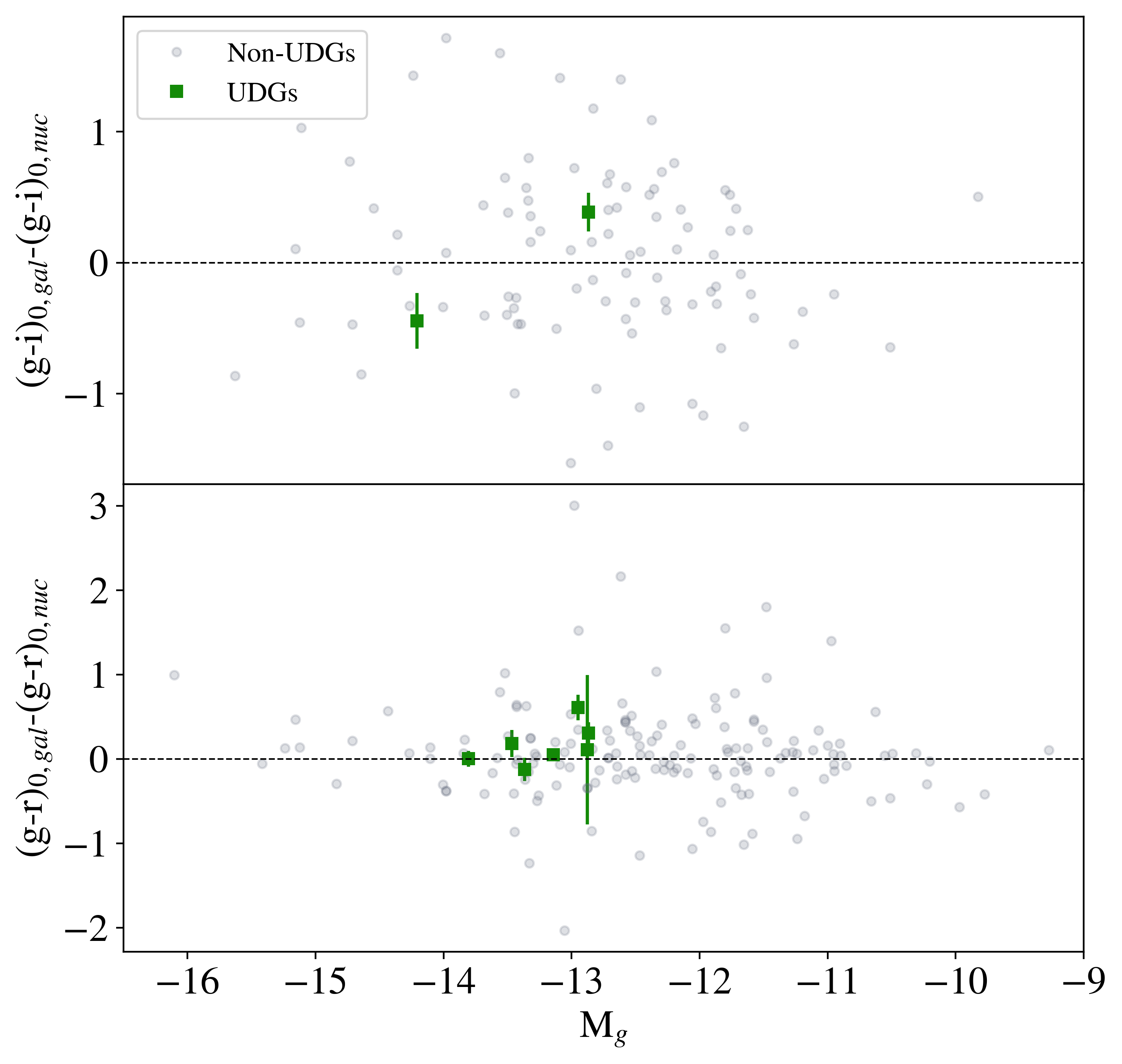}
\caption{Difference of color between the galaxy and the nucleus for the MATLAS non-UDGs ({\it gray dots}) and UDGs ({\it green squares}). {\it Top}: $(g-i)_0$ color. {\it Bottom}: ($g-r$)$_0$ color. The {\it black dashed lines} represent equal colors for the galaxy and nucleus. We see a tendency for the nucleus to be bluer than the galaxy in the case of the UDGs.}
\label{fig:nuccol}
\end{figure}

We investigate the properties of the UDG nuclei, as compared to the nuclei in the traditional dwarf (non-UDGs) galaxies. We look at the UDG nuclei offset from the host galaxy photometric center. We show in Figure~\ref{fig:nuc} this separation both in units arcseconds and fraction of the galaxy $R_e$ as a function of the galaxy $M_g$. All the UDG nuclei are located within $\sim$0.2 $R_e$ [$\sim$3\arcsec] with a median separation of 0.06 $R_e$ [1\arcsec]. The median angular offset is larger than the angular offset of non-UDGs of similar luminosity (0.36\arcsec), but in units of $R_e$, both samples are similar (0.06 $R_e$). We now focus on the nuclei luminosity by considering the contribution of the nucleus to the total luminosity of the galaxy and the nuclei $M_g$ (Figure \ref{fig:fractionfluxnuc}). Considering galaxies of similar luminosity ($-15.5<M_g<-11.5$), while we find a maximum contribution much larger for the traditional dwarf (non-UDGs) galaxies than for the UDGs (44.8\% compared to 5.5\%), the median contributions of the UDGs and non-UDGs nuclei are similar. Moreover, we observe on average brighter nuclei in the UDGs than in the non-UDGs. \citet{Forbes2020} report two single nucleated and one double nucleated UDGs in the IC1459 group which nuclei show redder colors than the galaxies. Figure~\ref{fig:nuccol} compares the $(g-i)_0$ and ($g-r$)$_0$ colors of the galaxy to the colors of the nuclei for the MATLAS non-UDGs and UDGs samples. While we find both bluer and redder nuclei in the traditional dwarf sample, we observe a tendency of the nuclei to be bluer than the galaxy in the case of the UDGs.

\subsection{Stellar masses}
\label{section:mstar}

\begin{figure}
\centering
\includegraphics[width=\linewidth]{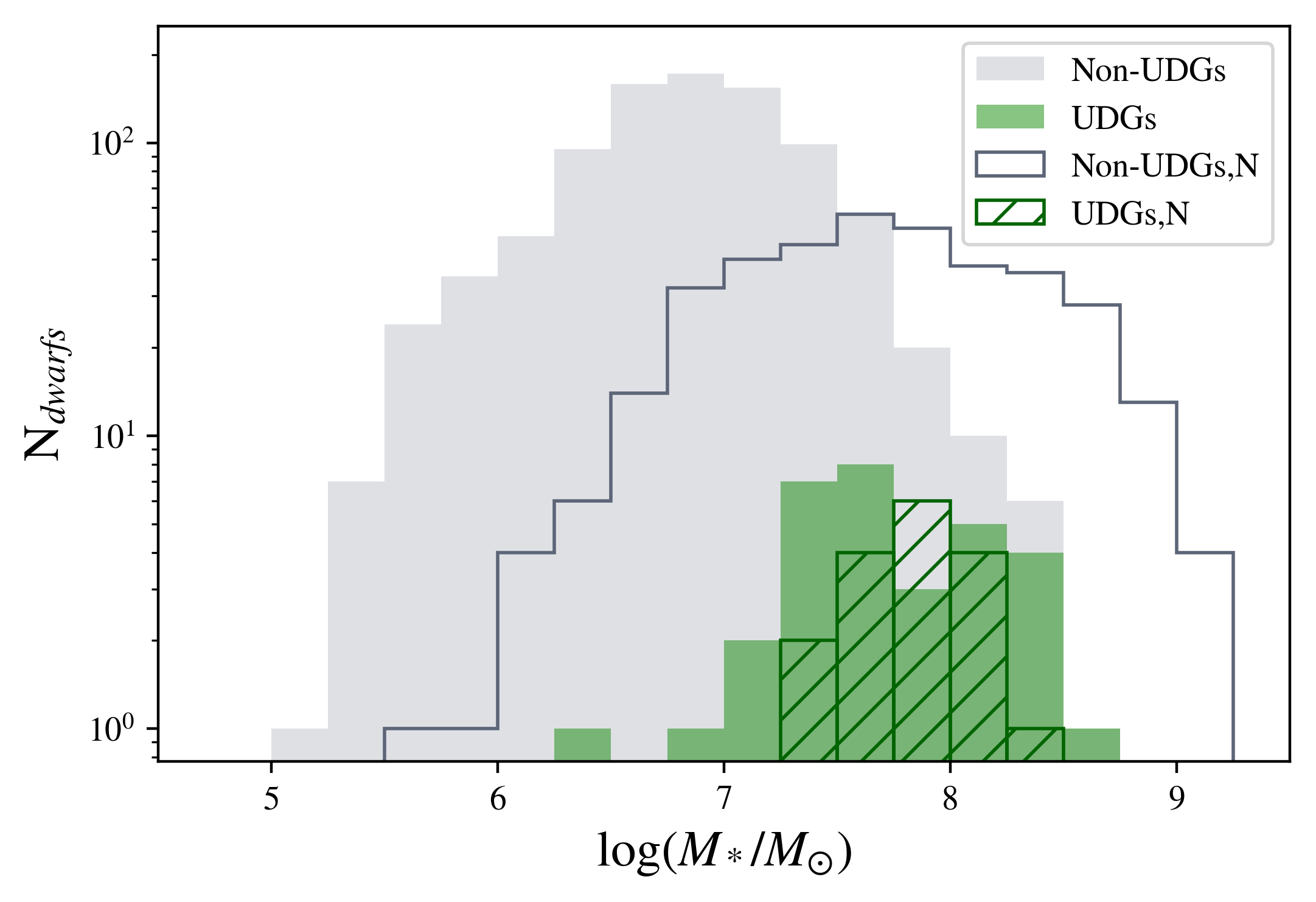}
\caption{Distribution of stellar masses for the 49 UDGs ({\it green}) and 1258 traditional dwarf (non-UDGs) galaxies ({\it gray}) with \textsc{Galfit} parameters and $(g-r)$ colors, based on the \citet{Bell2003} color-mass relation. The samples are divided by nucleation status: nucleated ({\it empty gray/hashed green}) and nonnucleated ({\it filled gray/green}).}
\label{fig:stellarmass}
\end{figure}

Using the colors above and the distance --- taken to be the dwarf distance when available, otherwise the ETG distance is assumed --- it is possible to estimate their stellar masses. We computed the stellar masses $M_\ast$ based on the stellar mass-to-light ratios from \citet{Bell2003} and the derived ($g-r$)$_0$ color\footnote{Formula: $log(M_*/M_{\odot})=-0.306+1.097(g-r)_0-0.4(M_r-4.77)$}. Stellar masses were measured for 1307 dwarfs, including 49 UDGs. The distributions of stellar masses for the MATLAS traditional dwarf (non-UDGs) galaxies and UDGs with \textsc{Galfit} parameters and $g-r$ colors are shown in Figure~\ref{fig:stellarmass}. As expected, the MATLAS UDGs represent a subsample of the massive dwarfs, but not the most massive ones which are all nucleated non-UDGs. With stellar masses ranging from $\log(M_\ast)\sim 6.5$ to $\log(M_\ast)\sim 8.7$, the MATLAS UDGs have similar masses to UDGs observed in the field, groups and clusters ($\log(M_\ast)\sim 6.8 - 8.6$, \citealt{Barbosa2020,Greco2018,Leisman2017,Forbes2020,Pina2019}).

\begin{figure*}
\centering
\includegraphics[width=0.48\textwidth]{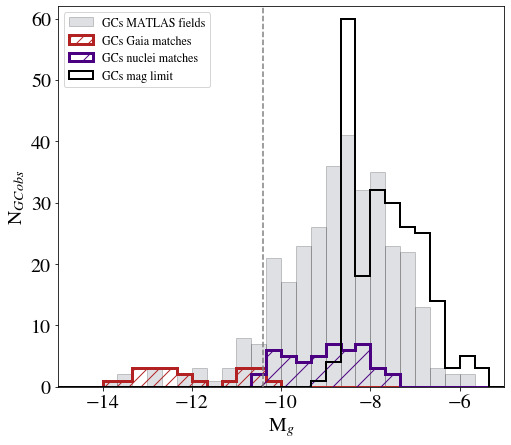}
\includegraphics[width=0.5\textwidth]{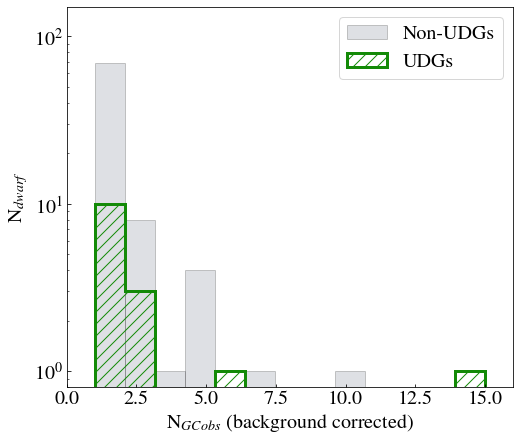}
\caption{{\it Left}: The histograms of the absolute magnitudes of all detected GCs in the MATLAS fields ({\it filled gray}), the GCs that are Gaia matches ({\it hashed red}) and the GCs that are nuclei matches ({\it hashed violet}). The Gaia matches were removed from the GC sample but the nuclei matches were kept since some may be real GCs seen in projection at the center of the galaxy. The GC detection limiting absolute magnitudes for each galaxy are displayed as the {\it black solid line} histogram. The {\it dashed gray vertical line} shows the absolute magnitude of the most luminous Milky Way globular cluster, Omega Centauri (M$_V = -10.4$). {\it Right}: The histogram of the GC counts $N_{GCobs}$, Gaia matches removed and background corrected, for the MATLAS non-UDGs ({\it filled gray}) and UDGs ({\it hashed green}). }
\label{fig:GCLF}
\end{figure*}

\section{UDG globular cluster populations}
\label{section:GCs}

\subsection{Catalog}

The GC system of UDGs provides important clues on their stellar population and the nature of their dark matter halo. In particular, the stellar population of GCs provides information about the early epoch of galaxy formation when intense star formation occurred and massive star clusters were formed. They can be used as a means of estimating the total mass of a galaxy based solely on photometric measures since the number of GCs in a galaxy has been found to correlate linearly with the total host halo mass (e.g., \citealt{Blakeslee1999,Spitler2009,Georgiev2010,Hudson2014,Harris2013,Harris2017,Burkert2020}). Assuming that the GCs trace the underlying gravitational potential, are in dynamical equilibrium, and are pressure-support dominated, their velocity dispersion can also be used to estimate the total mass of the system \citep{Doppel2021,Toloba2018,Forbes2017,Alabi2016,Zhu2014,Woodley2007}.

We studied the GC population of the MATLAS UDGs by selecting GC candidates on the basis of their colors and size information using the \textsc{SExtractor} package (see \citealt{Durrell2014,Munoz2014,Lim2017}). The GCs at the distances of MATLAS UDGs appear as point-like sources, so we selected point-like sources based on concentration indices ($\Delta_{4-8}$), that is, the difference between four- and eight-pixel diameter aperture-corrected magnitudes. We estimated $\Delta_{4-8}$ on the two best seeing filter images in each field and convolved the $\Delta_{4-8}$ values. If the seeing of the second best seeing image was greater than 1.3 times the one of the best seeing image, then we only used $\Delta_{4-8}$ from the best seeing image. We chose objects with $-0.08 < \Delta_{4-8} < 0.08$ range for most MATLAS dwarfs, but we used $-0.08 < \Delta_{4-8} < 0.16$ range for distances closer than 20~Mpc and seeing better than 1 arcsec. Among these point-like sources we chose GCs using their color information, ($u,g,i$), ($g,r,i$), or ($g,r$), depending on the bands available (see Section~\ref{section:cat}). The colors of spectroscopically confirmed GCs in M87 were used as references of GC color regions (see \citealt{Munoz2014,Lim2017} for details). We limited our GC candidates to brighter than $g-$band magnitudes 24.5 in most fields, and we used $m_g \leq 24.0$ for GC candidates in fields which only have two-band data. For some low image-quality fields (seeing greater than 1.1 arcsec), we limited our GC candidates to those with $m_g \leq 23.5$. 

In total, GCs were detected in 136 of the 150 fields, with only 14 fields missing due to problems in the images. 

We cross-checked our final GC catalog with the Gaia DR2 catalog \citep{GaiaDR22018a,GaiaDR22018b,Gaiaastro2018}, with a maximum separation of 1\arcsec. To ensure the nearby nature of the matched sources, we selected only the sources that have a parallax value larger than the parallax error. We find 10\% matches that we then remove from the GC catalog. We note that we also matched our catalog of dwarf nuclei to the GC catalog in order to identify possible nuclei contamination. The histograms of the absolute magnitudes of all detected GCs in the MATLAS fields,  the GCs that are Gaia matches and the GCs that are nuclei matches are shown in Figure~\ref{fig:GCLF} ({\it left}). Unlike the Gaia matches, we did not remove the nuclei matches since it is not possible to determine if the nucleus is real or is a GC projected near the galaxy center.

\subsection{Background and completeness correction}

In order to compute the total number of GCs associated with a traditional dwarf or a UDG, we first compute the number of GCs within $2 R_e$ (a reasonable radius to select; see \citealt{Mueller2021}), subsequently referred to as $N_{GCobs}$. A total of 1589 MATLAS dwarfs have a measured effective radius \citep{Poulain2021b}, including, by definition, all of our 59 UDGs. Of these, 1380 have no GC candidate within $2 R_e$, 36 of which are UDGs (i.e., 209 MATLAS dwarfs have $N_{GCobs} > 0$, 23 of which are UDGs). We then applied a background correction. The number of sources per unit area was computed for the surrounding region of size $3$ to $7 R_e$. This number was then scaled to the circular area of radius $2 R_e$ and removed from $N_{GCobs}$, resulting in a background corrected $N_{GCobs}$. The background region outer radius of $7 R_e$ extends beyond the MATLAS field of view for 21 MATLAS dwarfs, including one UDG (MATLAS-1177). In this case, the background counts were corrected to account for the missing coverage. Figure~\ref{fig:GCLF} ({\it right}) shows the distribution of  the background corrected $N_{GCobs}$ for both the normal dwarfs (non-UDGs) and our UDGs sample with \textsc{Galfit} parameters.

Finally, a completeness correction was applied by assuming a GCLF represented by a Gaussian with peak absolute magnitude M$_V = -7.6\pm0.1$ mag \citep{Rejkuba2012} and taking into account the magnitude limit of the GC catalog in each MATLAS field. A correcting factor was then applied to the background corrected $N_{GCobs}$ based on the part of the GCLF that is fainter than the observed limit, resulting in the quantity $N_{GC}$. In summary, a total of 23 UDGs and 186 traditional dwarf (non-UDGs) galaxies have \textsc{Galfit} parameter $R_e$ and $N_{GC} > 0$.

\subsection{Specific frequency}

\begin{figure*}
\centering
\includegraphics[scale=0.49]{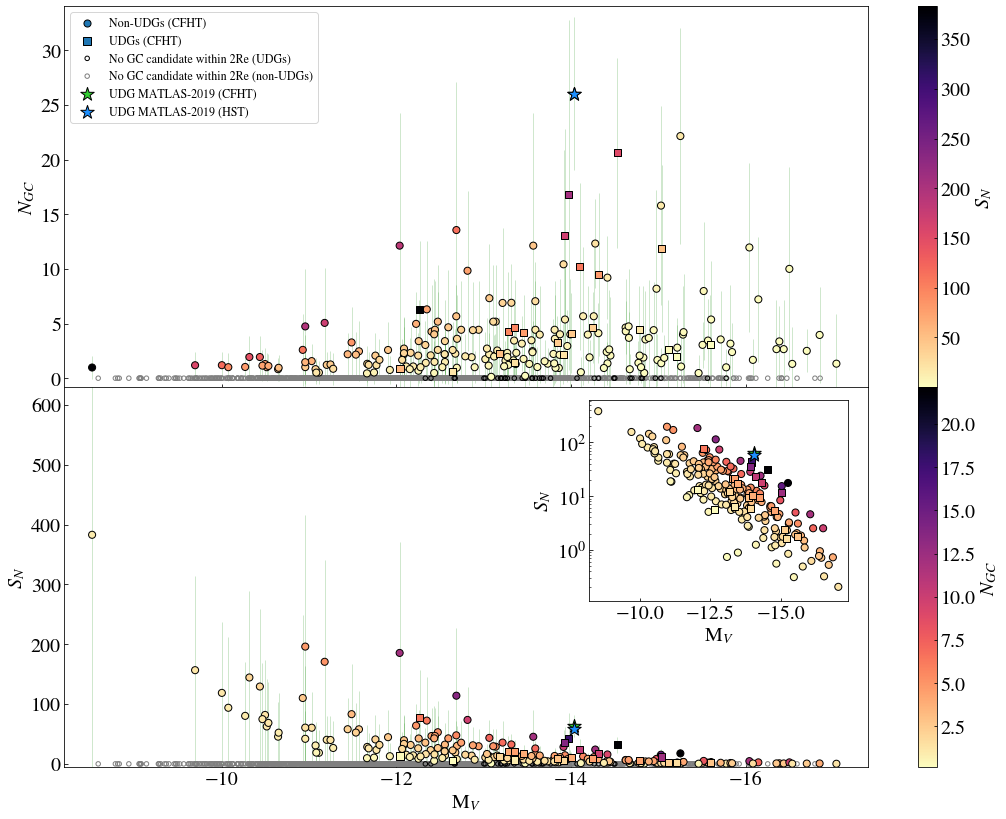}
\caption{{\it Top}: The background and completeness corrected $N_{GC}$ as a function of absolute magnitude for the 23 UDGs ({\it squares}) and 186 traditional dwarf (non-UDGs) galaxies ({\it circles}) with $N_{GC} > 0$. The color bar indicates the value of the specific frequency $S_N$. The UDG MATLAS-2019 observed with HST \citep{Mueller2021} is shown with the {\it star symbol}. {\it Bottom}: Same as left but now showing the specific frequency $S_N$ in the plot and the background and completeness corrected $N_{GC}$ in the color bar. The inset plot is with y-axis in log scale. In both plots, the 36 UDGs ({\it black open circles}) and 1344 non-UDGs ({\it gray open circles}) with a measured \textsc{Galfit} effective radius and no GC candidate within 2$R_e$ are shown. In total, 1380 out of the 1589 MATLAS dwarfs with a measured \textsc{Galfit} effective radius have no GC candidate within 2$R_e$.}
\label{fig:NgcSn}
\end{figure*}

As the number of GCs is a function of the brightness of the host galaxy, we show in Figure~\ref{fig:NgcSn} ({\it left}) the computed values of $N_{GC}$ as a function of the host galaxy absolute magnitude in the V-band $M_V$. The value for MATLAS-2019 is highlighted in the plot and compared to the one recently measured from HST observations and published in \citet{Mueller2021}. Our result based on the ground-based CFHT data is in very good agreement with the space-based HST data. 

The specific frequency $S_N$ \citep{Harris1981} was computed using the formula:
\begin{equation}
S_N=N_{GC}\cdot10^{0.4(M_V+15)}
\end{equation}

where $N_{GC}$ is the total, background and completeness corrected number of GCs. For MATLAS-2019, we obtain a value for $N_{GC}$ and $S_N$ of 26 and 63, respectively, in agreement with the value of $26\pm6$ and $58\pm14$ determined from the deep (sampling most of the GCLF) high resolution HST data \citep{Mueller2021}.

Prior studies of the GC population of UDGs in the Coma and Virgo clusters \citep{Lim2018,Lim2020} have claimed that the $S_N$ of these UDGs varies dramatically, with the mean $S_N$ being higher for UDGs than for classical dwarf (non-UDG) galaxies. Similarly, \citet{Forbes2020} finds that the richness of GCs in UDGs generally exceeds that found in normal dwarf galaxies of the same stellar mass. The specific frequencies of the MATLAS non-UDGs and UDGs are plotted in Figure~\ref{fig:NgcSn} ({\it right}). Considering galaxies of similar luminosity ($-15.5<M_g<-11.5$), the GC population of the traditional dwarf (non-UDGs) galaxies and UDGs span the same range of values. At the faint end, that is, for dwarfs with M$_V \gtrsim -11.5$, we find only traditional dwarfs since the UDGs that are that faint have no valid \textsc{Galfit} parameters due to their extremely low surface brightness. Hence, for dwarfs with $-15.5 \lesssim M_V \lesssim -11.5$ in low to moderate density environments, we find no evidence of a higher $S_N$ for UDGs than for traditional dwarf (non-UDG) galaxies, contrary to what is found in most clusters.

\subsection{Halo mass}

The virial mass of the MATLAS UDGs can be estimated using the number of GCs in a system \citep{Harris2013,Beasley2016b}. The virial mass of a galaxy scales linearly with the number of GCs over six orders of magnitude \citep{Harris2013}, the relation only flattens for halos with masses smaller than $10^{10}$\,M$_{\odot}$ or $N_{GC} \gtrsim 3$ (\citealt{Burkert2020}, see their Figure 1). According to \citet{Harris2017}, the virial mass $M_{halo}$ of a galaxy is connected to the total mass of the GC system $ M_{GC,tot}$ via the following formula:
\begin{equation}
    M_{GC,tot}/ M_{halo}=2.9\times10^{-5}
\end{equation}

The uncertainties associated with $M_{halo}$ are computed from the 0.28 dex scatter of the $M_{GC,tot}$-$M_{halo}$ relation \citep{Harris2017}. As in \citet{Mueller2021}, we assume a mean mass of a GC to be $1\times10^5$\,M$_{\odot}$ for dwarf galaxies \citep{Harris2017} and therefore multiply that number by $N_{GC}$ to compute $ M_{GC,tot}$.

\begin{figure}
\centering
\includegraphics[width=\linewidth]{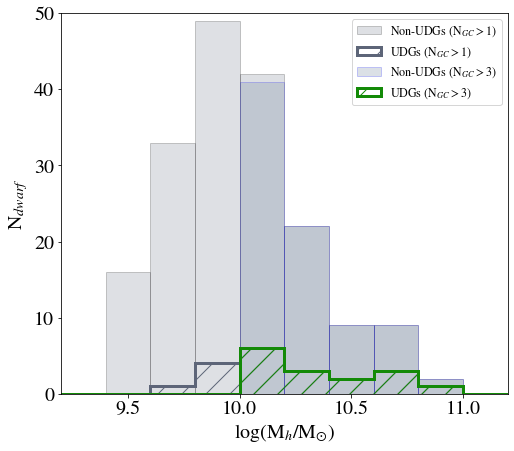}
\includegraphics[width=\linewidth]{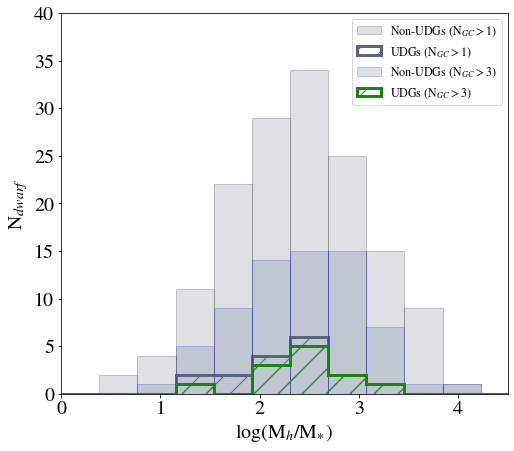}
\caption{{\it Top}: The distributions of halo masses for the 20 UDGs ({\it hashed green}) and 162 traditional dwarf (non-UDGs) galaxies ({\it filled gray}) with \textsc{Galfit} parameter $R_e$ and $N_{GC} > 1$. The {\it hashed green} (15 UDGs) and {\it filled blue} (68 non-UDGs) histograms are for the galaxies with $N_{GC} > 3$. {\it Bottom}: The distributions of halo-to-stellar mass ratios for the 17 UDGs ({\it hashed gray}) and 135 traditional dwarf (non-UDGs) galaxies ({\it filled gray}) with \textsc{Galfit} parameter $R_e$ and $N_{GC} > 1$ and stellar masses. The {\it hashed green} (12 UDGs) and {\it filled blue} (56 non-UDGs) histograms are for the galaxies with $N_{GC} > 3$ and stellar masses.}
\label{fig:halomass}
\end{figure}

The distributions of halo masses for the 23 UDGs and 186 traditional dwarf (non-UDGs) galaxies with \textsc{Galfit} parameter $R_e$ and $N_{GC} > 0$ are shown in Figure~\ref{fig:halomass} ({\it top}). The $N_{GC} > 3$ demarcation line is shown in the figure. The halo-to-stellar mass ratio distribution can be computed for those 19 UDGs and 151 traditional dwarf (non-UDGs) galaxies with estimated stellar masses (see Section~\ref{section:mstar}). As can be seen in Figure~\ref{fig:halomass} ({\it bottom}), the halo-to-stellar mass ratio distribution for the UDGs peaks at roughly the same value as for the traditional dwarf (non-UDGs) galaxies, but spans the smaller range of $M_h/M_\ast \sim 10-2000$. The distribution for $N_{GC} > 3$ is also shown in the figure.

For MATLAS-2019, we measure a GC system mass of $M_{GC,tot}=2.6\pm0.6\times10^{6}$\,M$_{\odot}$ and a halo mass of $M_{halo}=0.9\pm0.2\times10^{11}$\,M$_{\odot}$. A table of these derived properties of the MATLAS UDGs can be found in Appendix~\ref{section:appendix1}.


\section{Tidal features in UDGs}
\label{section:tidalfeatures}




\begin{figure*}
\centering
\includegraphics[scale=0.17]{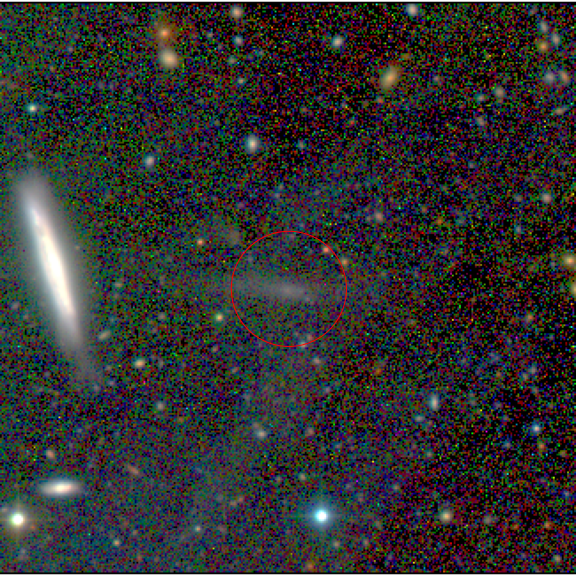}
\includegraphics[scale=0.17]{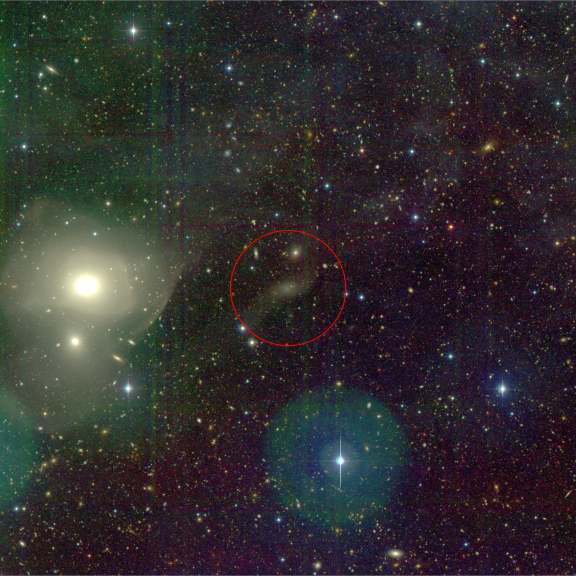}
\includegraphics[scale=0.17]{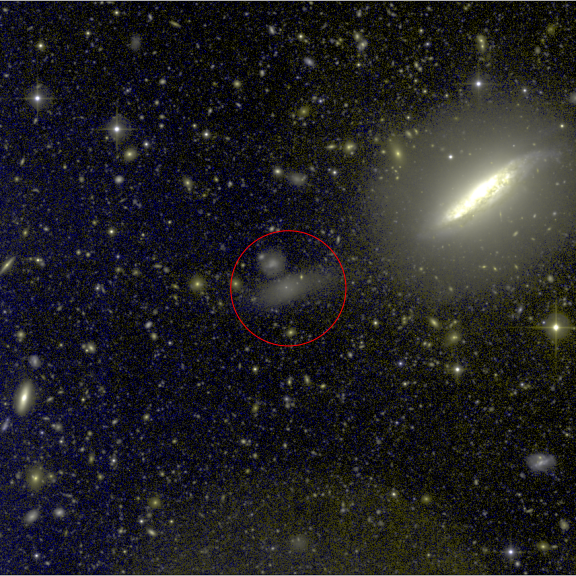}
\includegraphics[scale=0.17]{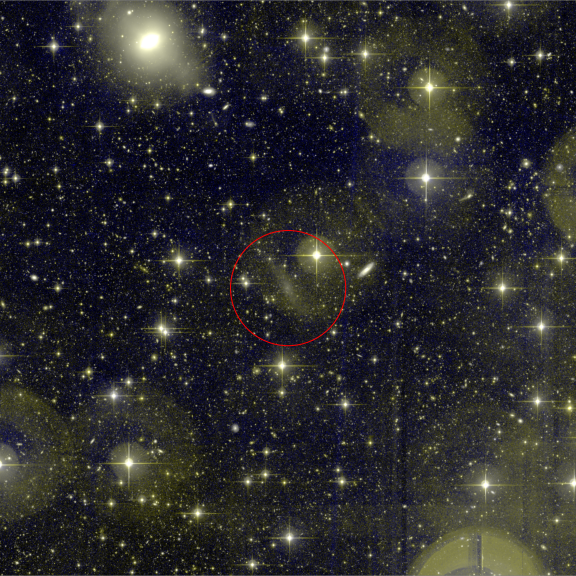}
\includegraphics[scale=0.17]{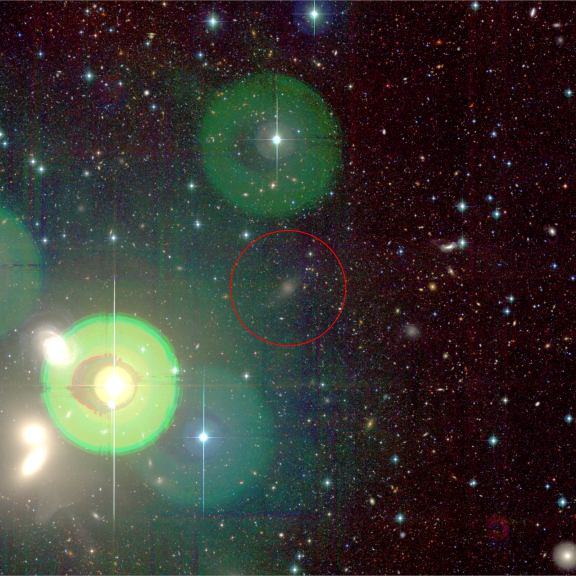}
\caption{Color ($g,r,i$) mages of the five UDGs that show signs of tidal disruption (from {\it left} to {\it right}: MATLAS-262, MATLAS-951, MATLAS-1059, MATLAS-1779 and MATLAS-1615). They show, from {\it left} to {\it right}: MATLAS-262 as a condensation within a long steam oriented in the east-west direction, close to perpendicular to an edge-on galaxy; the other filamentary structure close to the UDG with a NW-SE orientation is likely a foreground cirrus; MATLAS-951 has streams with a clear S-shape; MATLAS-1059 shows small tidal extensions; MATLAS-1779 has streams with a possible S-shape but there is contamination by a ghost stellar halo; and MATLAS-1615 displays streams oriented in the SE-NW direction but located in a high background region caused by multiple ghost stellar halos. From {\it left} to {\it right}, the image is 3, 25, 8, 25 and 25 sq.arcmin.\ in size. North is up and east is left. }
\label{fig:tidalfeatures}
\end{figure*}

\begin{figure*}
\centering
\includegraphics[scale=0.17]{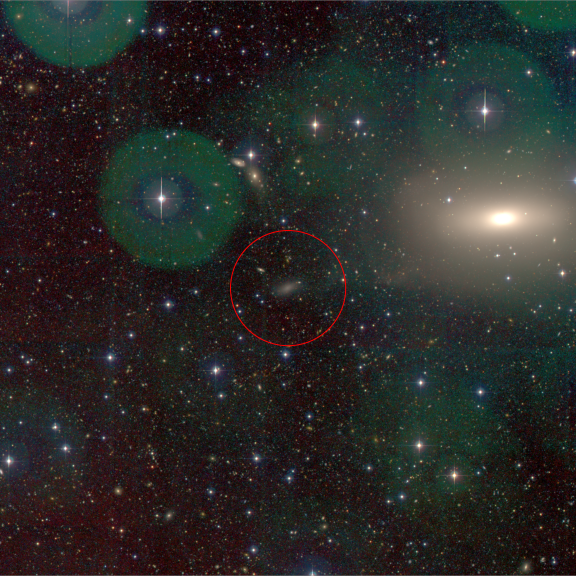}
\includegraphics[scale=0.17]{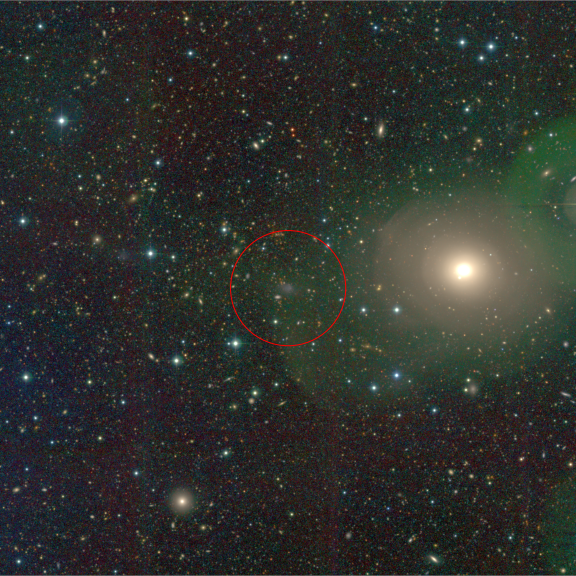}
\caption{Images of the two UDGs that are TDG candidates from {\it left} to {\it right}: MATLAS-478 and MATLAS-1824). They are found at the end of the tidal tail of a nearby massive galaxy, visible on the right in the image. From {\it left} to {\it right}, the image is 20 and 25 sq.arcmin.\ in size. North is up and east is left. }
\label{fig:tidalfeatures2}
\end{figure*}

One of the possible scenarios for UDG formation is that they are formed via tidal disruption \citep{Carleton2019,Roman2021,Gannon2021}. Another scenario is that some UDGs could be Tidal Dwarf Galaxies (TDGs), that is galaxies formed from gas expelled from a massive galaxy after an interaction. Therefore, searching for tidal features, the remnants of gravitational interactions between galaxies \citep{Bilek2020}, can be a powerful tool in constraining the formation scenario of UDGs. 

A search for tidal features in dwarfs in the Local Universe was presented by \citet{Paudel2018}. They visually identified tidal features and classified them in two main categories, dwarf-dwarf interaction/merger and dwarf-giant interactions, with a further grouping of their tidal features into three categories: interacting (ongoing interaction between two dwarf galaxies), shell (presence of shell features) and tidal tail (presence of amorphous tidal features such tidal streams or plumes which cannot be placed into the other classifications). Shell features in dwarfs can be produced by a merger origin \citep{Paudel2017} whereas an S-shaped elongated stellar envelope is likely due to tidal stretching from a nearby giant galaxy \citep{Paudel2013,Paudel2014}. 
Another nice example of a candidate UDG with an S-shape is the LSB dwarf HCC-087 in the Hydra I cluster \citep{Koch2012}, whose photometric properties are given in the works of \citet{Misgeld2008} and \citet{Misgeld2011}.

In our visual inspection of the UDGs, we follow the same classification scheme. We visually inspected both the \textit{g}-band and color images of the MATLAS UDGs and found that five out of the 59 MATLAS UDGs (MATLAS-262, MATLAS-951, MATLAS-1059, MATLAS-1779 and MATLAS-1615) show signs of tidal disruptions. As can be seen in Figure~\ref{fig:tidalfeatures}, all cases are dwarf-giant interactions and exhibit low surface brightness features extending in the direction of one or more massive galaxies (central ETG in the field: NGC~1248, NGC~3640, NGC~3674, NGC~5493 and NGC~5355, respectively). In particular, 1) MATLAS-262 is observed as a condensation within a long stream, oriented in the east-west direction, and appears to be close to perpendicular to a putative host edge-on galaxy; the other filamentary structure close to the UDG with a NW-SE orientation is likely a foreground cirrus as several other such structures with the same orientation are seen in the field; 2) MATLAS-951 has streams with a clear S-shape; 3) MATLAS-1059 shows small tidal extensions; 4) MATLAS-1779 has streams with a possible S-shape, however, there is contamination by a ghost stellar halo; and 5) MATLAS-1615 displays streams oriented in the SE-NW direction, but is located in a high background region caused by ghost stellar halos.

Another two UDGs (MATLAS-368 and MATLAS-2184) were found to show extended asymmetric isophotes (central ETG in the field: NGC~2577 and NGC~7454, respectively). However, MATLAS-2184 is in a region with many cirrus so the assymetric extension is possibly due to cirrus contamination. Optical cirrus exhibit on the CFHT images a range of colors (see \citealt{Miville2016}), and some have very similar colors as the stellar structures \citep{Sola2021}. Thus using colors in the optical regime is not enough to separate cirrus from tidal features. Having a broader range of wavelengths (from the ultraviolet to the infrared), may be tested, but data still lack to do such an experiment. In the near future, combining GALEX, Euclid and WISE data might be useful for that purpose.


For two of the UDGs (MATLAS-478 and MATLAS-1824), no tidal features associated with the dwarf were seen but they were found to be located at the end of the tidal tail of the nearby massive (host) galaxy (NGC~2768 and NGC~5557, respectively). MATLAS-478, is nucleated (dE,N), and a TDG candidate: the host galaxy is strongly disrupted and the tail emanates from the host in the direction of the UDG. MATLAS-1824 is a spectroscopically confirmed TDG \citep{Duc2014}.

Although the majority of the MATLAS UDGs do not appear to show evidence of recent tidal disruption or having formed as a TDG, one should note that we are able to recognize tidally disrupted dwarfs and TDGs only for a limited amount of time as tails of giant ellipticals live for only about 0.5 Gyr \citep{Mancillas2019} and the orbital time of a star 10~kpc away from a 10 billion solar masses galaxy is only 1 Gyr. Therefore a UDG that would be tidally perturbed 10 Gyr ago ($z\sim2$), when galaxy interactions were common, would most likely look relaxed today. We note that the majority (5/7) of the UDGs with tidal features or asymmetric isophotes are nucleated (dE,N), and do not show any difference in color than the other UDGs.

We now examine the local environment in which these UDGs, tidally disrupted dwarfs and TDGs, are found. One might expect the UDGs with tidal features to be located in more dense environments where they are more likely to have interactions. However, we find them in a range of environments. As seen in Figure~\ref{fig:locdensity}, the five UDGs that show signs of tidal disruptions are found at both low and high $\rho_{10}$ values. The two UDGs that are TDGs are located at similar, moderately low $\rho_{10}$ values, but this may be an observational bias; if the system is in a higher density environment, the tidal tail may not be as long lived. 

The individual systems where the tidally disrupted UDGs are located, shown in the maps in Appendix~\ref{section:appendix2}, further illustrate the diverse environments in which these galaxies are found. Some are projected to lie within the virial radii of various groups (e.g., NGC~3685, NGC~3640, NGC~2530), while other are not within any of the Kourkchi \& Tully optically identified groups (e.g., NGC~5493, NGC~1248). Some of these UDGs are located near galaxies with signatures of old major mergers (e.g., NGC~5493), while at least one is in close projection to a massive galaxy that has undergone a recent major merger (e.g., NGC~3641). The two TDG UDGs are both outside of the optically identified groups, although one is just outside the estimated boundary, such that the host ETG is considered a group member.

Here we have only considered the impact of the massive galaxies on the UDGs, but tidal features can also be caused by dwarf-dwarf interactions. A full analysis of the tidal features in the larger MATLAS dwarf sample is ongoing. 

\section{HI in UDGs}
\label{section:HI}




The detection of HI in a UDG allows us to not only measure its HI gas content but also to directly obtain a radial velocity and hence a distance estimate for the UDG. In the particular case of the MATLAS UDGs, this is an important tool given the uncertainties associated with assuming that the UDG is at the distance of the central massive ETG (presumed host). Past studies of HI-bearing UDGs located in the field \citep{Leisman2017}, groups \citep{Spekkens2018} and poor galaxy clusters \citep{Shi2017} have reported bluer colors, narrower line widths, larger gas fractions as compared to galaxies of similar HI-mass and environment.

We examined how many of the MATLAS UDGs are located in the regions observed for the ALFALFA and ATLAS$^{3D}$ HI surveys. As mentioned in Section~\ref{section:UDGselection}, of the 51 UDGs located in the observed regions, only three have a HI line detection (MATLAS-42, MATLAS-1337 in the ALFALFA catalog and MATLAS-1824 in our WRST catalog). These galaxies are located at HI distances of 33.5, 36 and 46.3~Mpc with HI masses $8.0\lesssim \log(M_{HI}) \lesssim 8.3$ \citep{Poulain2021b}. Based on the number of galaxies located in the observed regions and considering stellar masses high enough so we might detect HI (see \citealt{Poulain2021b}), we estimate a frequency of HI-bearing UDGs lower than for traditional dwarfs with $\sim$7\% and $\sim$10\%, respectively.

In \citet{Poulain2021b}, we compare the properties of these HI UDGs with those of traditional dwarf (non-UDGs) galaxies of similar HI mass. We find that the UDGs line widths are smaller than the median line width of the HI traditional dwarfs. Only two of the three HI UDGs have an available observation in the \textit{r}- and \textit{i}-band and thus an estimated stellar mass. Only one of these two HI UDGs has a gas fraction larger than the median of the traditional dwarf distribution and bluer colors than the traditional dwarfs of similar HI mass, the other one having redder colors than the median colors. We note that, given the low number statistics, the observed properties might not represent the overall HI-bearing UDG population.

One of the three HI UDGs, MATLAS-1824, is located along a prominent tidal tail of the massive galaxy NGC~5557, already mentioned in Section~\ref{section:tidalfeatures}. This UDG is also part of the six TDGs around ATLAS$^{3D}$ ETGs showing fine structures identified by \citet{Duc2014}, two of them having UDG properties based on their light profile fitting (see their Table 3). Another elongated tidal feature, perpendicular to the shell, passes through MATLAS-1824 and connects to another HI UDG, MATLAS-1830, which is not in our UDG sample due to the absence of successful two-dimensional surface brightness modeling caused by its extremely low surface brightness. 

\section{Conclusions}
\label{section:conclusion}

We have identified a population of UDGs in the MATLAS fields. Assuming the distance to the central ETG in the field and using the known distances when available, 59 UDGs were robustly identified using a two-dimensional surface brightness S\'ersic profile fit and a cut in surface brightness and effective radius. We find that they are at distances of $19.1 - 46.3$~Mpc and located predominantly in groups, with 32\% of the UDGs in groups with observed X-ray emission and 61\% within the estimated virial radii of optically identified groups detected via a galaxy linkage algorithm by \citet{Kourkchi2017}. The X-ray groups are all contained within the \citet{Kourkchi2017} catalog. Most of the X-ray groups have been well studied in the literature, and UDGs are associated with both dynamically young and dynamically relaxed systems.

Based on a detailed analysis of their photometric and structural properties, we find that the MATLAS UDGs do not belong to a separate and distinct group but instead are an extension to the traditional dwarf galaxies, making a clear-cut delineation/definition for a UDG arbitrary. The majority of the MATLAS UDGs are visually classified morphologically as dwarf ellipticals with log stellar masses in the range $\sim 6.5-8.7$. The fraction of nucleated UDGs ($\sim$34\%) is slightly above the one measured for all MATLAS dwarfs (visual: $\sim$23\%, \textsc{Galfit}: $\sim$27\%). However, looking at the dwarfs with $M_g$ in the range of the UDGs luminosity ($-15.5$ to $-11.5$), the nucleated fraction of the traditional dwarf (non-UDG) galaxies is $\sim$34\%, similar to the nucleated fraction of the UDGs. The MATLAS UDGs exhibit the same wide range of colors as the traditional dwarf galaxies, with ($g-r$)$_0=-0.1$ to $0.8$ and ($g-i$)$_0=-0.2$ to $1.2$. Their median color, ($g-r$)$_0=0.36$, is as red as the one measured in galaxy clusters such as Coma, ($g-r$)$_0=0.40$, but the range of colors is narrower.

Only five ($\sim$8\%) UDGs show signs of tidal disruptions and only two are tidal dwarf galaxy candidates. Although the majority of the MATLAS UDGs do not appear to show evidence of recent tidal disruption or having formed as a TDG, one should note that we are able to recognize tidally disrupted dwarfs and TDGs only for a limited amount of time. A UDG that would be tidally perturbed 10 Gyr ago ($z\sim2$), for example, would most likely look relaxed today. Three MATLAS UDGs are detected in HI, two in the ALFALFA catalog and one in our WRST catalog. One of these, MATLAS-1824, is a spectroscopically confirmed TDG located along a prominent tail of the massive galaxy NGC~5557. A study of globular cluster candidates selected in the CFHT multiband images finds, for the range of magnitude $-15.5 \lesssim M_V \lesssim -11.5$, no evidence of a higher GC specific frequency $S_N$ for UDGs than for traditional dwarf (non-UDG) galaxies, contrary to what is found in most clusters. The UDGs halo-to-stellar mass ratio distribution, as estimated from the GC count, peaks at roughly the same value as for the traditional dwarf (non-UDGs) galaxies, but spans the smaller range of $M_h/M_\ast \sim 10-2000$. We interpret these results to mean that the large majority of the field-to-group UDGs do not have a different formation scenario than traditional dwarf galaxies.

\section{Acknowledgements}
The authors would like to thank the referee for his/her thoughtful comments, which helped improve the work.
This research is based on observations obtained with MegaPrime/MegaCam, a joint project of CFHT and CEA/IRFU, at the Canada-France-Hawaii Telescope (CFHT) which is operated by the National Research Council (NRC) of Canada, the Institut National des Science de l'Univers of the Centre National de la Recherche Scientifique (CNRS) of France, and the University of Hawaii. This work is based in part on data products produced at Terapix available at the Canadian Astronomy Data Centre as part of the Canada-France-Hawaii Telescope Legacy Survey, a collaborative project of NRC and CNRS. M.P. acknowledges the Vice Rector for Research of the University of Innsbruck for the granted scholarship. S.P. acknowledges support from the New Researcher Program (Shinjin grant No. 2019R1C1C1009600) through the National Research Foundation of Korea. S.L. acknowledges the support from the Sejong Science Fellowship Program through the National Research Foundation of Korea (NRF-2021R1C1C2006790). M.B. acknowledges the support from the Polish National Science Centre under the grant 2017/26/D/ST9/00449. This work has made use of data from the European Space Agency (ESA) mission {\it Gaia} (\url{https://www.cosmos.esa.int/gaia}), processed by the {\it Gaia} Data Processing and Analysis Consortium (DPAC,
\url{https://www.cosmos.esa.int/web/gaia/dpac/consortium}). Funding for the DPAC
has been provided by national institutions, in particular the institutions
participating in the {\it Gaia} Multilateral Agreement. 

Funding for the Sloan Digital Sky Survey IV has been provided by the Alfred P. Sloan Foundation, the U.S. Department of Energy Office of Science, and the Participating Institutions. SDSS-IV acknowledges support and resources from the Center for High Performance Computing  at the University of Utah. The SDSS website is www.sdss.org. SDSS-IV is managed by the Astrophysical Research Consortium for the Participating Institutions of the SDSS Collaboration including the Brazilian Participation Group, the Carnegie Institution for Science, Carnegie Mellon University, Center for Astrophysics Harvard \& Smithsonian, the Chilean Participation Group, the French Participation Group, Instituto de Astrof\'isica de Canarias, The Johns Hopkins University, Kavli Institute for the Physics and Mathematics of the Universe (IPMU) / University of Tokyo, the Korean Participation Group, Lawrence Berkeley National Laboratory, Leibniz Institut f\"ur Astrophysik Potsdam (AIP), Max-Planck-Institut f\"ur Astronomie (MPIA Heidelberg), Max-Planck-Institut f\"ur Astrophysik (MPA Garching), Max-Planck-Institut f\"ur Extraterrestrische Physik (MPE), National Astronomical Observatories of China, New Mexico State University, New York University, University of Notre Dame, Observat\'ario Nacional / MCTI, The Ohio State University, Pennsylvania State University, Shanghai Astronomical Observatory, United Kingdom Participation Group, Universidad Nacional Aut\'onoma de M\'exico, University of Arizona, University of Colorado Boulder, University of Oxford, University of Portsmouth, University of Utah, University of Virginia, University of Washington, University of Wisconsin, Vanderbilt University, and Yale University.

\section{Data availability}
The data underlying this article are available at the CDS.

\bibliographystyle{aa}
\bibliography{bibliography}

\appendix

\section{Potential foreground hosts}
\label{section:appendix_hosts}

The distance estimates of UDG candidates are central to their identification and classification. This is perfectly illustrated by the debate in the literature over the distance to the UDG candidate NGC~1052-DF2. The galaxy was originally assumed to be a satellite of NGC~1052 at a distance of 20~Mpc \citep{vanDokkum2018b}, though it lies in close projection to multiple galaxies: it has a projected separation of 13.7{\arcmin} ($\sim$80~kpc) from NGC~1052, but also lies in close projection to NGC~1042 (20.8{\arcmin}; $\approx$50 -- 80~kpc), and is $<1\deg$ from the center of the NGC~988 group \citep{Trujillo2019}, using the coordinates from  \citet{Kourkchi2017}. These three host candidates have distances of 20~Mpc, $\approx$8--13~Mpc, and 15.1~Mpc, respectively, and the UDG would be a typical dwarf at the lower distances. Subsequent independent distance estimates of the galaxy (e.g., spectra, tip of the red giant branch, surface brightness fluctuation, GCLF and planetary nebula luminosity function) have not ended the debate, as they also span the entire range, from $\sim$13~Mpc -- $\sim$22~Mpc (e.g., \citealt{vanDokkum2018b,Shen2021,Blakeslee2018,Trujillo2019,Fensch2019a}).

We have assumed throughout the text that the UDGs are satellites of nearby ETGs from the ATLAS$^{3D}$ sample, but this has yet to be proven for the majority of the MATLAS dwarfs. Projection effects can easily confuse which massive galaxy, if any --- we cannot rule out the possibility of some fraction of the dwarfs being truly isolated systems ---, is the most likely host. In particular, we looked for massive foreground galaxies that could be potential hosts, and which would suggest that the UDGs are closer than we originally assumed.

We constructed a catalog of massive galaxies from the  ATLAS$^{3D}$ parent catalog (ETG $+$ LTG galaxies) and supplemented this with galaxies from the NEARGALCAT and the Kourkchi \& Tully catalog, after removing duplicate galaxies within a 10\arcsec\ match radius. We further estimated stellar masses from the available $K_s$ magnitudes, removing any galaxy with $M_*/M_\odot < 10^{9.5}$, the approximate mass of the LMC \citep{Marel04}, as such low mass galaxies are unlikely to host UDGs. Finally, we removed potential host galaxies at distances $\leq1$~Mpc: our UDG candidates do not show the graininess that would be expected if they are within this distance, and this also prevents the nearest massive galaxies (such as M31) from being flagged as a potential host for almost every dwarf in the sample. These restrictions result in a catalog of 2411 potential host galaxies with distances $2.8$~Mpc $< d <$ 130.2~Mpc and stellar masses $10^{9.5} < M_*/M_\odot < 10^{11.8}$.

For each UDG in the sample, we then calculated the physical separation between the UDG and each of the 2411 potential hosts, assuming in each instance that the UDG is at the same distance as the potential host galaxy. For this analysis, we kept any potential host with a projected separation $<800$~kpc, based on the largest dwarf separation in Cen~A \citep{Mueller2018b}; this value (800~kpc) is also consistent with the separations of backsplash UDG galaxies associated with the galactic-mass host halos simulated in \citet{Benavides2021}, although they find that these separations increase to $>3$~Mpc in massive groups (see their Figure~7).

\begin{figure*}
\centering
\includegraphics[width = 0.9\textwidth]{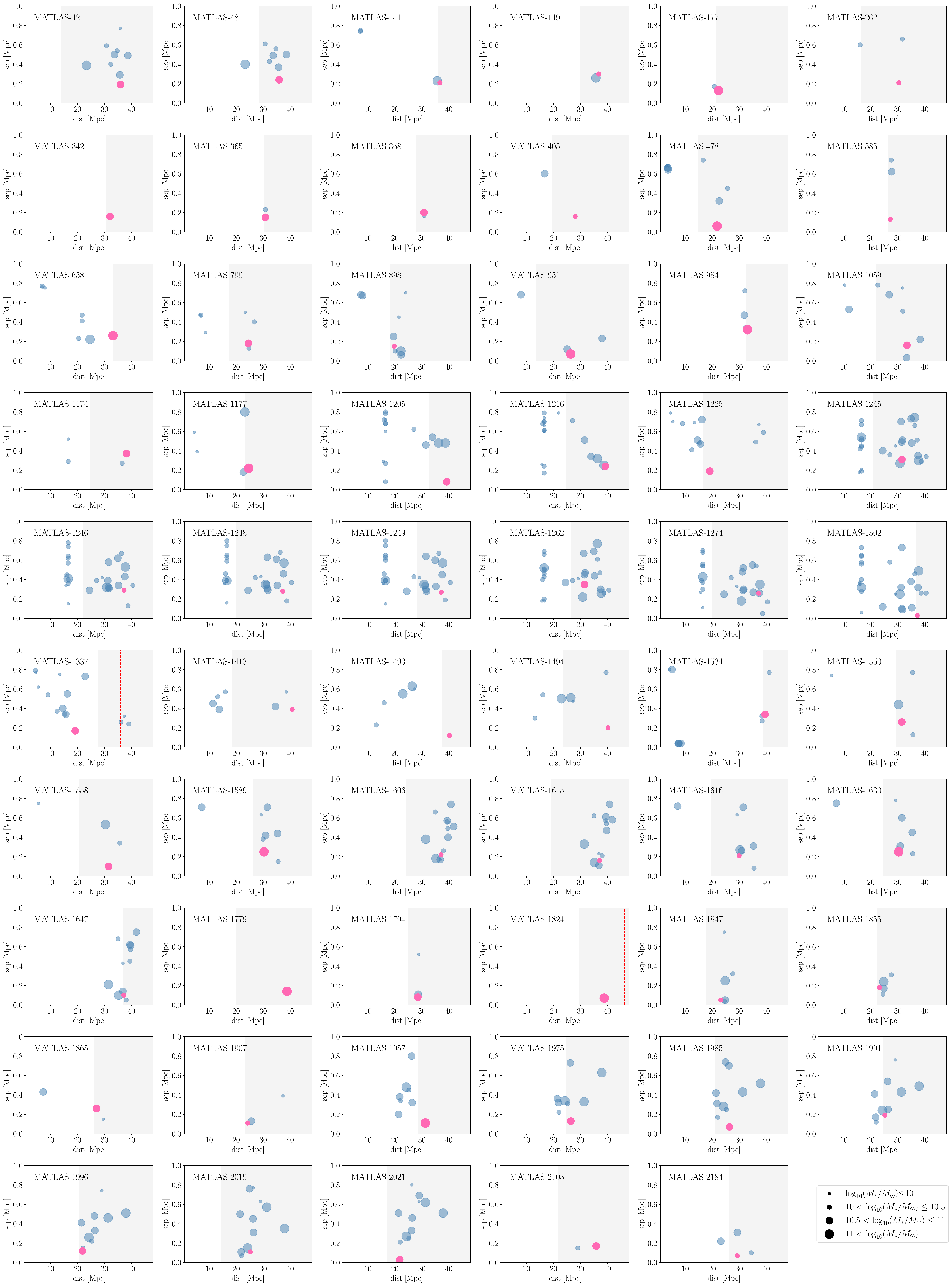}
\caption{Projected separations in physical units versus the distance to the potential host galaxy for each UDG candidate in the sample. We note that the data has been truncated at separations $<0.8$~Mpc and distances $\lesssim 50$~Mpc. The potential hosts have been separated into four mass bins, which is reflected by the size of the data points. The galaxy we have assumed is the host in this work is plotted in {\it pink}, while the distances at which the dwarf would be considered a UDG are shaded in {\it gray}. The {\it dashed red vertical lines} indicate the measured distances, when an estimate is available.}
\label{figure:host_dist}
\end{figure*}

The identified potential host galaxies were then plotted versus the distance of the host in Figure~\ref{figure:host_dist}, with each subplot corresponding to a single UDG candidate. We note that we also truncated the sample to those galaxies with distances  $\lesssim50$~Mpc in order to focus on foreground hosts. The host galaxies were separated into four mass bins --- $<10^{10}$,$10^{10}$--$10^{10.5}$, $10^{10.5}$--$10^{11}$, and $>10^{11}$ --- and the size of the datapoints increases with increasing mass. We note that the galaxy that we have assumed is the host in this work is marked in pink. Furthermore, the shaded region corresponds to distances at which the effective radius is $\geq 1.5$~kpc, meaning that the galaxy would be classified as a UDG anywhere in the shaded region, and dashed red vertical lines mark the independent distances estimates for the three UDGs that have such measurements.

The subplots demonstrate how difficult it is to associate dwarfs/UDGs with a massive host, when no independent distance estimates for the dwarf/UDG exist. Nevertheless, we can make a few broad generalizations. There are a number of systems (19) where the UDG classification appears to be quite secure. This includes subplots where there is a single potential host (e.g., MATLAS-1779), or where there are multiple hosts at roughly similar distances, and the dwarf would be classified as a UDG if it were a satellite of any of these hosts (e.g., MATLAS-42, MATLAS-1855, MATLAS-1996). We still cannot exclude, however, the possibility that these are isolated dwarfs at much lower distances, which would be then be interesting for their isolation rather than their physical diameters.

The remaining systems are more difficult to interpret. There are two systems where all potential foreground hosts (for the remainder of this discussion, `foreground hosts' will be used to indicate any host candidate at distances below the gray shaded `UDG region') that have estimated stellar masses $M_*/M_\odot \leq 10^{10}$; this number jumps to seven if we make a cut at $M_*/M_\odot \leq 10^{10.5}$. Such low mass galaxies are expected to have a very low probability of hosting a UDG, so these dwarfs are likely satellites of more massive, and more distant hosts. If we speculate that the most massive galaxy in each subplot is the most likely potential host, 29/40 of the hosts (59 - 19 = 40; from the paragraph above) lie within the gray shaded `UDG region'. It is worth noting, however, that there is one system (MATLAS-262) where \textit{every} potential host has $M_*/M_\odot \leq 10^{10.5}$; the galaxy that we assumed is the host has the highest mass (10.47) of all the potential hosts as well as the smallest physical separation, suggesting that UDGs may be found around some lower mass hosts.

It is readily apparent that some of the assumed distances are suspect, however. As discussed in Section~\ref{section:environment}, there is confusion between the membership of the NGC~4261 group and the Virgo cluster, which is easily identified by the vertical alignment of points at $d\approx16$~Mpc (e.g., MATLAS-1246); none of our UDG candidates in the NGC~4261 group would remain UDGs if their distances were revised downward to $d\approx16$~Mpc. We note that the groups discussed in the main text and Appendix~\ref{section:appendix1} are not as well defined in these plots as the Virgo cluster. As discussed in \citet{Mueller2020}, the NGC~5846 group is well defined in velocity space, but independent distance estimates to the most massive members taken from the literature are discrepant by approximately $\pm 5$~Mpc, and this spread can be observed in our plots (e.g., MATLAS-2019); this is likely also a problem for the NGC~4261 group, which additionally suffers from confusion with the foreground Virgo cluster.

Although some of our UDG candidates have several potential foreground hosts --- including high mass host candidates and hosts with smaller separations than our tentatively assigned host --- it is not always the case that the UDG has a bad distance estimate. MATLAS-1337 has an independent distance estimate based on an HI ALFALFA detection that indicates it is either at a much larger distance than we first assumed, or that it has an inordinately large velocity dispersion ($>1000$~km~s$^{-1}$) for a host at a lower distance.

\section{Local environment of UDGs}
\label{section:appendix2}

Spatial maps illustrating the local environment of all UDGs in the MATLAS dwarfs sample are included here, to highlight variations in their environments and complement the discussion of their stacked properties in Section~\ref{section:environment}. MATLAS pointings that overlap have been combined into a single plot and a buffer of 10\arcmin\ around each footprint is included in order to identify any massive galaxies and/or structures that are just outside the MATLAS field of view. 

The plots contain:

The ATLAS$^{3D}$ ETG sample: the MATLAS targets were pulled from the ATLAS$^{3D}$ ETG sample, and an ETG is typically centered in (or near the center of) the MegaCam field of view. Plotted as circles, these galaxies are color-coded according to the heliocentric velocities published in \citet{Cappellari2011}. Each galaxy is labeled in the maps.

LTGs in the ATLAS$^{3D}$ parent sample: these galaxies, which meet all of the same selection criteria as the ATLAS$^{3D}$ ETG sample except for morphology, are plotted as diamonds. Like the ETGs, the points are color-coded according to the heliocentric velocities published in \citet{Cappellari2011} and each galaxy is labeled in the plots.

The MATLAS footprint: The regions imaged as part of the MATLAS survey are shaded pale gray. Where fields overlap, the image size has been adjusted to show all of the contiguous footprints, and the plot bounds were adjusted so that the MATLAS footprint is centered.
 
The MATLAS dwarf sample: the dwarf galaxies identified by the MATLAS team \citep{Habas2020,Poulain2021b} are plotted as small `x' symbols. The distribution of the dwarfs and the highlighted UDG subpopulations can be compared in each map. It should be noted, however, that bright stars and cirrus are not marked; an apparent lack of dwarfs may mean that there are no dwarfs in that region or that they were not visible due to contamination from bright stellar halos or cirrus. The sources of contamination in the MATLAS fields are discussed in \citet{Bilek2020}, while \citet{Heesters2021} examines the distribution of dwarfs in the fields and the impact of the stellar halos and cirrus on the dwarf identification.
    
Ultra diffuse galaxies: the subsample of UDGs identified in this work are plotted as squares. Green squares indicate the UDGs that are not associated with any tidal features, while the turquoise squares indicate UDGs with tidal features --- shells, tidal tails, or tidal streams/plumes --- and the yellow squares indicate UDGs that are at the end of a tidal tail that is associated with a nearby massive galaxy (see Section~\ref{section:tidalfeatures} for more details).
    
Groups from \citet{Kourkchi2017}: Kourkchi \& Tully created a catalog of groups with recessional velocities $< 3500$~km~s$^{-1}$ using a galaxy linkage program and expected scaling relations. They published a catalog that includes the luminosity-weighted group center and virial radius for each group, which we have over-plotted in pale blue. Their galaxy linkage program, however, also links groups of galaxies into larger systems. To focus on the local environment, we plotted only the groups with radii $<5\deg$.
    
Galaxies from \citet{Kourkchi2017}: Plotting just the galaxies from the ATLAS$^{3D}$ parent sample (ETGs + LTGs) and the MATLAS dwarfs leaves a mass gap ($9 \lesssim \log(M_{stellar}/M_\odot) \lesssim 10$) in the plots, and misses the massive galaxies at distances $d<10$~Mpc and $d\gtrsim40$~Mpc. To fill in these gaps, we have also plotted the full catalog of galaxies used by Kourkchi \& Tully to identify the nearby groups (described above), after removing duplicates from the ATLAS$^{3D}$ sample (260/260 ETGs, 606/611 LTGs) and the MATLAS dwarfs catalog (234/2210 dwarfs). These galaxies are displayed as pentagons. When a distance is available, the pentagons are color-coded according to the velocity, which was calculated assuming only the Hubble flow --- and adopting $H_0 = 75$~km~s$^{-1}$ as in \citet{Kourkchi2017} --- else they are solid gray.

The color bar on each map spans a unique velocity range. Every massive galaxy (the ATLAS$^{3D}$ ETG + LTG sample and the Kourkchki \& Tully catalog) with a measured velocity that falls within the spatial boundaries of the map was taken into account, and the min and max values were rounded down and up, respectively, to the nearest 100~km~s$^{-1}$. Regardless of how wide the spread in velocities is, however, the color bar remains separated into eight colored blocks to better distinguish galaxies in adjoining velocity bins. Velocities for the 325 MATLAS dwarfs with prior distance estimates are not included, to keep the plots cleaner. 

While some of the MATLAS fields contain a number of galaxies with a wide spread in velocities ($\Delta v > 2000$~km~s$^{-1}$; for example, the field that contains the galaxy pair NGC~5194/NGC~5195, the group including NGC~5169/NGC~5173/NGC~5198, plus a few more distant galaxies) many of the groups are spatially separated from one another. This is not true however for the W Cloud, a.k.a. the NGC~4261 group, which abuts the Virgo cluster. The sharp increase in the number of galaxies in the field is immediately obvious in the relevant subplot below. Properties of this group are uncertain, due to the confusion in separating the group members from the cluster members.

It can be seen in the figures that the UDGs populate a range of environments. As discussed in  Section~\ref{section:environment}, 61\% of the UDG sample, as well as two-thirds of the full dwarf sample, falls within the virial radii of one of the \citet{Kourkchi2017} groups. The dynamical states of the groups is also varied, and UDGs are associated with both dynamically young and dynamically mature groups.

There are also a few potential galaxy pairs/associations that were not identified in the Kourkchi \& Tully catalog (e.g., NGC~5506 \& NGC~5507) with UDGs nearby, as well as a few relatively isolated hosts (e.g., IC~0560, NGC~5493).

\begin{figure*}
\centering
\includegraphics[width=0.99\textwidth]{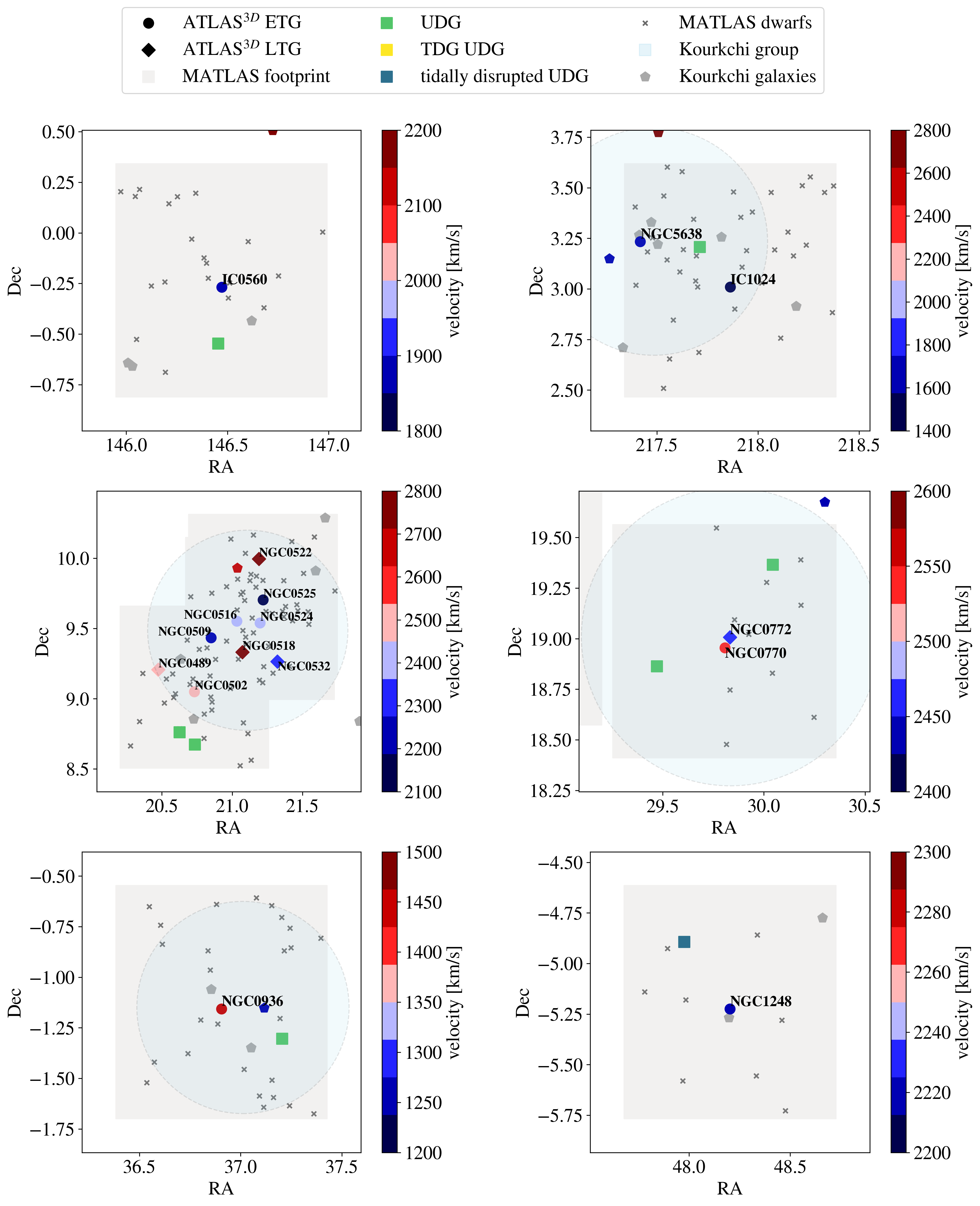}
\caption{ Maps of MATLAS fields containing UDGs.}
\label{fig:group_maps1}
\end{figure*}

\begin{figure*}
\centering
\includegraphics[width=0.99\textwidth]{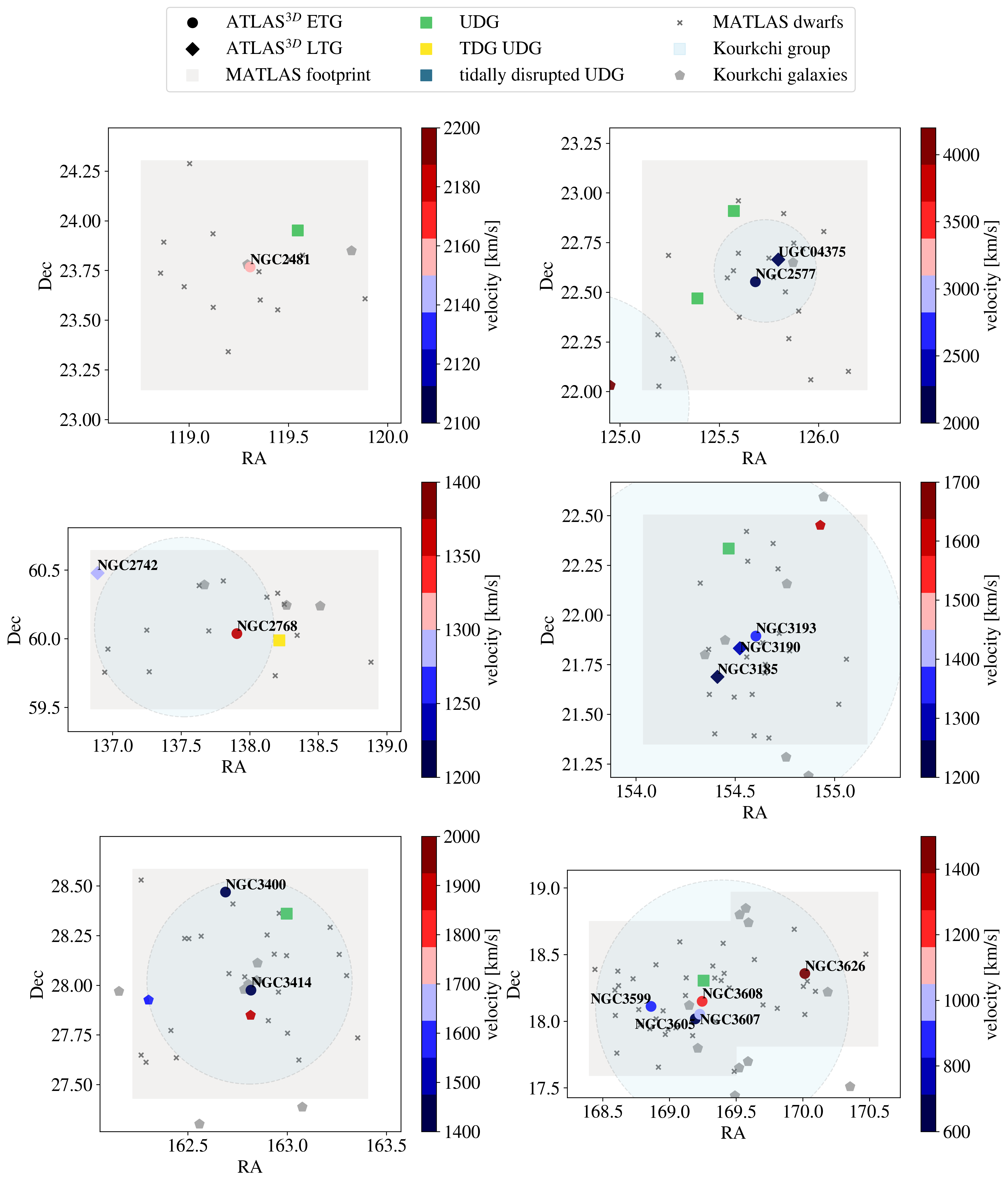}
\caption{ Maps of MATLAS fields containing UDGs (continued).}
\label{fig:group_maps2}
\end{figure*}

\begin{figure*}
\centering
\includegraphics[width=0.99\textwidth]{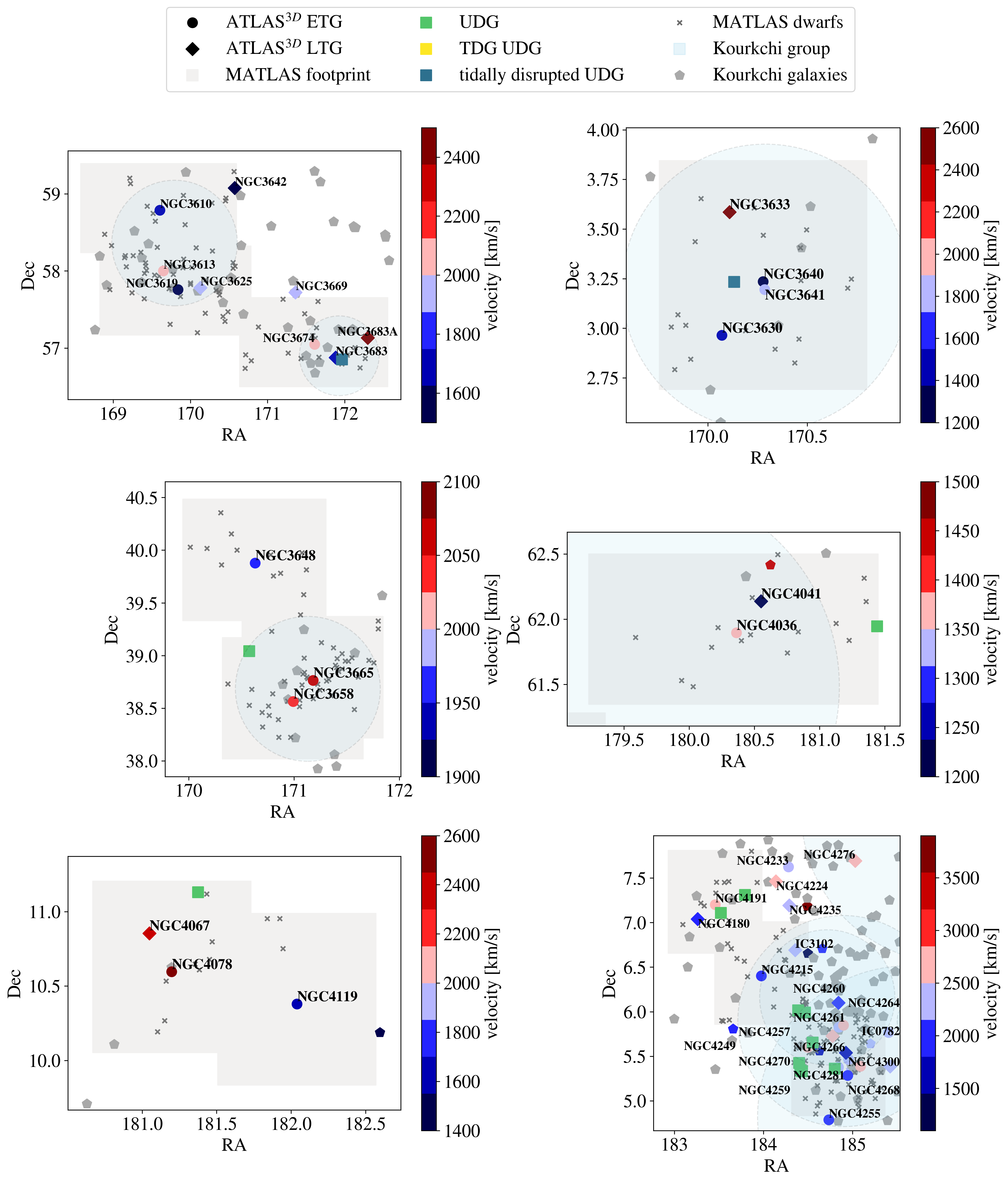}
\caption{ Maps of MATLAS fields containing UDGs (continued).}
\label{fig:group_maps3}
\end{figure*}

\begin{figure*}
\centering
\includegraphics[width=0.99\textwidth]{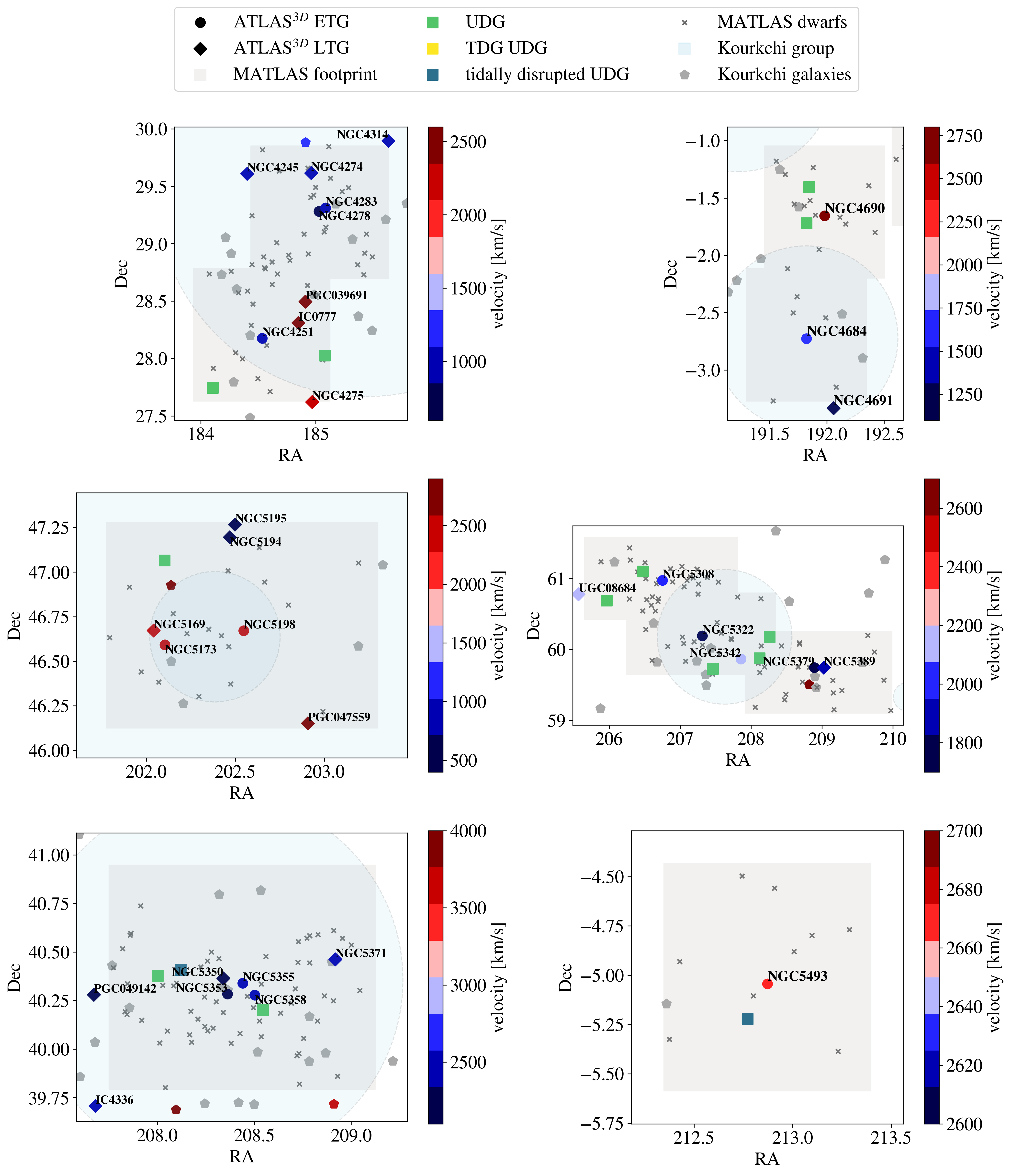}
\caption{ Maps of MATLAS fields containing UDGs (continued).}
\label{fig:group_maps4}
\end{figure*}

\begin{figure*}
\centering
\includegraphics[width=0.99\textwidth]{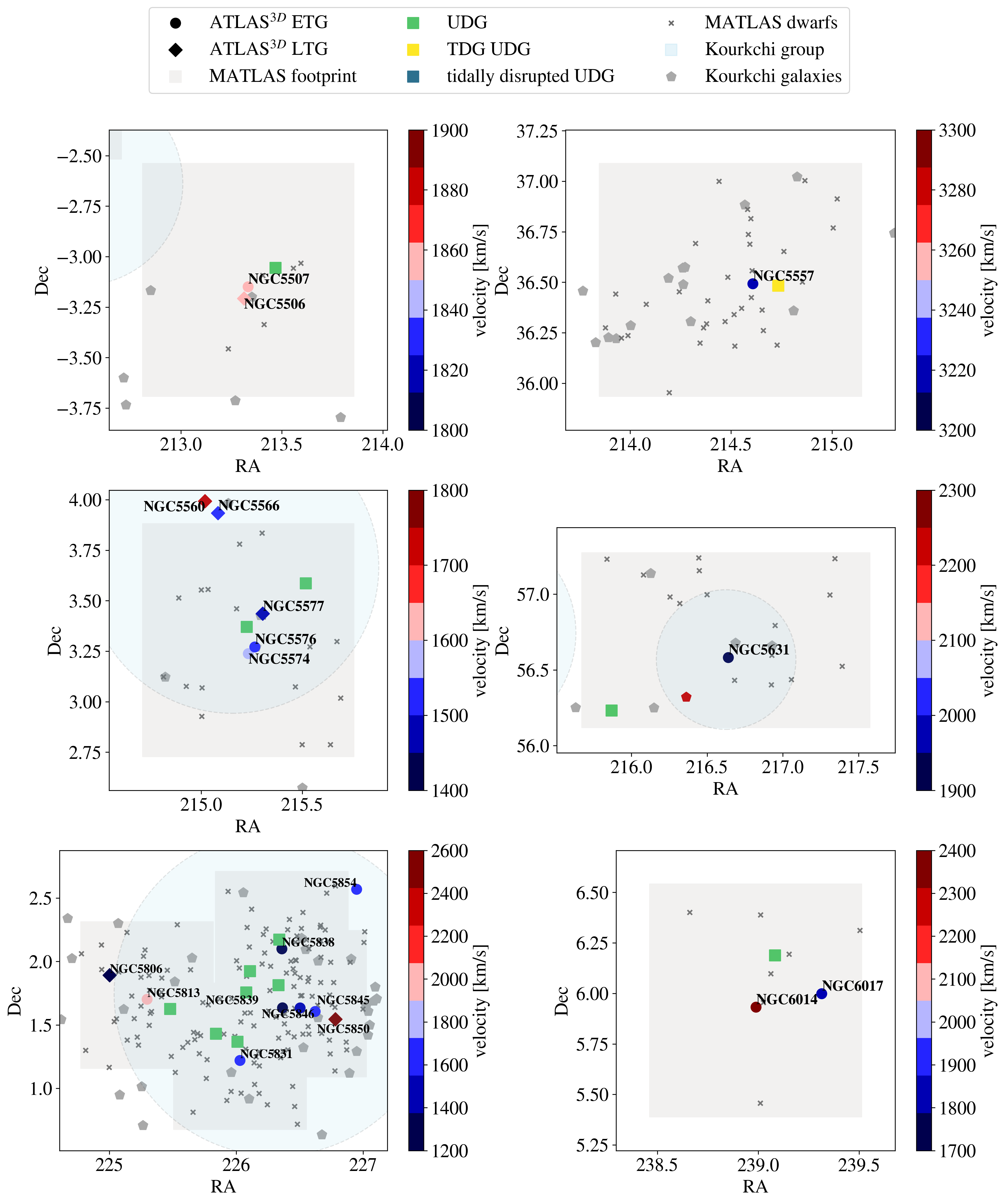}
\caption{ Maps of MATLAS fields containing UDGs (continued).}
\label{fig:group_maps5}
\end{figure*}

\begin{figure*}
\centering
\includegraphics[width=0.99\textwidth]{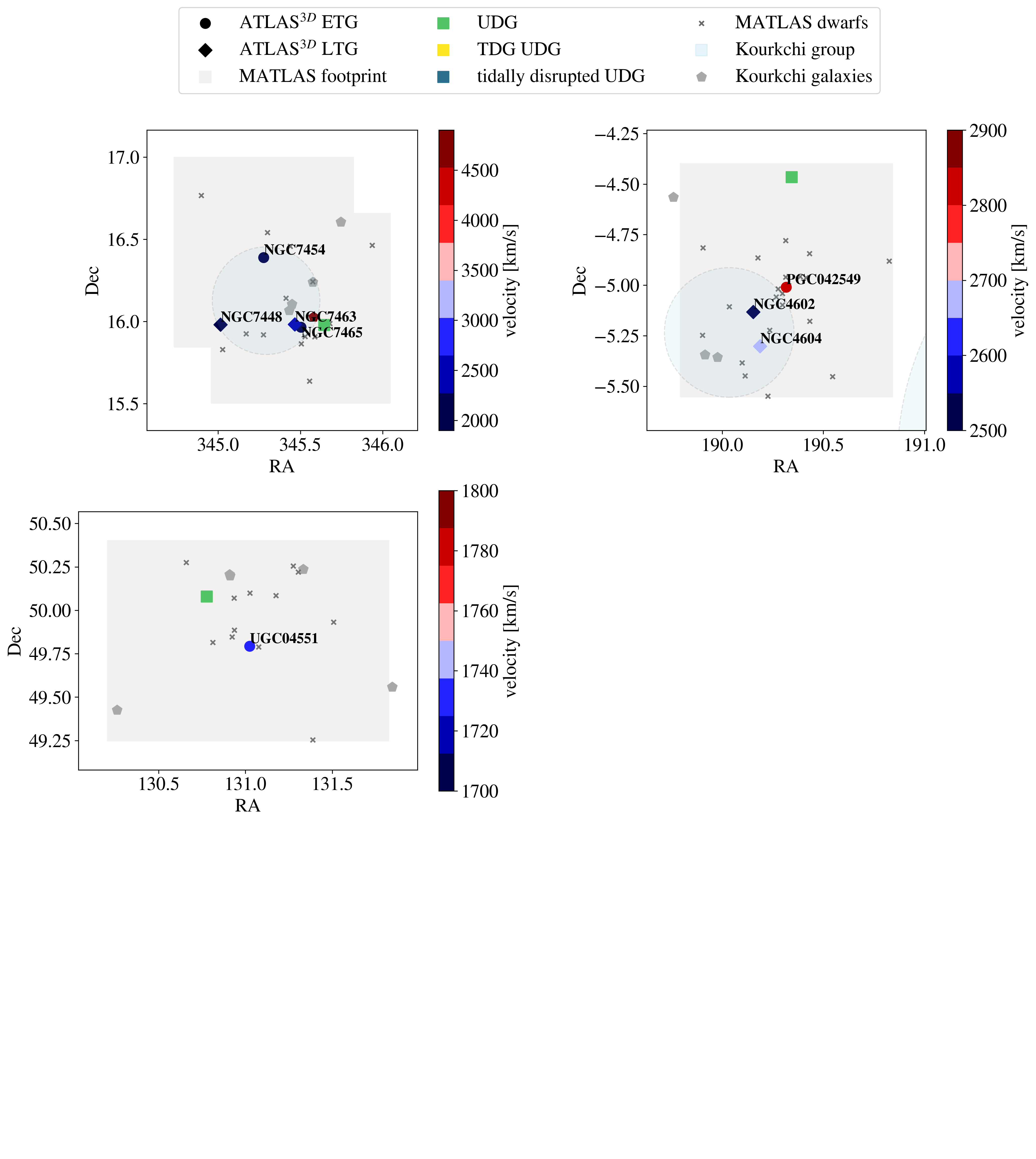}
\caption{ Maps of MATLAS fields containing UDGs (continued).}
\label{fig:group_maps6}
\end{figure*}

\section{Tables}
\label{section:appendix1}

The tables in this appendix summarize the photometric properties (morphologies, magnitudes, colors) of the UDG sample as well as additional derived quantities discussed throughout this work.

\newpage

\begin{table*}
\centering
\caption{Photometric properties of the MATLAS UDGs.}
\begin{tabular}{lrrlccccc} 
\hline
ID & RA & Dec & Morph & Distance & $g$   & ($g-r$)$_0$   & ($g-i$)$_0$   & $M_g$ \\
   &    &     &       & (Mpc)             & (mag) & (mag)     & (mag)     & (mag) \\
(1)   & (2)   & (3)    & (4)   & (5)   & (6)   & (7)     & (8)     & (9) \\
\hline
\\
MATLAS-42    &  20.6260 &   8.7614 &dI    &   35.9,33.5 &   $17.296\pm0.006$ &    $0.40\pm0.02$ &    $0.55\pm0.03$ &  $-15.326\pm0.006$ \\
MATLAS-48    &  20.7340 &   8.6737 &dE    &   35.9,-- &   $18.76\pm0.01$ &    $0.50\pm0.01$ &    $0.72\pm0.01$ &  $-14.01\pm0.01$ \\
MATLAS-141   &  29.4714 &  18.8628 &dE    &   36.7,-- &   $20.12\pm0.02$ &    $0.55\pm0.03$ &    $0.90\pm0.04$ &  $-12.71\pm0.02$ \\
MATLAS-149   &  30.0435 &  19.3647 &dI    &   36.7,-- &   $20.46\pm0.02$ &    $0.67\pm0.03$ &    $1.01\pm0.04$ &  $-12.36\pm0.02$ \\
MATLAS-177   &  37.2057 &  $-1.3041$ &dE    &   22.4,-- &   $19.15\pm0.01$ &    $0.56\pm0.02$ &    $0.83\pm0.03$ &  $-12.61\pm0.01$ \\
MATLAS-262   &  47.9769 &  $-4.8931$ &dE    &   30.4,-- &   $20.04\pm0.05$ &    $0.51\pm0.05$ &             -- &  $-12.37\pm0.05$ \\
MATLAS-342   & 119.5476 &  23.9516 &dE    &   32.0,-- &   $18.56\pm0.02$ &              -- &    $0.49\pm0.06$ &  $-13.97\pm0.02$ \\
MATLAS-365   & 125.3900 &  22.4683 &dE,N  &   30.8,-- &   $19.58\pm0.10$ &    $0.72\pm0.10$ &    $1.16\pm0.11$ &  $-12.87\pm0.10$ \\
MATLAS-368   & 125.5731 &  22.9086 &dE,N  &   30.8,-- &   $18.83\pm0.10$ &    $0.72\pm0.02$ &    $0.82\pm0.03$ &  $-13.62\pm0.10$ \\
MATLAS-405   & 130.7772 &  50.0784 &dE,N  &   28.0,-- &   $18.87\pm0.02$ &    $0.48\pm0.02$ &    $0.68\pm0.02$ &  $-13.37\pm0.02$ \\
MATLAS-478   & 138.2168 &  59.9870 &dE,N  &   21.8,-- &   $17.486\pm0.006$ &    $0.526\pm0.007$ &    $0.85\pm0.01$ &  $-14.207\pm0.006$ \\
MATLAS-585   & 146.4547 &  $-0.5469$ &dE    &   27.2,-- &   $18.789\pm0.005$ &    $0.36\pm0.02$ &             -- &  $-13.384\pm0.005$ \\
MATLAS-658   & 154.4670 &  22.3337 &dE    &   33.1,-- &   $20.63\pm0.17$ &              -- &              -- &  $-11.97\pm0.17$ \\
MATLAS-799   & 162.9980 &  28.3601 &dE    &   24.5,-- &   $17.28\pm0.006$ &    $0.577\pm0.007$ &              -- &  $-14.669\pm0.006$ \\
MATLAS-898   & 169.2570 &  18.3027 &dE    &   19.8,-- &   $17.94\pm0.02$ &    $0.35\pm0.05$ &    $0.61\pm0.05$ &  $-13.55\pm0.02$ \\
MATLAS-951   & 170.1326 &   3.2332 &dE,N  &   26.3,-- &   $18.023\pm0.006$ &    $0.679\pm0.007$ &    $1.052\pm0.009$ &  $-14.076\pm0.006$ \\
MATLAS-984   & 170.5741 &  39.0399 &dE    &   33.1,-- &   $19.26\pm0.02$ &    $0.39\pm0.02$ &    $0.68\pm0.03$ &  $-13.34\pm0.02$ \\
MATLAS-1059  & 171.9652 &  56.8550 &dE,N  &   33.4,-- &   $18.099\pm0.006$ &    $0.488\pm0.008$ &              -- &  $-14.520\pm0.006$ \\
MATLAS-1174  & 181.3757 &  11.1305 &dE    &   38.1,-- &   $19.73\pm0.02$ &    $0.16\pm0.06$ &              -- &  $-13.17\pm0.02$ \\
MATLAS-1177  & 181.4412 &  61.9456 &dE,N  &   24.6,-- &   $18.357\pm0.005$ &    $0.55\pm0.01$ &             -- &  $-13.597\pm0.005$ \\
MATLAS-1205  & 183.5229 &   7.1086 &dE,N  &   39.2,-- &   $19.500\pm0.005$ &    $0.48\pm0.01$ &              -- &  $-13.467\pm0.005$ \\
MATLAS-1216  & 183.7925 &   7.3112 &dE,N  &   39.2,-- &   $20.02\pm0.03$ &    $0.52\pm0.04$ &              -- &  $-12.95\pm0.03$ \\
MATLAS-1225  & 184.1047 &  27.7437 &dE    &   19.1,-- &   $18.50\pm0.02$ &              -- &              -- &  $-12.90\pm0.01$ \\
MATLAS-1245  & 184.3903 &   6.0166 &dE    &   31.5,-- &   $19.44\pm0.02$ &    $0.29\pm0.06$ &              -- &  $-13.05\pm0.02$ \\
MATLAS-1246  & 184.4013 &   5.4247 &dE    &   37.2,-- &   $18.04\pm0.01$ &    $0.61\pm0.01$ &    $0.86\pm0.01$ &  $-14.82\pm0.01$ \\
MATLAS-1248  & 184.4155 &   5.3419 &dE    &   37.2,-- &   $17.56\pm0.005$ &    $0.64\pm0.01$ &    $0.89\pm0.02$ &  $-15.292\pm0.005$ \\
MATLAS-1249  & 184.4336 &   5.3309 &dE    &   37.2,-- &   $19.57\pm0.07$ &    $0.79\pm0.09$ &    $1.17\pm0.11$ &  $-13.29\pm0.07$ \\
MATLAS-1262  & 184.4664 &   5.9891 &dE,N  &   31.5,-- &   $18.69\pm0.02$ &    $0.55\pm0.03$ &              -- &  $-13.81\pm0.02$ \\
MATLAS-1274  & 184.5520 &   5.6525 &dE    &   37.2,-- &   $17.586\pm0.006$ &    $0.517\pm0.009$ &    $0.70\pm0.01$ &  $-15.276\pm0.006$ \\
MATLAS-1302  & 184.8020 &   5.3615 &dE    &   37.2,-- &   $20.82\pm0.03$ &    $0.55\pm0.04$ &    $0.82\pm0.05$ &  $-12.03\pm0.03$ \\
MATLAS-1337  & 185.0780 &  28.0257 &dI,N  &   19.1,35.9&   $18.514\pm0.006$ &              -- &              -- &  $-14.264\pm0.006$ \\
MATLAS-1413  & 190.3429 &  $-4.4653$ &dE    &   40.7,-- &   $19.39\pm0.02$ &    $0.74\pm0.05$ &              -- &  $-13.65\pm0.02$ \\
MATLAS-1493  & 191.8227 &  $-1.7204$ &dE,N  &   40.2,-- &   $19.88\pm0.02$ &    $0.51\pm0.02$ &              -- &  $-13.14\pm0.02$ \\
MATLAS-1494  & 191.8468 &  $-1.4045$ &dE    &   40.2,-- &   $18.833\pm0.009$ &    $0.50\pm0.01$ &              -- &  $-14.188\pm0.009$ \\
MATLAS-1534  & 202.1029 &  47.0652 &dE    &   39.6,-- &   $20.27\pm0.02$ &    $0.43\pm0.04$ &              -- &  $-12.72\pm0.02$ \\
MATLAS-1550  & 205.9650 &  60.6890 &dE    &   31.5,-- &   $19.52\pm0.01$ &    $0.49\pm0.02$ &              -- &  $-12.97\pm0.01$ \\
MATLAS-1558  & 206.4767 &  61.0976 &dE,N  &   31.5,-- &   $19.51\pm0.02$ &    $0.65\pm0.02$ &              -- &  $-12.98\pm0.02$ \\
MATLAS-1589  & 207.4632 &  59.7278 &dE    &   30.3,-- &   $19.95\pm0.02$ &    $0.47\pm0.02$ &              -- &  $-12.46\pm0.02$ \\
MATLAS-1606  & 208.0022 &  40.3767 &dE,N  &   37.1,-- &   $18.453\pm0.002$ &              -- &              -- &  $-14.394\pm0.002$ \\
MATLAS-1615  & 208.1200 &  40.4081 &dE,N  &   37.1,-- &   $18.11\pm0.01$ &    $0.60\pm0.01$ &    $0.89\pm0.01$ &  $-14.74\pm0.01$ \\
MATLAS-1616  & 208.1220 &  59.8762 &dE,N  &   30.0,-- &   $18.68\pm0.05$ &    $0.58\pm0.05$ &              -- &  $-13.71\pm0.02$ \\
MATLAS-1630  & 208.2604 &  60.1730 &dE    &   30.3,-- &   $19.23\pm0.02$ &    $0.78\pm0.03$ &    $1.14\pm0.03$ &  $-13.17\pm0.02$ \\
MATLAS-1647  & 208.5436 &  40.2012 &dE    &   37.1,-- &   $20.75\pm0.02$ &    $0.81\pm0.02$ &    $1.08\pm0.03$ &  $-12.10\pm0.02$ \\
MATLAS-1779  & 212.7722 &  $-5.2221$ &dE,N  &   38.8,-- &   $19.27\pm0.02$ &    $0.49\pm0.12$ &             -- &  $-13.67\pm0.02$ \\
MATLAS-1794  & 213.4693 &  $-3.0559$ &dE    &   28.5,-- &   $19.02\pm0.02$ &              -- &              -- &  $-13.25\pm0.02$ \\
MATLAS-1824  & 214.7337 &  36.4832 &dE    &   38.8,46.3 &   $18.44\pm0.01$ &    $0.06\pm0.02$ &    $0.07\pm0.03$ &  $-14.89\pm0.01$ \\
MATLAS-1847  & 215.2256 &   3.3710 &dE    &   23.2,-- &   $18.77\pm0.05$ &    $0.56\pm0.05$ &              -- &  $-13.06\pm0.04$ \\
MATLAS-1855  & 215.5181 &   3.5866 &dE,N  &   23.2,-- &   $18.67\pm0.05$ &    $0.36\pm0.05$ &              -- &  $-13.16\pm0.05$ \\
MATLAS-1865  & 215.8694 &  56.2322 &dE    &   27.0,-- &   $18.53\pm0.01$ &    $0.22\pm0.01$ &    $0.31\pm0.02$ &  $-13.628\pm0.008$ \\
MATLAS-1907  & 217.7121 &   3.2066 &dE    &   24.2,-- &   $18.150\pm0.004$ &    $0.492\pm0.005$ &              -- &  $-13.769\pm0.004$ \\
MATLAS-1957  & 225.4779 &   1.6258 &dE,N  &   31.3,-- &   $18.46\pm0.10$ &              -- &              -- &  $-14.01\pm0.02$ \\
MATLAS-1975  & 225.8389 &   1.4309 &dE,N  &   26.4,-- &   $19.23\pm0.07$ &    $0.43\pm0.17$ &              -- &  $-12.88\pm0.07$ \\
MATLAS-1985  & 226.0105 &   1.3671 &dE    &   26.4,-- &   $18.850\pm0.006$ &              -- &              -- &  $-13.258\pm0.006$ \\
\\
\hline
\end{tabular}
\begin{tablenotes}
\footnotesize
\item \textbf{Notes.} This table (continued next page) is available at the CDS. Columns meanings: (1) UDG ID; (2) and (3) Right ascension and declination; (4) Morphology; (5) Assumed host ETG distance, dwarf distance; (6) Apparent magnitude in the \textit{g}-band; (7) $g-r$ color corrected for Galactic extinction; (8) $g-i$ color corrected for Galactic extinction; (9) Absolute magnitude in the \textit{g}-band computed using the UDG distance when available, otherwise the assumed host ETG distance is used.
\end{tablenotes}
\label{tab:UDGprop1}
\end{table*}

\begin{table*}
\centering
\caption{Photometric properties of the MATLAS UDGs (continued).}
\begin{tabular}{lrrlccccc} 
\hline
ID & RA & Dec & Morph & Distance & $g$   & ($g-r$)$_0$   & ($g-i$)$_0$   & $M_g$ \\
   &    &     &       & (Mpc)             & (mag) & (mag)     & (mag)     & (mag) \\
(1)   & (2)   & (3)    & (4)   & (5)   & (6)   & (7)     & (8)     & (9) \\
\hline
\\
MATLAS-1991  & 226.0792 &   1.7555 &dE    &   25.2,-- &   $18.061\pm0.004$ &    $0.59\pm0.02$ &    $0.87\pm0.03$ &  $-13.946\pm0.004$ \\
MATLAS-1996  & 226.1090 &   1.9246 &dE    &   21.8,-- &   $19.95\pm0.13$ &              -- &              -- &  $-11.75\pm0.13$ \\
MATLAS-2019  & 226.3340 &   1.8127 &dE    &   25.2,20.0 &   $17.79\pm0.01$ &    $0.59\pm0.02$ &    $0.85\pm0.02$ &  $-13.74\pm0.01$ \\
MATLAS-2021  & 226.3400 &   2.1722 &dE    &   21.8,-- &   $19.35\pm0.08$ &    $-0.03\pm0.12$ &   $-0.19\pm0.34$ &  $-12.34\pm0.08$ \\
MATLAS-2103  & 239.0837 &   6.1882 &dE    &   35.8,-- &   $18.037\pm0.006$ &              -- &              -- &  $-14.733\pm0.006$ \\
MATLAS-2184  & 345.6448 &  15.9791 &dE    &   29.3,-- &   $19.39\pm0.05$ &    $0.12\pm0.05$ &    $0.18\pm0.06$ &  $-12.95\pm0.05$ \\
\\
\hline
\end{tabular}
\begin{tablenotes}
\footnotesize
\item \textbf{Notes.} This table (including previous page) is available at the CDS. Columns meanings: (1) UDG ID; (2) and (3) Right ascension and declination; (4) Morphology; (5) Assumed host ETG distance, dwarf distance; (6) Apparent magnitude in the \textit{g}-band; (7) $g-r$ color corrected for Galactic extinction; (8) $g-i$ color corrected for Galactic extinction; (9) Absolute magnitude in the \textit{g}-band computed using the UDG distance when available, otherwise the assumed host ETG distance is used.
\end{tablenotes}
\label{tab:UDGprop2}
\end{table*}

\begin{table*}
\centering
\caption{Additional properties of the MATLAS UDGs.}
\begin{tabular}{lcccccc} 
\hline
ID & $\mu_{0,g}$    & $\langle \mu_{e,g} \rangle$  & $R_{e,g}$   & $R_{e,g}$  & $log(M_*/M_{\odot})$  & $M_{halo}/M_\ast$  \\
    & (mag/arcsec$^2$) & (mag/arcsec$^2$)              & (arcsec)  & (kpc) &  &  \\
(1)   & (2)   & (3)    & (4)   & (5)   & (6)   & (7) \\
\hline
\\
MATLAS-42    &   $24.39\pm0.02$ &   $25.17\pm0.02$ &   $21.19\pm0.13$ &    $3.46\pm0.02$ &    $8.32\pm0.16$ &      -- \\
MATLAS-48    &   $24.38\pm0.04$ &   $25.17\pm0.04$ &   $10.75\pm0.12$ &    $1.89\pm0.02$ &    $7.95\pm0.17$ &      -- \\
MATLAS-141   &   $25.22\pm0.08$ &   $26.00\pm0.08$ &    $8.48\pm0.22$ &    $1.52\pm0.04$ &    $7.51\pm0.14$ &      -- \\
MATLAS-149   &   $25.97\pm0.09$ &   $26.75\pm0.09$ &   $10.24\pm0.31$ &    $1.84\pm0.06$ &    $7.55\pm0.14$ &      -- \\
MATLAS-177   &   $25.36\pm0.05$ &   $26.15\pm0.05$ &   $14.17\pm0.22$ &    $1.55\pm0.02$ &    $7.49\pm0.15$ &  $158.5_{-88.4}^{+155.2}$ \\
MATLAS-262   &   $26.84\pm0.17$ &   $27.63\pm0.17$ &   $18.56\pm1.04$ &    $2.75\pm0.16$ &    $7.32\pm0.13$ &      -- \\
MATLAS-342   &   $24.02\pm0.07$ &   $24.80\pm0.07$ &   $10.02\pm0.22$ &    $1.57\pm0.04$ &      -- &      -- \\
MATLAS-365   &   $25.05\pm0.30$ &   $25.84\pm0.30$ &   $10.08\pm0.88$ &    $1.52\pm0.13$ &    $7.83\pm0.06$ &      -- \\
MATLAS-368   &   $24.50\pm0.14$ &   $25.28\pm0.14$ &   $11.03\pm0.22$ &    $1.66\pm0.03$ &    $8.13\pm0.15$ &   $55.6_{-31.1}^{+54.5}$ \\
MATLAS-405   &   $25.33\pm0.05$ &   $26.11\pm0.05$ &   $15.86\pm0.27$ &    $2.17\pm0.04$ &    $7.66\pm0.16$ &      -- \\
MATLAS-478   &   $24.55\pm0.02$ &   $25.33\pm0.02$ &   $20.94\pm0.13$ &    $2.22\pm0.01$ &    $8.07\pm0.17$ &      -- \\
MATLAS-585   &   $24.59\pm0.02$ &   $25.37\pm0.02$ &   $11.68\pm0.08$ &    $1.55\pm0.01$ &    $7.50\pm0.15$ &      -- \\
MATLAS-658   &   $25.93\pm0.67$ &   $26.71\pm0.67$ &    $9.28\pm1.93$ &    $1.50\pm0.31$ &       -- &      -- \\
MATLAS-799   &   $23.98\pm0.02$ &   $24.76\pm0.02$ &   $17.74\pm0.10$ &    $2.12\pm0.01$ &    $8.33\pm0.17$ &      -- \\
MATLAS-898   &   $24.54\pm0.06$ &   $25.32\pm0.06$ &   $16.92\pm0.34$ &    $1.63\pm0.03$ &    $7.54\pm0.13$ &  $325.6_{-174.3}^{+314.7}$ \\
MATLAS-951   &   $25.23\pm0.02$ &   $26.02\pm0.02$ &   $22.39\pm0.18$ &    $2.87\pm0.23$ &    $8.25\pm0.17$ &   $38.9_{-22.2}^{+38.4}$ \\
MATLAS-984   &   $24.59\pm0.06$ &   $25.37\pm0.06$ &    $9.40\pm0.18$ &    $1.52\pm0.03$ &    $7.52\pm0.16$ &      -- \\
MATLAS-1059  &   $24.30\pm0.02$ &   $25.08\pm0.02$ &   $14.05\pm0.09$ &    $2.29\pm0.02$ &    $8.14\pm0.17$ &      -- \\
MATLAS-1174  &   $25.66\pm0.08$ &   $26.44\pm0.08$ &   $12.38\pm0.33$ &    $2.31\pm0.06$ &    $7.10\pm0.12$ &      -- \\
MATLAS-1177  &   $24.42\pm0.02$ &   $25.21\pm0.02$ &   $13.22\pm0.09$ &    $1.59\pm0.01$ &    $7.86\pm0.16$ &  $208.3_{-118.3}^{+205.2}$ \\
MATLAS-1205  &   $24.817\pm0.006$ &   $25.599\pm0.006$ &    $9.362\pm0.007$ &    $1.795\pm0.001$ &    $7.71\pm0.16$ &      -- \\
MATLAS-1216  &   $25.94\pm0.10$ &   $26.72\pm0.10$ &   $12.35\pm0.38$ &    $2.37\pm0.07$ &    $7.56\pm0.14$ &      -- \\
MATLAS-1225  &   $25.29\pm0.05$ &   $26.07\pm0.05$ &   $18.43\pm0.27$ &    $1.71\pm0.03$ &      -- &      -- \\
MATLAS-1245  &   $25.74\pm0.08$ &   $26.53\pm0.08$ &   $14.75\pm0.39$ &    $2.27\pm0.06$ &    $7.26\pm0.10$ &      -- \\
MATLAS-1246  &   $24.22\pm0.03$ &   $25.01\pm0.03$ &   $13.98\pm0.13$ &    $2.54\pm0.02$ &    $8.45\pm0.17$ &   $32.9_{-18.7}^{+32.4}$ \\
MATLAS-1248  &   $23.95\pm0.02$ &   $24.73\pm0.02$ &   $15.35\pm0.09$ &    $2.79\pm0.02$ &    $8.68\pm0.17$ &   $22.4_{-12.8}^{+22.1}$ \\
MATLAS-1249  &   $25.21\pm0.21$ &   $26.00\pm0.21$ &   $10.91\pm0.70$ &    $1.98\pm0.13$ &    $8.10\pm0.06$ &      -- \\
MATLAS-1262  &   $24.46\pm0.07$ &   $25.24\pm0.07$ &   $11.55\pm0.24$ &    $1.78\pm0.04$ &    $7.95\pm0.14$ &  $390.9_{-214.5}^{+380.8}$ \\
MATLAS-1274  &   $23.96\pm0.02$ &   $24.75\pm0.02$ &   $15.26\pm0.09$ &    $2.78\pm0.02$ &    $8.48\pm0.17$ &      -- \\
\\
\hline
\end{tabular}
\begin{tablenotes}
\footnotesize
\item \textbf{Notes.} This table (continued next page) is available at the CDS. Columns meanings: (1) UDG ID; (2) Central surface brightness in the \textit{g}-band; (3) Average surface brightness within $R_{e}$ in the \textit{g}-band; (4) Effective radius in arcsec in the \textit{g}-band; (5) Effective radius in kpc in the \textit{g}-band computed using the UDG distance when available, otherwise the assumed host distance is used; (6) Stellar mass computed based on the stellar mass-to-light ratios from \citet{Bell2003} and the derived ($g-r$)$_0$ color using the formula $log(M_*/M_{\odot})=-0.306+1.097(g-r)_0-0.4(M_r-4.77)$; (7) Halo-to-stellar mass ratio computed based on the mass ratio from \citet{Harris2017} and the derived GC count using the formula $M_{halo}=M_{GC,tot}/2.9\times10^{-5}$.
\end{tablenotes}
\label{tab:UDGmass1}
\end{table*}

\begin{table*}
\centering
\caption{Additional properties of the MATLAS UDGs (continued).}
\begin{tabular}{lcccccc} 
\hline
ID & $\mu_{0,g}$    & $\langle \mu_{e,g} \rangle$  & $R_{e,g}$   & $R_{e,g}$  & $log(M_*/M_{\odot})$  & $M_{halo}/M_\ast$  \\
    & (mag/arcsec$^2$) & (mag/arcsec$^2$)              & (arcsec)  & (kpc) &  &  \\
(1)   & (2)   & (3)    & (4)   & (5)   & (6)   & (7) \\
\hline
\\
MATLAS-1302  &   $25.90\pm0.13$ &   $26.68\pm0.13$ &    $8.36\pm0.35$ &    $1.52\pm0.06$ &    $7.24\pm0.13$ &      -- \\
MATLAS-1337  &   $24.20\pm0.03$ &   $24.98\pm0.03$ &   $11.08\pm0.10$ &    $1.95\pm0.02$  &  --  &      -- \\
MATLAS-1413  &   $25.94\pm0.07$ &   $26.72\pm0.07$ &   $16.47\pm0.40$ &    $3.28\pm0.08$ &    $8.17\pm0.10$ &      -- \\
MATLAS-1493  &   $24.88\pm0.06$ &   $25.67\pm0.06$ &    $8.12\pm0.16$ &    $1.60\pm0.03$ &    $7.62\pm0.16$ &      -- \\
MATLAS-1494  &   $24.88\pm0.03$ &   $25.66\pm0.03$ &   $13.09\pm0.14$ &    $2.57\pm0.03$ &    $8.03\pm0.17$ &      -- \\
MATLAS-1534  &   $25.21\pm0.06$ &   $26.00\pm0.06$ &    $7.88\pm0.16$ &    $1.53\pm0.03$ &    $7.34\pm0.13$ &      -- \\
MATLAS-1550  &   $25.09\pm0.04$ &   $25.87\pm0.04$ &   $10.50\pm0.15$ &    $1.62\pm0.02$ &    $7.53\pm0.16$ &      -- \\
MATLAS-1558  &   $25.83\pm0.05$ &   $26.61\pm0.05$ &   $14.84\pm0.25$ &    $2.28\pm0.04$ &    $7.77\pm0.15$ &  $250.3_{-139.6}^{+245.1}$ \\
MATLAS-1589  &   $25.73\pm0.06$ &   $26.51\pm0.06$ &   $11.61\pm0.24$ &    $1.72\pm0.04$ &    $7.29\pm0.15$ &      -- \\
MATLAS-1606  &   $24.435\pm0.003$ &   $25.217\pm0.003$ &   $12.713\pm0.008$ &    $2.306\pm0.002$ &      -- &      -- \\
MATLAS-1615  &   $24.58\pm0.04$ &   $25.36\pm0.04$ &   $15.90\pm0.17$ &    $2.88\pm0.03$ &    $8.39\pm0.17$ &  $167.4_{-95.3}^{+165.0}$ \\
MATLAS-1616  &   $25.12\pm0.08$ &   $25.90\pm0.08$ &   $15.67\pm0.22$ &    $2.30\pm0.03$ &    $7.95\pm0.16$ &  $160.1_{-90.3}^{+157.3}$ \\
MATLAS-1630  &   $25.18\pm0.08$ &   $25.97\pm0.08$ &   $12.52\pm0.31$ &    $1.85\pm0.05$ &    $8.04\pm0.15$ &      -- \\
MATLAS-1647  &   $25.81\pm0.07$ &   $26.59\pm0.07$ &    $8.30\pm0.21$ &    $1.51\pm0.04$ &    $7.65\pm0.15$ &      -- \\
MATLAS-1779  &   $25.65\pm0.06$ &   $26.44\pm0.06$ &   $15.30\pm0.22$ &    $2.90\pm0.04$ &    $7.80\pm0.02$ & $912.3_{-432.4}^{+852.0}$ \\
MATLAS-1794  &   $24.95\pm0.07$ &   $25.73\pm0.07$ &   $12.39\pm0.29$ &    $1.72\pm0.04$ &      -- &      -- \\
MATLAS-1824  &   $24.36\pm0.04$ &   $25.14\pm0.04$ &   $12.37\pm0.14$ &    $2.81\pm0.03$ &    $7.65\pm0.17$ &      -- \\
MATLAS-1847  &   $25.40\pm0.11$ &   $26.18\pm0.11$ &   $17.12\pm0.49$ &    $1.94\pm0.06$ &    $7.66\pm0.12$ &  $352.0_{-187.5}^{+339.7}$ \\
MATLAS-1855  &   $24.84\pm0.13$ &   $25.63\pm0.13$ &   $13.90\pm0.48$ &    $1.57\pm0.05$ &    $7.41\pm0.13$ &  $556.2_{-297.5}^{+537.5}$ \\
MATLAS-1865  &   $24.35\pm0.03$ &   $25.13\pm0.03$ &   $11.80\pm0.13$ &    $1.55\pm0.02$ &    $7.38\pm0.18$ & $1897.1_{-1095.8}^{+1879.3}$ \\
MATLAS-1907  &   $24.21\pm0.02$ &   $24.99\pm0.02$ &   $13.15\pm0.07$ &    $1.552\pm0.009$ &    $7.85\pm0.18$ &      -- \\
MATLAS-1957  &   $24.07\pm0.14$ &   $24.85\pm0.14$ &   $10.67\pm0.22$ &    $1.63\pm0.03$ &      -- &      -- \\
MATLAS-1975  &   $25.18\pm0.24$ &   $25.96\pm0.24$ &   $12.54\pm0.91$ &    $1.61\pm0.12$ &    $7.40\pm0.04$ &  $316.5_{-152.8}^{+297.0}$ \\
MATLAS-1985  &   $25.11\pm0.03$ &   $25.90\pm0.03$ &   $14.48\pm0.13$ &    $1.87\pm0.02$ &      -- &      -- \\
MATLAS-1991  &   $24.01\pm0.02$ &   $24.79\pm0.02$ &   $12.52\pm0.07$ &    $1.538\pm0.009$ &    $8.06\pm0.15$ &  $140.8_{-78.7}^{+137.9}$ \\
MATLAS-1996  &   $26.27\pm0.19$ &   $27.05\pm0.19$ &   $14.87\pm0.45$ &    $1.58\pm0.05$ &      -- &      -- \\
MATLAS-2019  &   $24.42\pm0.04$ &   $25.21\pm0.04$ &   $17.16\pm0.21$ &    $1.69\pm0.02$ &    $7.99\pm0.15$ &  $930.7_{-519.9}^{+911.8}$ \\
MATLAS-2021  &   $26.06\pm0.19$ &   $26.84\pm0.19$ &   $17.79\pm0.91$ &    $1.89\pm0.10$ &    $6.49\pm0.04$ & $727.6_{-351.0}^{+682.7}$ \\
MATLAS-2103  &   $24.27\pm0.02$ &   $25.05\pm0.02$ &   $14.25\pm0.10$ &    $2.49\pm0.02$ &      -- &      -- \\
MATLAS-2184  &   $25.17\pm0.16$ &   $25.96\pm0.16$ &   $11.61\pm0.54$ &    $1.66\pm0.08$ &    $6.96\pm0.13$ &      -- \\
\\
\hline
\end{tabular}
\begin{tablenotes}
\footnotesize
\item \textbf{Notes.} This table (including previous page) is available at the CDS. Columns meanings: (1) UDG ID; (2) Central surface brightness in the \textit{g}-band; (3) Average surface brightness within $R_{e}$ in the \textit{g}-band; (4) Effective radius in arcsec in the \textit{g}-band; (5) Effective radius in kpc in the \textit{g}-band computed using the UDG distance when available, otherwise the assumed host distance is used; (6) Stellar mass computed based on the stellar mass-to-light ratios from \citet{Bell2003} and the derived ($g-r$)$_0$ color using the formula $log(M_*/M_{\odot})=-0.306+1.097(g-r)_0-0.4(M_r-4.77)$; (7) Halo-to-stellar mass ratio computed based on the mass ratio from \citet{Harris2017} and the derived GC count using the formula $M_{halo}=M_{GC,tot}/2.9\times10^{-5}$.
\end{tablenotes}
\label{tab:UDGmass2}
\end{table*}

\end{document}